# DEVELOPMENT OF NOVEL LIGHT PROPAGATION ALGORITHMS IN TURBID MEDIA WITH VARYING OPTICAL HETEROGENEITY

*A Philosophiae Doctoral Thesis:*

## Daniele Ancora

*May 22nd 2017*
Final Version

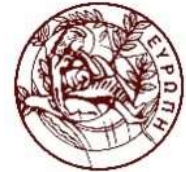

**University of Crete - Πανεπιστήμιο Κρήτης**
**Department of Materials Science and Technology**

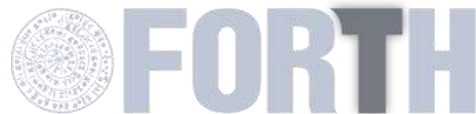

**Foundation for Research and Technology – Hellas**
**Institute of Electronic Structure and Laser**

**Academic Supervisors:**
*Maria Kafesaki* (University of Crete)
*Jorge Ripoll* (Universidad Carlos III de Madrid)
*Chrysoula Tsogka* (Stanford University)

**Scientific Supervisor:**
*Giannis Zacharakis*

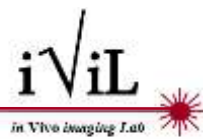

Supplementary title:

# LIGHT PROPAGATION IN EXTREME CONDITIONS:

## THE ROLE OF OPTICALLY CLEAR TISSUES AND SCATTERING LAYERS IN OPTICAL BIOMEDICAL IMAGING

*A Philosophiae Doctoral Thesis:*

### Daniele Ancora

*May 22nd 2017*
Final Version

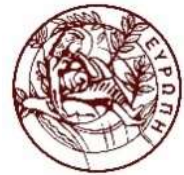

**University of Crete - Πανεπιστήμιο Κρήτης**
**Department of Materials Science and Technology**

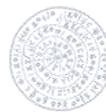

**Foundation for Research and Technology – Hellas**
**Institute of Electronic Structure and Laser**

**Academic Supervisors:**
*Maria Kafesaki* (University of Crete)
*Jorge Ripoll* (Universidad Carlos III de Madrid)
*Chrysoula Tsogka* (Stanford University)

**Scientific Supervisor:**
*Giannis Zacharakis*

**Supervising Members of the Committee:**

Maria Kafesaki (University of Crete)

Jorge Ripoll Lorenzo (Universidad Carlos III de Madrid)

Chrysoula Tsogka (Stanford University)

Giannis Zacharakis (Foundation for Research and Technology - Hellas)

Dimitris Papazoglou (University of Crete)

Charalambidis Dimitris (University of Crete)

Vangelis Sakkalis (Foundation for Research and Technology - Hellas)

To Sara,

the brightest light I have ever met

and to my Family,

that always trusted me more than what I was able of.

# Contents

# ABSTRACT

The field of biomedical imaging has experiences a rapid growth in recent years driven by i) the increased demand for better disease detection and therapy monitoring and ii) the desire to visualize biology even at the nanoscopic level. This growth has been supported by the implementation of ad-hoc designed experimental systems and related theoretical and computational/numerical support methods. In this dynamic environment, the continuous medical request for harmless imaging probes and higher resolution, has ultimately pushed the imaging research community towards the developing of novel techniques in the optical wavelength regime. The high resolution, especially in microscopy, and the flexibility in the realization of optical setups favored the kick-start of optical imaging techniques, which have finally met their main challenge into the highly scattering of light in biological tissue. Especially for biological samples, the numerous scattering events occurring during the photon propagation process limit the penetration depth and the possibility to perform direct imaging in thicker and not transparent samples. To overcome this limitation, numerous theoretical strategies where proposed to isolate the scattering contribution, minimize the image blurring and reduce the speckled noise due to the random light-path scrambling induced by the complex variation of refractive index in biological tissues. In this thesis, we will examine theoretically and experimentally the scattering process from two opposite points of view, tackling at the same time specific challenges in optical imaging science.

We start by examining the light propagation in diffusive biological tissues considering the particular case of the presence of optically transparent regions enclosed in a highly scattering environment. We will point out how, the correct inclusion of this information, can ultimately lead to higher resolution reconstructions and especially aiming at brain tumor neuroimaging. We examined in details the increased accuracy in the forward modelling of the fluorescent emission of spherical tumor distributions in a mouse head, in particular if compared with other currently used techniques.

We then examine the extreme case of the three-dimensional imaging of a totally hidden sample, in which the phase has been scrambled by a random scattering layer. By using appropriate numerical methods, we prove the possibility to perform such hidden reconstructions in a very efficient way, opening the path toward the unexplored field of three-dimensional hidden imaging. We present how, the properties described while addressing these challenges, lead us to the development of a novel alignment-free three-dimensional tomographic technique that we refer to as Phase-Retrieved Tomography. We have proved this method theoretically and used it for the study of the fluorescence distribution in a three-dimensional spherical tumor model, the multicellular cancer cell spheroid, one of the most important biological models for the study of such a complex disease. We finally conclude our study, by imaging spherical tumors under two extremely different experimental conditions, improving the depth to resolution ratio of the current state of the art in live microscopic imaging, as defined by Light Sheet Fluorescence Microscopy.

Throughout the whole doctoral period, these studies have been stimulating and creating new questions and ideas, which will be discussed in the following and that form the natural continuation of the projects exposed in the present thesis.

# INTRODUCTION

Biomedical optical imaging is one of the research fields that have undergone the fastest growing process in recent years. The reason for this can be found in a very broad variety of factors that stimulated the developing of novel optical imaging techniques. Among others, the need for nearly non-invasive imaging probes that can be exploited for in-vivo functional and disease medical detection, favorited the usage of optical wavelength sources. In fact optical photons, having energies in the range of $1 - 10\ eV$, deposit very little energy to the investigated tissues, if for example compared with x-rays in the range of $10 - 100\ keV$, thus strongly reducing tissue damage or sample alteration due to the measuring modality itself. Among all the plethora of available light-generating devices, laser sources are dominating the most recent biomedical imaging applications where high intensity, coherence, and tunability of the sources are needed to accomplish imaging at regimes not accessible before. The interest for the optical imaging modalities was further increased, by the discovery of least toxic fluorescent molecules that can specifically label a certain class of biological molecules and processes. However, even if, the availability of good sources and imaging probes is strictly indispensable for imaging purposes, nearly nothing could have been done without the support of modern computer architectures and numerical approaches. Since the invention of the microscope, which allowed direct observation of transmitting specimens (usually sandwiched in thin glass slides), everything has been revolutionized with the advent of high performance computers. One of the most important possibilities offered by the use of advanced numerical methods for imaging purposes, is the possibility of reinterpreting and analyzing the optical signals detected after it experienced propagation through the specimen of interest. By doing so and in coordination with appropriate theory developed ad-hoc, it has been possible to develop three-dimensional, functional and even sub-diffraction limited imaging modalities also in the case of non-transparent, or even turbid, samples. In fact, the main obstacle in the process of the translation from the bench to the bedside for many of the optical imaging modalities is the scattering phenomena.

Scattering, at optical wavelength strongly limits (or for thick samples impedes) the possibility of outperforming direct observation of the specimen of interest and that is why the usage of numerical methods comes into play. By exploiting the theoretical knowledge of the laws that rule the photon propagation under different experimental situations, it is possible to infer what source generated the signal detected under a given excitation. This is the modern paradigm of computational imaging techniques, in which the sample can be seen as a sort of unknown entity that acts as a generic transformation of the input light signal; thus returning a modified output signal that we can record with specific devices. The signal detected then, contains (only in some cases) enough information that can be exploited to obtain a meaningful reconstruction of the object itself. This approach often classifies modern imaging methodologies under the category of inverse problems, in which starting from the results we calculate the source that has generated them. The main difficulty behind this approach is that due to the intense scattering and diffusion the problem to define is far from being well-posed. According to Hadamard definition, a problem is well-posed if it admits a solution, if this solution is unique and changes continuously with the changes of the initial conditions. Light scattering, in fact, in many cases makes (even one of) those assumptions to fail, thus pushing the problem towards the ill-posed space. The most important task in modern biomedical optical imaging research, then, is to tackle the complexity of the problem by finding new ways to reduce the ill-posedness ultimately leading to accurate, high resolution and quantitative reconstructions. Throughout the following thesis, in fact, we will examine how we the ill-posedness can be reduced in some of the most interesting imaging scenarios. We will accomplish this by modelling the light diffusion with more accurate and general approaches, by relaxing the strictness on the uniqueness of the solution and exploiting this fact at our advantage, increasing the resolution or obtaining high quality tomographic reconstructions even in highly difficult experimental conditions. In fact, the word "tomography" (from τόμος "slice" and γράφω "to write") will be the leading point along the whole text, the ultimate goal of our research path.

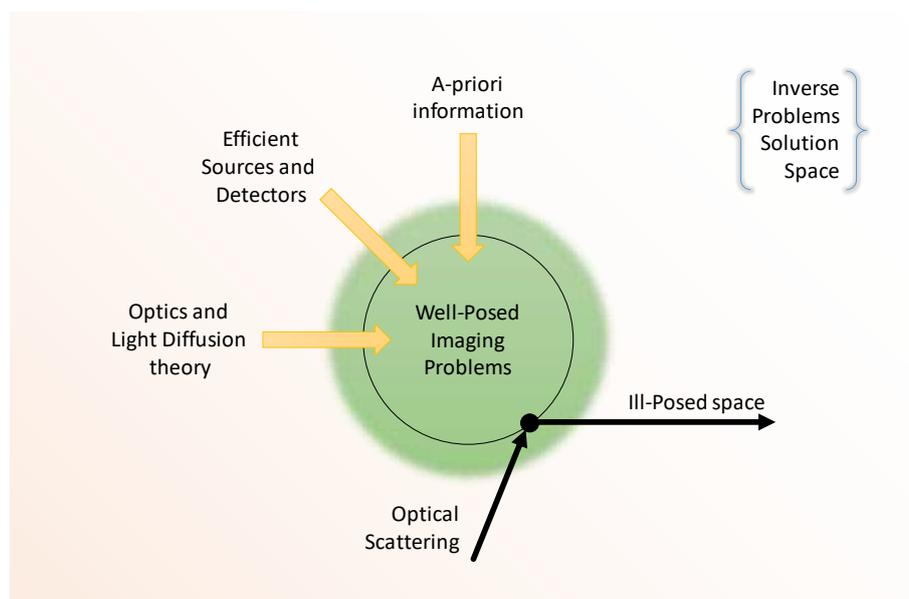

*Figure 0-1 | A representation of the optical imaging problem functional to the concepts behind the present thesis. Throughout the whole text, we will focus our attention, in fact, in the optical scattering phenomena and how this creates difficulties in the computed tomography imaging process. Then, we will find novel stratagems to overcome or even exploit what seems to be a not wanted phenomena, unlocking novel concepts in biomedical imaging modalities.*

In this thesis we will focus mainly on two different tomographic approaches, situated at the extreme poles of the light propagation in biological tissues. We will examine in detail the condition of imaging in highly diffusive media in presence of transparent tissues and we will also treat the case of three-dimensional imaging behind highly scattering layers, analyzing its important implications. Due to the duality of the work presented, we will firstly introduce their underling theory in two self-consistent and separated chapters. In each of them, we will derive or recall some of the fundamentals behind the principles of light propagation in biological tissues and autocorrelation imaging modalities, thus opening the path for the presentation of the original work produced in this dissertation. Two chapters are dedicated to the theory and further two are dedicated to the original research carried out during the doctoral period. The following text can be read in two different, but complementary, ways: linearly from the top down to the last page or by reading the four chapters in an alternated fashion. In fact, *Chapter I* and *Chapter III* form the bulk of the work related to the study of transparent layers embedded in scattering tissues, while *Chapter II* and *Chapter IV* present the hidden imaging modalities and the tomographic imaging technique that was developed as a consequence of this study. The reason behind this choice has to be searched in the fact that the following thesis is presented in the exact same order as those studies were approached, with the explicit intention to retrace the natural evolution of three years of stimulating work.

In *Chapter III* we will present the work related to the neuroimaging of a mouse brain. We will study how the presence of an optically transparent tissue surrounding the brain, the Cerebral Spinal Fluid, importantly affects the light propagation. We will find how the correct modelling of this tissue by using the Monte Carlo Photon Propagation technique can result into higher resolution imaging, if compared to that of normally used Diffusion Equation. In fact, such structure is commonly neglected in Diffuse Optical Tomography reconstructions, due to the fact that in it the diffusion approximation fails. We will point out how, only recently, diffuse tomography can make use of the Monte Carlo methods thanks to modern GPU paradigms that allow fastest simulation, otherwise too slow for practical imaging applications. An ideal neuroimaging experiment was virtually reproduced to study the fluorescent emission of a spherical brain tumor in presence of the clear layer, comparing the results with classical approaches that would neglect such structure. We found this structure to be of extremely high importance, thus making the clear layers fundamental to aim high quality reconstructions. Not only this, but we realized also that its thickness influenced the results, opening the paths towards a very interesting project related to early Alzheimer's disease detection. In fact, in dementia related diseases there is effectively a brain mass loss and an increment of the thickness for the Cerebral Spinal Fluid, which can potentially make this study of high interest for monitoring of the disease itself. Even if a very first work around such a study is ready to be submitted to the scientific community, we will not talk about this in this text, we will just mention it as a natural consequence of this work.

In *Chapter IV* instead, we will present our work related to the hidden three-dimensional imaging and its consequence. We started approaching this problem after some very interesting works related to hidden bi-dimensional imaging where recently released. Imaging through scattering layer, in fact, could be of extremely high interest for bioimaging applications, and this is what stimulated us to fulfill the gap of hidden object tomography. We made use of novel autocorrelation imaging applied in a three-dimensional fashion, firstly numerically proving the feasibility of three-dimensional imaging and then stepping forward towards the creation of another, yet complementary, tomographic application. We noticed in fact that even if autocorrelation imaging underlines a not unique phase-retrieved solution, it is always possible to automatically produce perfectly aligned three-dimensional reconstructions. We have proven this theoretically, numerically and experimentally, finally imaging the fluorescence emission of a tumor spheroid with a new technique that we named Phase-Retrieved Tomography, which we will extensively describe in the following.

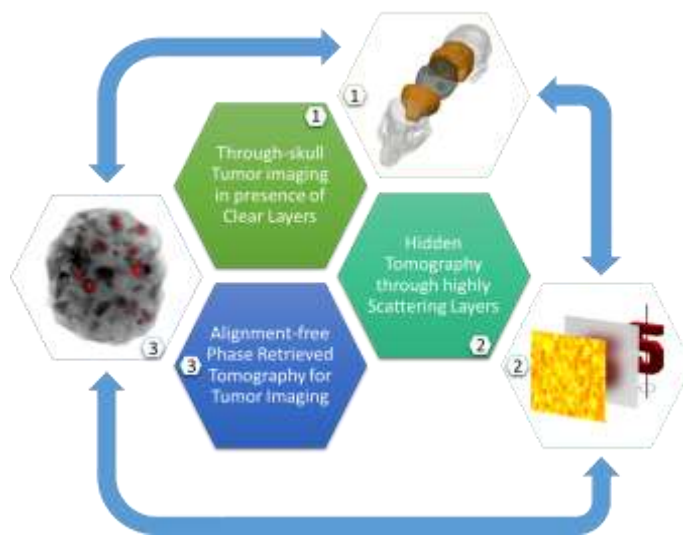

*Figure 0-2 | Schematics showing the link between the different part of the work presented in chapter III and IV. With this illustration we want to emphasize the natural evolution of the work behind this thesis project.*

All the studies presented in this thesis are published either as peer-reviewed conference proceedings, journal articles or arXiv preprints. Two works briefly treated in Chapter II were published, as result of the collaboration with my colleague Diego Di Battista, in the journal of **Applied Physics Letters** [Di Battista et al., "Tailoring non-diffractive beams from amorphous light speckles," Appl. Phys. Lett. **109**, 121110 (2016), doi: 10.1063/1.4962955] and in the **Optica** open-access journal [Di Battista et al., "Tailored light sheets through opaque cylindrical lenses," Optica **3**, 11 (2016), doi: 10.1364/OPTICA.3.001237] under the form of a letter. For both of the works above mentioned, I have contributed by performing experimental measurements and its consequent data analysis, as well as developing appropriate numerical methods and image processing. More importantly for the framework of the present thesis, the work presented in *Chapter III* was published in the peer-reviewed journal **IEEE *Transaction on Medical Imaging*** [Ancora et al., "Fluorescence Diffusion in the Presence of Optically Clear Tissues in a Mouse Head Model", IEEE Trans. Med. Imag. **36**, 5, (2017) doi: 10.1109/TMI.2016.2646518] and a further article, conceived after these findings, is going to be submitted to ***NeuroImage*** thanks to the collaboration with Prof. Antonio Pifferi from Politecnico di Milano. The results in *Chapter IV* are in the **e-Print – *arXiv*** repository [Ancora et al., "Phase-Retrieved Tomography enables imaging of a Tumor Spheroid in Mesoscopy Regime"**,** arXiv: 1610.06847 (2016)] and currently under final review process in the peer-reviewed journal of ***Scientific Reports***. Moreover, these results presented in an oral talk in the international conference **SPIE - *Photonics West 2017*** [Ancora et al., "Optical projection tomography via phase retrieval algorithms for hidden three-dimensional imaging," Proc. SPIE **10074**, doi: 10.1117/12.2252894 (2017)] awarded an invitation to write an article in the invited-only journal ***Methods***, which we are going to submit in the very near future. *Chapter III* and *IV*, then, constitute the bulk of the present thesis and for which I have carried out most of the work: numerical simulations, data analysis, experimental measurements, image processing and article writing, in fact, were done by myself, constantly tuning them accordingly to continuous feedback exchange with the other researches involved.

Even if apparently not closely related, the two parts (*Chapters III* and *IV*) of this work are intrinsically connected by the implicit will to advance through-skull imaging. A visual guide for this can be found in Figure 0-2. In fact, the skull can be seen as a scattering layer surrounding the brain and so, in a certain way, we approached two modalities to image through it. Firstly focusing on spherical tumors in the brain boxed in the outer skull, then considering a three-dimensional scattering layer enclosing a generic hidden fluorescent object and, finally, using these results to realize an innovative technique that we experimentally proven with the imaging of a tumor spheroid. Of course, we are aware that at this stage of the work it is not possible to merge the two approaches; in fact, although there is an optically clear layer tissue behind the skull, the brain is itself a highly scattering structure which will definitely affect the photon propagation. This work, then, does not have the ambition to be a closed research path, but instead leaves opened some new interesting challenges that we will try to explore in the future years of research activity. We are now ready to start covering our journey, giving a first description on how light gets diffused in biological media and the general concepts behind the approximations made.

# Chapter I
# THEORY OF LIGHT DIFFUSION IN BIOLOGICAL TISSUES

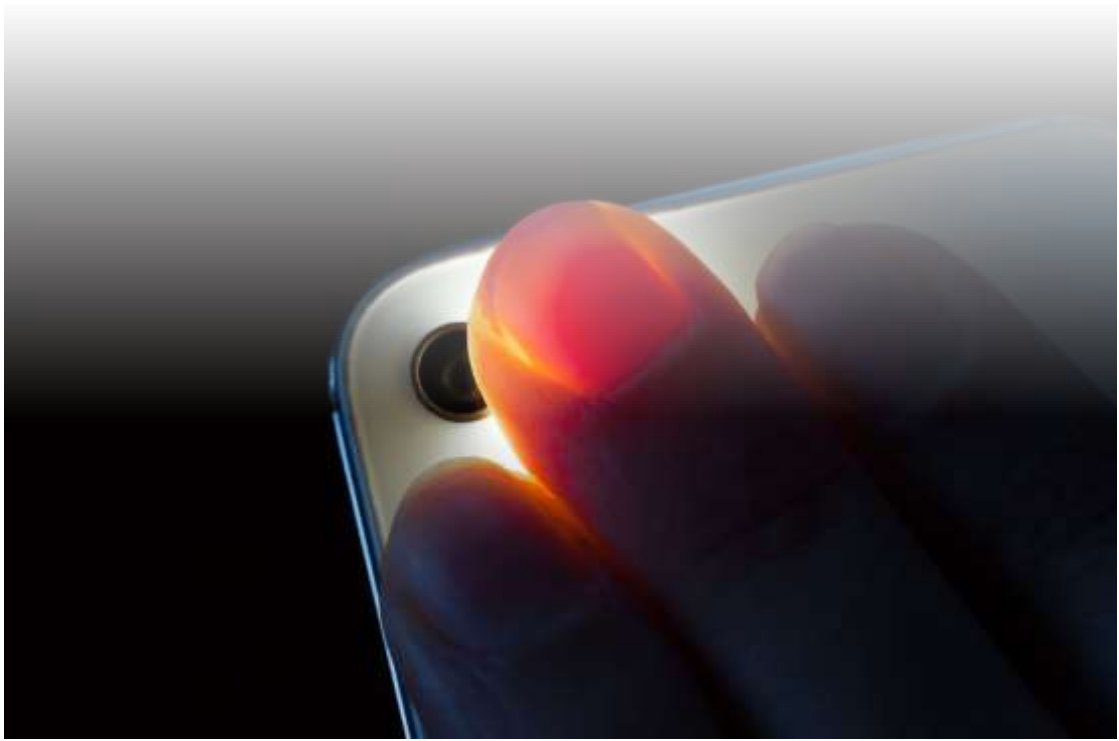

# I.1 Light Diffusion in Biological Tissues

When an electromagnetic excitation is generated at a certain location in space and time, a field starts its journey in the form of a propagating electromagnetic radiation. The entities that carry such a field are commonly referred to as *photons* and they are responsible for the interaction between the field and the matter. Among all the possible wavelengths $\lambda$ that a photon can carry, we, as human beings, are pushed to consider as light that part of the electromagnetic spectrum which our visual system can detect and interpret. Such innate idea is somehow confined in a very narrow window among the vastness of all the possible radiations that can be propagated by the photons. In fact, the region of the electromagnetic spectrum that we can appreciate, the visible light, ranges from wavelengths of around $400\ nm$ up to $700\ nm$ (Figure I-1). Different wavelengths, at this regime, produce in our eyes the perception of different colors, which is tightly bounded with the different kind of light interactions with matter: absorption, emission and scattering. Without going deeper into further details, of which beautiful explanations can be found in [1] and [2], we will start our discussion focusing on how the light propagates through biological tissues and how the field of biomedical imaging aims at its description.

Initially, we have to briefly introduce a few important quantities that will guide us through the reading of the present chapter. Let us start our essay considering the generic electric $\mathbf{E}$ and magnetic $\mathbf{H}$ vector fields defined in a locally isotropic medium, characterized by its dielectric constant $\epsilon$ and magnetic permeability $\mu$, together with the density current $\mathbf{j}$. The formulations that rule their relations are described by the classical formulation of the Maxwell's equations [3]:

$$\nabla \times \mathbf{H} - \frac{\epsilon}{c}\frac{\partial \mathbf{E}}{\partial t} = \frac{4\pi}{c}\mathbf{j} \tag{1}$$

$$\nabla \times \mathbf{E} + \frac{\mu}{c}\frac{\partial \mathbf{H}}{\partial t} = 0 \tag{2}$$

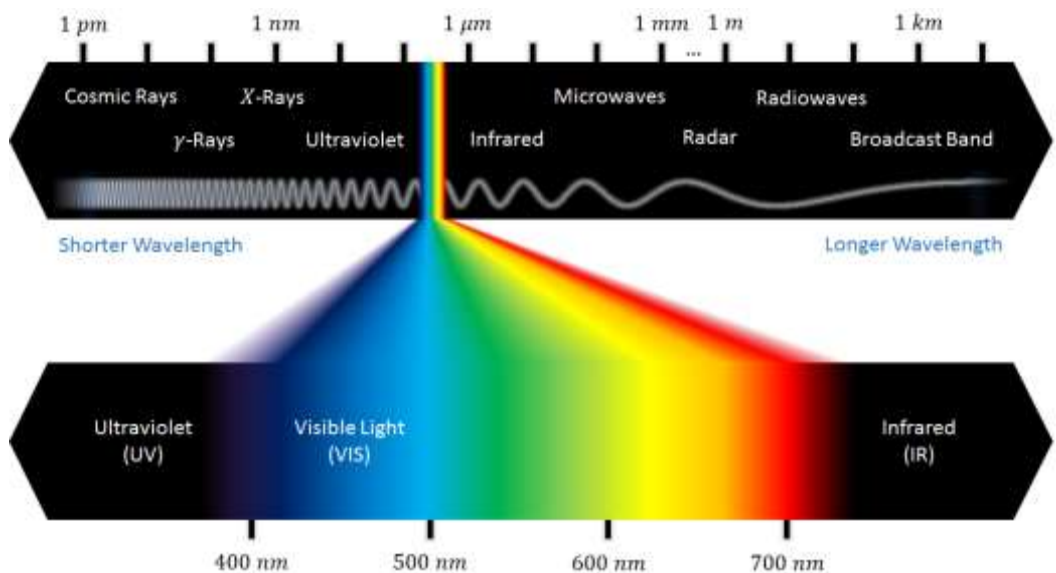

*Figure I-1 | Electromagnetic spectrum at different ranges of wavelengths. The visible range of optical wavelengths is enhanced in the lower part. In this regime we will conduct our work.*

of which their sum can be rewritten in the form of:

$$\frac{1}{4\pi}\left(\epsilon\mathbf{E}\cdot\frac{\partial\mathbf{E}}{\partial t}+\mu\mathbf{H}\cdot\frac{\partial\mathbf{H}}{\partial t}\right)+\mathbf{j}\cdot\mathbf{E}+\frac{c}{4\pi}\nabla\cdot(\mathbf{E}{\times}\mathbf{H})=0. \qquad (3)$$

Now it worth a little consideration to the quantities present in the above formulation. The first term it can be rewritten as:

$$\frac{1}{4\pi}\left(\epsilon\mathbf{E}\cdot\frac{\partial\mathbf{E}}{\partial t}+\mu\mathbf{H}\cdot\frac{\partial\mathbf{H}}{\partial t}\right)=\frac{1}{8\pi}\frac{\partial}{\partial t}(\epsilon\mathbf{E}^2+\mu\mathbf{H}^2)=\frac{\partial}{\partial t}W \qquad (4)$$

where we defined the *energy density* ($Joules/cm^3$) in the form of

$$W=\frac{1}{8\pi}(\epsilon\mathbf{E}^2+\mu\mathbf{H}^2). \qquad (5)$$

In these terms, the first part represents the variation in time of the energy density, constituted by the sum of two contribution dependent on the electric and magnetic fields. The second term, instead, represents the absorbed energy per unit volume ($Watts/cm^3$) or equivalently the Joule's heat, which in fact describes the resistive energy dissipation:

$$\frac{dP_{abs}}{dV}=\mathbf{j}\cdot\mathbf{E}. \qquad (6)$$

Finally, the last term represents the (local) net *energy flow* and usually we write it in the form of:

$$\frac{c}{4\pi}\nabla\cdot(\mathbf{E}{\times}\mathbf{H})=\nabla\cdot\mathbf{S}, \qquad (7)$$

where S is commonly referred to as the *Poynting vector* ($Watts/cm^2$):

$$\mathbf{S}=\frac{c}{4\pi}\mathbf{E}{\times}\mathbf{H}. \qquad (8)$$

We can now discuss a few points concerning this very important quantity. Among the other considerations that we might notice, the Poynting vector indeed represents tightly what we are used to associate with an experimental intensity measurement. In fact, what we commonly measure in the lab, it is not directly the electric or the magnetic field (and neither their phase, as we will discuss in the next chapter) but instead the flow of energy that is passing through the detector area $S$ during a finite time-interval. What we see by a camera detection is, then, the integral of the Poynting vector over the detector area, allowing us to consider **S** as the amount of energy that crosses a surface having its normal vector parallel to the direction of **E**×**H**.

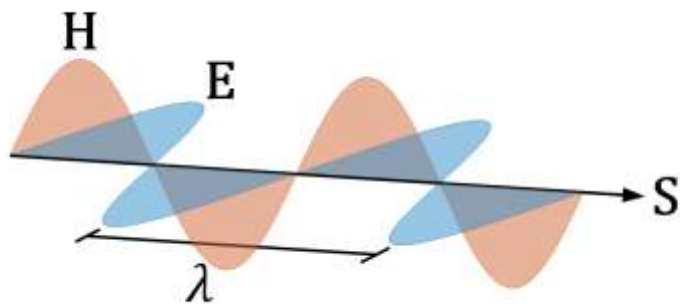

*Figure I-2 | Representation of the Poynting vector in space. If we consider E and H perpendicular to each other, the Poynting vector (i.e. the directionality of the flow of energy in space) is perpendicular to both the electric and magnetic field.*

With the above considerations in mind, we can rewrite the energy conservation equation as:

$$\frac{\partial W}{\partial t} + \frac{dP_{abs}}{dV} + \nabla \cdot \mathbf{S} = 0, \tag{9}$$

which is referred to as the *Poynting's Theorem*. Such formulation can also be found by analyzing the energy balance and from it we can roll back to the definition of the Poynting vector. In the latter the Poynting vector assumes another interesting interpretation: in these terms, **S** represents the energy transfer generated by the oscillation of the electric and magnetic fields, being such transfer perpendicular to the fields that generate it.

Such vector is, then, a very important quantity and it oscillates at frequency equal to that of the electromagnetic field. Since it does not exist a measuring device that can follow such frequency (having a period of $T = 2\pi/\omega$), the quantity to which we have access in direct measurements is its time-average integral:

$$\langle \mathbf{S} \rangle = \frac{1}{2T'} \int_{-T'}^{T} \frac{c}{4\pi} \left[ \mathbf{E}(\boldsymbol{r}, t) \times \mathbf{H}(\boldsymbol{r}, t) \right] dt. \tag{10}$$

It is worth to notice that, due to the time-averaging, $\langle \mathbf{S} \rangle$ has inexorably lost any phase-related information; we will get back to this point in the following chapter, while at the moment proceed further with the treatise.

By making an assumption on the time-harmonic dependence for the fields in the form of $\mathbf{E}_t(\boldsymbol{r}, t) = \mathbf{E}(\boldsymbol{r})e^{-i\omega t}$ and $\mathbf{H}_t(\boldsymbol{r}, t) = \mathbf{H}(\boldsymbol{r})e^{-i\omega t}$ and considering our measurement to be in the far-field (where the electromagnetic wave propagates as a planar wave) we can arrive to write explicitly:

$$\langle \mathbf{S} \rangle = \frac{\epsilon c_0}{8\pi} |\mathbf{E}|^2 \hat{\mathbf{s}}. \tag{11}$$

While writing this formulation, we have considered the electric and magnetic field to be mutually orthogonal, which allows us to write $\sqrt{\mu}H = \sqrt{\epsilon}E$, and with $c_0 = c/n_0$ being the speed of light in the medium of refractive index $n_0 = \sqrt{\epsilon\mu}$. In this case, then, we may write the fields as $\mathbf{H} = H\hat{\mathbf{h}}$ and $\mathbf{E} = E\hat{\mathbf{e}}$, so that we can express the vector that defines the directionality of **S**:

$$\hat{\mathbf{s}} = \hat{\mathbf{e}} \times \hat{\mathbf{h}}. \tag{12}$$

At this stage, we have everything to define two important quantities that will characterize the medium: the absorption and scattering cross-section. We will not derive them explicitly, for this we refer to a more appropriate reading [2], but we will present their qualitative description.

First of all, let us consider a particle that can absorb and/or scatter part of the incident energy $\left| \langle \mathbf{S}^{(inc)} \rangle \right|$, and which has a geometrical cross-section of area $A$. The absorption cross-section $\sigma_a$, then, is defined as the ratio between the effectively absorbed power $\bar{P}_{abs}$ and the total power incident onto the particle per unit area $\left| \langle \mathbf{S}^{(inc)} \rangle \right| = \bar{P}_{abs}/A$:

$$\sigma_a = \frac{\bar{P}_{abs}}{|\langle \mathbf{S}^{(inc)} \rangle|} = 4\pi k \int_V \frac{\epsilon^{(i)}(\mathbf{r})|E(\mathbf{r})|^2}{\epsilon|E_0|^2}. \tag{13}$$

In this formulation, we have made use of $k = \omega/c_0$ and $\epsilon^{(i)}(\mathbf{r})$ which is the imaginary part of the permittivity $\epsilon(\mathbf{r}) = \epsilon^{(r)}(\mathbf{r}) + i\epsilon^{(i)}(\mathbf{r})$. The absorption cross-section just defined has units of $(cm^2)$ and it represents the effective size that the electromagnetic field sees during the interaction with it (being attenuated) compared to its geometrical dimension. As an example, we could say that a transparent particle will have $\sigma_a \ll A$. In complete analogy with what we described for absorption, we can define the scattering cross-section as the ratio between the scattered power $\bar{P}_{sc}$ and the total incident power per unit area:

$$\sigma_s = \frac{\bar{P}_{sc}}{|\langle \mathbf{S}^{(inc)} \rangle|} = \int_V \frac{\nabla \cdot \langle \mathbf{S}^{(sc)} \rangle}{|\langle \mathbf{S}^{(inc)} \rangle|} dV = \int_S \frac{\langle \mathbf{S}^{(sc)} \rangle \cdot \mathbf{n}}{|\langle \mathbf{S}^{(inc)} \rangle|} dS, \tag{14}$$

where $\langle \mathbf{S}^{(sc)} \rangle$ is the scattered flow of energy. Also the scattering cross-section has units of $(cm^2)$ and it represent the effective size of the particle that deviates the incident electromagnetic field from its original direction. We can also define the *total extinction cross-section* as the sum of the above definition: $\sigma_t = \sigma_a + \sigma_s$.

Before to conclude this section, it is worth to define other two quantities which will be of high interest throughout the following work. First of all, let us spend a few words about the function that is connected with the probability of a photon, travelling in a certain direction $\hat{\mathbf{s}}_0$, to be scattered towards a generic direction in space $\hat{\mathbf{s}}$. This function it is called the *phase function $p(\hat{\mathbf{s}}, \hat{\mathbf{s}}_0)$* and gives important information regarding the directionality of the scattering event. Among others, there are a few explicit formulas which are commonly used in literature to describe some specific classes of scattering events:

- *Isotropic phase function*, when the scattering direction is uniformly distributed along the whole solid angle:

$$p(\hat{\mathbf{s}}, \hat{\mathbf{s}}_0) = \frac{W_0}{4\pi}. \tag{15}$$

- *Rayleigh's phase function*, particularly useful when treating the scattering due to molecules and small aerosol particles:

$$p(\hat{\mathbf{s}}, \hat{\mathbf{s}}_0) = p(\hat{\mathbf{s}} \cdot \hat{\mathbf{s}}_0) = \frac{3}{16\pi}(1 + (\hat{\mathbf{s}} \cdot \hat{\mathbf{s}}_0)^2)W_0. \tag{16}$$

- *Henyey-Greenstein's (HG) phase function*, originally derived for interstellar scattering but then adopted also by the biomedical optics community, since it describes quite well the light-tissue interaction:

$$p(\hat{\mathbf{s}}, \hat{\mathbf{s}}_0) = p(\hat{\mathbf{s}} \cdot \hat{\mathbf{s}}_0) = \frac{1}{4\pi} \frac{W_0(1 - g^2)}{\left(1 + g^2 - 2g(\hat{\mathbf{s}} \cdot \hat{\mathbf{s}}_0)\right)^{3/2}}. \tag{17}$$

We can notice, however that the scalar product present in the phase functions is $\hat{\mathbf{s}} \cdot \hat{\mathbf{s}}_0 = \cos\theta$, being $\theta$ the angle at which the photon has been scattered. In the HG phase function, the term $g$ is called anisotropy coefficient and it can be related to the average cosine of the scattering angle at which the incident photons are deflected:

$$g = \langle \cos\theta \rangle = \langle \hat{\mathbf{s}} \cdot \hat{\mathbf{s}}_0 \rangle \tag{18}$$

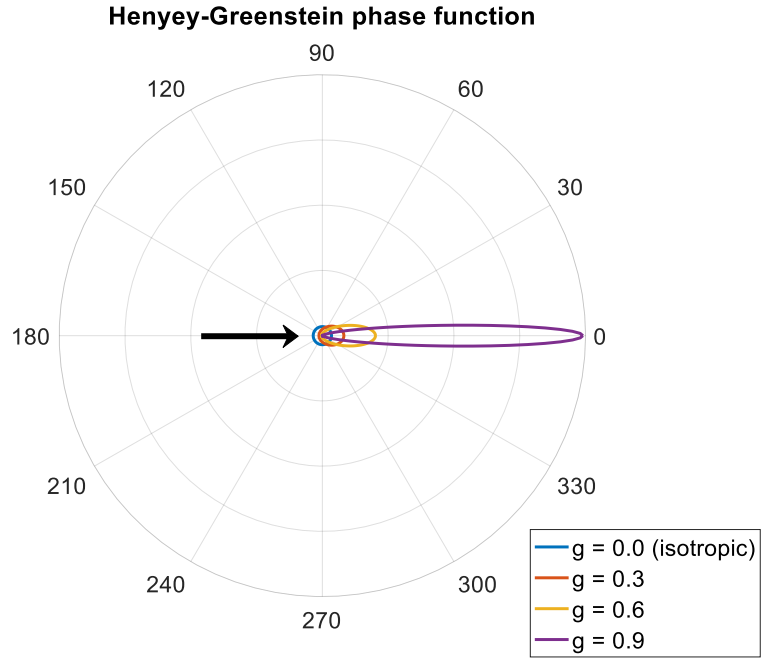

*Figure I-3| **Light Scattering phase function.** In this plot the HG phase function is shown with different values for the anisotropy coefficient. We can notice how for g=0, the HG phase function reduces to the isotropic diffusion, while for typical values of 0.9 commonly found for biological tissues the scattering probability is maximized in the forward direction.*

and can be directly calculated by the ensemble average of the phase function itself:

$$g = \frac{\int_{(4\pi)} p(\hat{s}, \hat{s}_0) \hat{s} \cdot \hat{s}_0 \, d\Omega}{\int_{(4\pi)} p(\hat{s}, \hat{s}_0) \, d\Omega} = \frac{1}{W_0} \int_{(4\pi)} p(\hat{s}, \hat{s}_0) \hat{s} \cdot \hat{s}_0 \, d\Omega. \tag{19}$$

Thus, $g$ is a number that represents the average scattering angle to which the particle deflects the incident radiation. Lastly, it is very important to mention that we are not going to use the absorption and scattering cross-section during the following text, instead we will make use of the more useful optical parameters. Let us consider a small region $\delta V$ in which a statistically relevant number of particles are present and let us also assume that all of them are characterized by the same absorption and scattering cross-sections. We can define in this volume the *particle density*:

$$\rho = \frac{N}{\delta V} \tag{20}$$

which can be used to calculate the *absorption coefficient* $(cm^{-1})$:

$$\mu_a = \rho \sigma_a \tag{21}$$

and equivalently the *scattering coefficient* $(cm^{-1})$:

$$\mu_s = \rho \sigma_s \tag{22}$$

Such coefficients will characterize the medium when we do not have access to its detailed structural information (related to the particle's number and cross-sections) and we want to approach the problem with a statistical significance. We will make large use of such coefficient throughout the text, discussing the relative implications and approximations made. With the quantities just introduced we are now ready to approach a more detailed description of how light propagates and interacts with biological tissues. A key point on this will be the introduction of the equation that explains how the photons travel through the diffusive media, the important radiative transfer equation and its equivalent counterpart.

# I.2 Radiative Transfer Equation

The *radiative transfer equation* (RTE) was firstly introduced by Plank in his work [4] aiming at the description of how heat radiates through materials. Although the equation was efficiently proven to work under different areas of research, it is important to recall that its scalar formulation is purely phenomenological. Although a vector formulation of the theory (VRTE) [5] was proven possible to be derived directly from the Maxwell's equations, in the present text we will consider only the original scalar theory, which is general enough to explain the results we will present in the following. Two major approximations are needed to justify the reduction to the scalar problems: the addition of intensities rather than fields will hold within a small differential volume, and that the electric and magnetic field vectors are mutually orthogonal which requires all distances to be considered in the far-field.

Let us define a small volume $\delta V$ spatially located at the generic position $\mathbf{r}$, of which we are interested in the calculation of the *energy flow* $w_{\mathbf{r}}$ pointing towards a general direction in space $\hat{\mathbf{s}}_{\mathbf{j}}$. If we now consider the Poynting vector averaged over such a small volume $\langle \mathbf{S} \rangle = S\hat{\mathbf{s}}$, we have that we can explicitly write the *energy flow* as the following integral:

$$w_{\mathbf{r}}(\hat{\mathbf{s}}_{\mathbf{j}}) = \frac{1}{\delta V \|\mathbf{S}(\mathbf{r})\|_{\delta V}} \int_{\delta V} S(\mathbf{r} - \mathbf{r}')(\hat{\mathbf{s}}' \cdot \hat{\mathbf{s}}_{\mathbf{j}}) \, d\mathbf{r}'. \tag{23}$$

As a starting point for the formulation of the scalar RTE, we will consider the average Poynting vector defined in such a small volume as constituted by a magnitude $\|\mathbf{S}(\mathbf{r})\|_{\delta V}$, corresponding to the average value of the Poynting vector in the volume, and its average directionality in space $\langle \hat{\mathbf{s}}_{\mathbf{r}} \rangle_{\delta V}$ so that it can be rewritten as:

$$\langle \mathbf{S}(\mathbf{r}) \rangle_{\delta V} = \|\mathbf{S}(\mathbf{r})\|_{\delta V} \, \langle \hat{\mathbf{s}}_{\mathbf{r}} \rangle_{\delta V}. \tag{24}$$

To calculate the total averaged flow of energy within the small volume we can write explicitly the integral:

$$\|\mathbf{S}(\mathbf{r})\|_{\delta V} = \frac{1}{\delta V} \int_{\delta V} S(\mathbf{r} - \mathbf{r}') \, d\mathbf{r}'. \tag{25}$$

To calculate, instead, the contribution of the average directionality of such a vector we can notice that $w_{\mathbf{r}}(\hat{\mathbf{s}}_{\mathbf{j}})$ is also related to the probability that, at a certain specific point in space $\mathbf{r}$, the energy flow has a certain directionality in space $\hat{\mathbf{s}}_{\mathbf{j}}$. Due to this fact, its integral over the solid angle is normalized to unity:

$$\frac{1}{4\pi} \int_{(4\pi)} w_{\mathbf{r}}(\hat{\mathbf{s}}_{\mathbf{j}}) \, d\Omega = 1 \tag{26}$$

and so we can define the average direction in space as:

$$\langle \hat{\mathbf{s}}_{\mathbf{r}} \rangle_{\delta V} = \frac{1}{4\pi} \int_{(4\pi)} w_{\mathbf{r}}(\hat{\mathbf{s}}_{\mathbf{j}}) \, \hat{\mathbf{s}}_{\mathbf{j}} \, d\Omega. \tag{27}$$

This formulation for $\langle \hat{\mathbf{s}}_{\mathbf{r}} \rangle_{\delta V}$ implicitly admits that the spectrum of all possible directionalities is associated to a certain probability distribution. It is now useful to introduce the *specific intensity*, which represent the amount of power flowing in a certain direction in the solid angle. We can define it in terms of the volume average of the energy flow (i.e. the average Poynting vector) as:

$$I(\mathbf{r}, \hat{\mathbf{s}}) = \frac{1}{4\pi} \|\mathbf{S}(\mathbf{r})\|_{\delta V} \, w_{\mathbf{r}}(\hat{\mathbf{s}}). \tag{28}$$

Instead, if we rewrite it in terms of the actual values for the Poynting vector, its definition results to be:

$$I(\mathbf{r}, \hat{\mathbf{s}}_j) = \frac{1}{4\pi} \int_{\delta V} \langle \mathbf{S}(\mathbf{r} - \mathbf{r}') \rangle \cdot \hat{\mathbf{s}}_j \, d\mathbf{r}'. \tag{29}$$

By integrating the specific intensity around the full solid angle, we obtain the quantity called *average intensity*:

$$U(\mathbf{r}) = \int_{(4\pi)} I(\mathbf{r}, \hat{\mathbf{s}}) \, d\Omega, \tag{30}$$

and if we consider the previously obtained expression for the specific intensity, this yields to:

$$U(\mathbf{r}) = \frac{1}{4\pi} \|\mathbf{S}(\mathbf{r})\|_{\delta V} \int_{(4\pi)} w_{\mathbf{r}}(\hat{\mathbf{s}}) \, d\Omega = \|\mathbf{S}(\mathbf{r})\|_{\delta V}. \tag{31}$$

The average intensity is then equal to the magnitude of the average Poynting vector in the small volume $\delta V$, which is an expected result. Another important quantity to define is the *energy density*, which results to be expressed as:

$$u(\mathbf{r}) = \frac{1}{c_0 \delta V} \int_{\delta V} |\langle \mathbf{S}(\mathbf{r} - \mathbf{r}') \rangle| \, d\mathbf{r}' \tag{32}$$

that can be equivalently written as:

$$u(\mathbf{r}) = \frac{1}{c_0} \|\mathbf{S}(\mathbf{r})\|_{\delta V} = \frac{U(\mathbf{r})}{c_0}. \tag{33}$$

All the above defined quantities do not take into account the directionality of the flow of energy. To take into consideration we have to define the *total flux density* (or *flux*):

$$\mathbf{J}(\mathbf{r}) = \int_{(4\pi)} I(\mathbf{r}, \hat{\mathbf{s}}) \hat{\mathbf{s}} \, d\Omega \tag{34}$$

that is very useful because it results to be the power that a detector measures per unit area. In terms of the Poynting vector it can be written as:

$$\mathbf{J}(\mathbf{r}) = \frac{1}{4\pi} \|\mathbf{S}(\mathbf{r})\|_{\delta V} \int_{(4\pi)} w_{\mathbf{r}}(\hat{\mathbf{s}}) \hat{\mathbf{s}} \, d\Omega = \|\mathbf{S}(\mathbf{r})\|_{\delta V} \, \langle \hat{\mathbf{s}}_{\mathbf{r}} \rangle_{\delta V} \tag{35}$$

In this case $\mathbf{J}(\mathbf{r})$ is a vector quantity, but let us consider the specific case in which we want to measure the power that flows on the detector surface $A$ having normal vector $\hat{\mathbf{n}}$. The total flux that flows toward a specific direction in space $\hat{\mathbf{n}}$ is the projection of $\mathbf{J}(\mathbf{r})$ onto that direction, so that we have:

$$J_n(\mathbf{r}) = \mathbf{J}(\mathbf{r}) \cdot \hat{\mathbf{n}} = \int_{(4\pi)} I(\mathbf{r}, \hat{\mathbf{s}}) \hat{\mathbf{s}} \cdot \hat{\mathbf{n}} \, d\Omega \tag{36}$$

And equivalently expressed in the form of the Poynting vector:

$$J_n(\mathbf{r}) = \|\mathbf{S}(\mathbf{r})\|_{\delta V} \langle \hat{\mathbf{s}}_{\mathbf{r}} \rangle_{\delta V} \cdot \hat{\mathbf{n}}. \tag{37}$$

We can also notice that it is possible to separate the flux into two different directionalities with respect to the measurement vector $\hat{\mathbf{n}}$, the *forward flux* and the *backward flux*, which respectively are:

$$J^+(\mathbf{r}) = \int_{(2\pi)^+} I(\mathbf{r}, \hat{\mathbf{s}}) \hat{\mathbf{s}} \cdot \hat{\mathbf{n}} \, d\Omega = \frac{1}{4\pi} \|\mathbf{S}(\mathbf{r})\|_{\delta V} \int_{(2\pi)^+} w_{\mathbf{r}}(\hat{\mathbf{s}}) \hat{\mathbf{s}} \cdot \hat{\mathbf{n}} \, d\Omega \tag{38}$$

$$J^-(\mathbf{r}) = \int_{(2\pi)^-} I(\mathbf{r}, \hat{\mathbf{s}}) \hat{\mathbf{s}} \cdot (-\hat{\mathbf{n}}) \, d\Omega = \frac{1}{4\pi} \|\mathbf{S}(\mathbf{r})\|_{\delta V} \int_{(2\pi)^-} w_{\mathbf{r}}(\hat{\mathbf{s}}) \hat{\mathbf{s}} \cdot (-\hat{\mathbf{n}}) \, d\Omega \tag{39}$$

and which are related to the total flux contribution as:

$$J_n(\mathbf{r}) = J^+(\mathbf{r}) - J^-(\mathbf{r}). \tag{40}$$

Let us consider explicitly the case of the power intercepted by a detector surface A, of which we can write as a sum of infinitesimal surfaces $dS$:

$$A = \int_A dS. \tag{41}$$

The total power traversing the differential area is then given by:

$$dP(\mathbf{r}) = J_n(\mathbf{r}) \, dS \tag{42}$$

It is easy to understand that this formulation takes into account both directions of the flow. In our case, we are interested to measure the flux that flows *into* the detector, while the other contribution in this term represent a non-physical quantity. We have that:

$$P_{det} = \int_A J^-(\mathbf{r}) \, dS = \int_A dS \int_{(2\pi)^-} I(\mathbf{r}, \hat{\mathbf{s}}) \hat{\mathbf{s}} \cdot \hat{\mathbf{n}} \, d\Omega \tag{43}$$

and more explicitly if we want to consider the Poynting vector:

$$P_{det} = \frac{1}{4\pi} \int_A \|\mathbf{S}(\mathbf{r})\|_{\delta V} \, dS \int_{(2\pi)^-} w_{\mathbf{r}}(\hat{\mathbf{s}}) \hat{\mathbf{s}} \cdot \hat{\mathbf{n}} \, d\Omega. \tag{44}$$

At this stage, it is worth a consideration. The medium that we have considered so far can be continuous and homogeneous (or inhomogeneous), but whenever we consider a detector located somewhere inside it we are effectively perturbing its continuity. By inserting the detector, we are in fact removing the contribution given by the forward flux $J^+(\mathbf{r})$, due to the fact that the device is physically present in the medium. In fact, the correct formulation for this situation is assuming that the *boundary condition* at the detector surface are represented by $J_n(\mathbf{r}) = J^-(\mathbf{r})$.

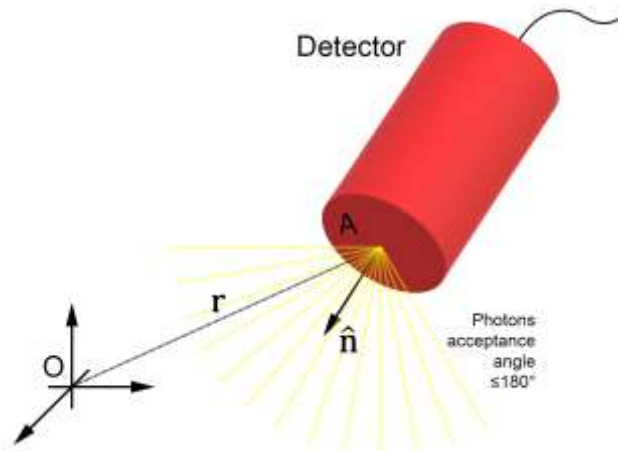

*Figure I-4 | A generic representation of a detector.* In this case is shown its sensitive area A that has normal vector $\hat{\mathbf{n}}$ and position in space $\mathbf{r}$. The image is taken from [2].

We can express the final power detected by a detector of surface $A$ by noticing its relation with the average Poynting vector and the total flux intensity expression:

$$P_{det} = \int_A \|\mathbf{S}(\mathbf{r})\|_{\delta V} \langle \hat{\mathbf{s}}_\mathbf{r} \rangle_{\delta V} \cdot \hat{\mathbf{n}} \, dS = \int_A \mathbf{J}(\mathbf{r}) \cdot \hat{\mathbf{n}} \, dS \tag{45}$$

which has final units of ($Watts$). In the present study, we are not interested in considering in details the angular acceptance contribution of the detector, even though we know it depends on the *numerical aperture $NA$* of the detector itself. Such quantity is defined as:

$$NA = n \sin \theta_A \tag{46}$$

where $n$ is the refractive index of the surrounding medium and $\theta_A$ the maximum acceptance angle at which the photon can be intercepted by the detector. By taking this into account, it is possible to write the detected power in the form of:

$$P_{det} = \int_A dS \int_{(2\pi)^-} I(\mathbf{r}, \hat{\mathbf{s}}) f(\hat{\mathbf{s}} \cdot \hat{\mathbf{n}}) \hat{\mathbf{s}} \cdot \hat{\mathbf{n}} \, d\Omega \tag{47}$$

where $f(\hat{\mathbf{s}} \cdot \hat{\mathbf{n}})$ is the function that defines the acceptance curve and $\hat{\mathbf{s}} \cdot \hat{\mathbf{n}} = \cos \theta$. Ripoll in his text [2] proposes the Hanning function as a good choice for taking into account the numerical aperture effects:

$$f(\theta) = \frac{1}{2} \left[ 1 + \cos \left[ \pi \left( \frac{\theta}{2\theta_A} \right)^k \right] \right]. \tag{48}$$

Although this is found to be important in particular when the measurement is taken in free space, outside the media of interest, we consider it a negligible perturbation for the scope of this work.

### I.2.1 DERIVATION OF THE RADIATIVE TRANSFER EQUATION

We are now ready to derive the *Radiative Transfer Equation* (RTE) and to do that we will make use of the energy conservation law found in the previous chapter Eq. (9), assuming a temporal modulation of the energy density. The only implicit assumption that we make is that the variation of such modulation, for example the source, is that its frequency is slower compared to the electromagnetic oscillation. In this case, we can use a time-averaged formulation for the energy conservation:

$$\frac{1}{c_0} \frac{\partial \langle \mathbf{S}(\mathbf{r}) \rangle_{\delta V} \cdot \hat{\mathbf{s}}_\mathrm{J}}{\partial t} + \langle \frac{dP_{abs}}{dV}(\mathbf{r}) \rangle \left( \hat{\mathbf{s}} \cdot \hat{\mathbf{s}}_\mathrm{J} \right) + \hat{\mathbf{s}}_\mathrm{J} \cdot \nabla (\langle \mathbf{S}(\mathbf{r}) \rangle \cdot \hat{\mathbf{s}}_\mathrm{J}) = 0. \tag{49}$$

To proceed further in the derivation of the RTE we have to calculate the above expression in the small differential volume $\delta V$. To do so we have to implicitly consider this small volume big enough to include $N$ particles having absorption cross-section $\sigma_a$ and scattering cross-section $\sigma_s$. Thus, we can integrate the expression above in the small volume:

$$\frac{1}{c_0}\frac{\partial}{\partial t}\int_{\delta V}(\hat{\mathbf{s}}\cdot\hat{\mathbf{s}}_{\mathsf{J}})S(\mathbf{r}-\mathbf{r}')\,d\mathbf{r}' + \int_{\delta V}(\hat{\mathbf{s}}\cdot\hat{\mathbf{s}}_{\mathsf{J}})\langle\frac{dP_{abs}}{dV}(\mathbf{r}-\mathbf{r}')\rangle\,d\mathbf{r}'$$
$$+\int_{\delta V}(\hat{\mathbf{s}}\cdot\hat{\mathbf{s}}_{\mathsf{J}})\hat{\mathbf{s}}_{\mathsf{J}}\cdot\nabla_{\mathbf{r}'}S(\mathbf{r}-\mathbf{r}')\,d\mathbf{r}' = 0 \tag{50}$$

having in mind that $\langle\mathbf{S}(\mathbf{r})\rangle = S(\mathbf{r})\hat{\mathbf{s}}$. Now it is worth to examine each term, noticing that they can be written as interesting quantities:

- Volume Averaged Change in Energy Density

$$\frac{1}{c_0}\frac{\partial}{\partial t}\int_{\delta V}(\hat{\mathbf{s}}\cdot\hat{\mathbf{s}}_{\mathsf{J}})S(\mathbf{r}-\mathbf{r}')\,d\mathbf{r}' \;\;\rightarrow\;\; \frac{1}{c_0}\frac{\partial}{\partial t}\|\mathbf{S}(\mathbf{r})\|_{\delta V}\,w_{\mathbf{r}}(\hat{\mathbf{s}}_{\mathsf{J}})\delta V \tag{51}$$

- Volume Averaged Absorbed Power

$$\int_{\delta V}(\hat{\mathbf{s}}\cdot\hat{\mathbf{s}}_{\mathsf{J}})\langle\frac{dP_{abs}}{dV}(\mathbf{r}-\mathbf{r}')\rangle\,d\mathbf{r}' \;\;\rightarrow\;\; N\sigma_a\|\mathbf{S}(\mathbf{r})\|_{\delta V}\,w_{\mathbf{r}}(\hat{\mathbf{s}}_{\mathsf{J}}) \tag{52}$$

- Volume Averaged Change in the Energy Flow

$$\int_{\delta V}(\hat{\mathbf{s}}\cdot\hat{\mathbf{s}}_{\mathsf{J}})\hat{\mathbf{s}}_{\mathsf{J}}\cdot\nabla_{\mathbf{r}'}S(\mathbf{r}-\mathbf{r}')\,d\mathbf{r}'$$
$$\rightarrow\;\; \hat{\mathbf{s}}_{\mathsf{J}}\cdot\nabla\big[\|\mathbf{S}(\mathbf{r})\|_{\delta V}\,w_{\mathbf{r}}(\hat{\mathbf{s}}_{\mathsf{J}})\big]\delta V + N\sigma_{tot}\|\mathbf{S}(\mathbf{r})\|_{\delta V}\,w_{\mathbf{r}}(\hat{\mathbf{s}}_{\mathsf{J}}) \tag{53}$$
$$- N\sigma_{tot}\int_{(4\pi)}\|\mathbf{S}(\mathbf{r})\|_{\delta V}\,w_{\mathbf{r}}(\hat{\mathbf{s}}')p(\hat{\mathbf{s}}_{\mathsf{J}},\hat{\mathbf{s}}')\,d\Omega'$$

We are not interested in deriving those equations, for further details we refer to [2], but we will make use of those results to arrive to the final formulation for the RTE. Now we have to take into account that the density of the particles in the small volume $\delta V$ can be expressed in the form of:

$$\rho = \frac{N}{\delta V} \tag{54}$$

and the relations between cross sections and relative average definitions:

$$\mu_a = \rho\sigma_a, \quad \mu_s = \rho\sigma_s, \quad \mu_{tot} = \rho\sigma_{tot}. \tag{55}$$

Recognizing the presence of the specific intensity in the expressions above, we can finally write the common form for the *Radiative Transfer Equation*:

$$\frac{1}{c_0}\frac{\partial}{\partial t}I(\mathbf{r},\hat{\mathbf{s}}) + \hat{\mathbf{s}}\cdot\nabla I(\mathbf{r},\hat{\mathbf{s}}) + (\mu_a+\mu_s)I(\mathbf{r},\hat{\mathbf{s}}) - \mu_t\int_{(4\pi)}I(\mathbf{r},\hat{\mathbf{s}}')p(\hat{\mathbf{s}}_{\mathsf{J}},\hat{\mathbf{s}}')\,d\Omega' = 0. \tag{56}$$

Written in these terms, we can recognize some important quantities that have important physical meanings in the overall energy balance expressed by the RTE. In fact, if we consider the energy flow at position $\mathbf{r}$ towards a specific direction $\hat{\mathbf{s}}_{\mathsf{J}}$, we have the following contributions:

- A temporal change in $I(\mathbf{r},\hat{\mathbf{s}}_{\mathsf{J}})$, $\rightarrow \frac{1}{c_0}\frac{\partial}{\partial t}I(\mathbf{r},\hat{\mathbf{s}})$
- A contribution due to the Absorption
- A contribution due to the Scattering, which can be divided into:
  - A spatial change in $I(\mathbf{r},\hat{\mathbf{s}}_{\mathsf{J}}) \rightarrow \hat{\mathbf{s}}\cdot\nabla I(\mathbf{r},\hat{\mathbf{s}})$
  - Flow energy loss due to scattering

　　　o　Flow energy gain due to scattering

So far, we have not considered any source term being present in the RTE, which implicitly assumes that the source is placed at infinite distance, but at this stage it is worth to make a further consideration. First of all, we can distinguish between two different classes of sources: a first class is the one that include external illumination devices (fiber, laser) or that does not need an external stimulus to produce the light (bioluminescence); a second class is the one comprehending all of those sources which need an external stimulus to be able to radiate (fluorescence or phosphorescence).

In the first case, we can simply add the contribution of the source $\epsilon(\mathbf{r}, \hat{\mathbf{s}})$ to the energy balance of the RTE, which now can be written in the more general form:

$$\frac{1}{c_0}\frac{\partial}{\partial t}I(\mathbf{r}, \hat{\mathbf{s}}) + \hat{\mathbf{s}} \cdot \nabla I(\mathbf{r}, \hat{\mathbf{s}}) + \mu_t I(\mathbf{r}, \hat{\mathbf{s}}) - \mu_t \int_{(4\pi)} I(\mathbf{r}, \hat{\mathbf{s}}')p(\hat{\mathbf{s}}, \hat{\mathbf{s}}')\,d\Omega' = \epsilon(\mathbf{r}, \hat{\mathbf{s}}). \quad (57)$$

In the second case, specifically when we consider the source term being the fluorescence, we need to couple the equation of the fluorescence emission to the equation that account for its excitation. If we consider, in fact, a general excitation term at a certain wavelength we have:

$$\begin{aligned}\frac{1}{c_0}\frac{\partial}{\partial t}I(\mathbf{r}, \hat{\mathbf{s}}; \lambda_{ex}) &+ \hat{\mathbf{s}} \cdot \nabla I(\mathbf{r}, \hat{\mathbf{s}}; \lambda_{ex}) + \mu_t I(\mathbf{r}, \hat{\mathbf{s}}; \lambda_{ex}) \\ &- \mu_t \int_{(4\pi)} I(\mathbf{r}, \hat{\mathbf{s}}'; \lambda_{ex})p(\hat{\mathbf{s}}, \hat{\mathbf{s}}')\,d\Omega' = \epsilon(\mathbf{r}, \hat{\mathbf{s}}; \lambda_{ex}).\end{aligned} \quad (58)$$

and for the excited fluorescent emission:

$$\begin{aligned}\frac{1}{c_0}\frac{\partial}{\partial t}I(\mathbf{r}, \hat{\mathbf{s}};\ \lambda_{em}) &+ \hat{\mathbf{s}} \cdot \nabla I(\mathbf{r}, \hat{\mathbf{s}};\ \lambda_{em}) + \mu_t I(\mathbf{r}, \hat{\mathbf{s}};\ \lambda_{em}) \\ &- \mu_t \int_{(4\pi)} I(\mathbf{r}, \hat{\mathbf{s}}';\ \lambda_{em})p(\hat{\mathbf{s}}, \hat{\mathbf{s}}')\,d\Omega' = \mu_a^{fl}(\mathbf{r})\Phi(\lambda_{em})I(\mathbf{r}, \hat{\mathbf{s}}; \lambda_{ex}).\end{aligned} \quad (59)$$

With the above formulations we conclude the section dedicated to the study of the RTE, which as we said represents the most accurate way to describe the light diffusion within any (also biological) medium. Before continuing further, it is useful to recall all the approximations done so far, that allowed us to describe the RTE in its compact form:

- *Far-field approximation*, implicit when we consider each scatterer contribution in the form of an outgoing spherical wave. This allowed us to assume that electric and magnetic fields are perpendicular to each other and, not less important, that the Poynting vector is varying slowly within the small volume.
- *Average of the flow of energy*, again the Poynting vector was considered constant in the volume $\delta V$ and it was possible to calculate an average. It meant that every particle in the volume received the same amount of energy.
- *Average incident flow of energy much greater than local average scattered flow of energy*
- *Incoherent scattering*, we neglected effects due to interference and this allowed us to sum intensities rather than fields.
- *Statistically equivalent optical properties in the whole medium*, we assumed the optical properties of $\delta V$ to be equivalent to those outside it.
- *Neglect the depolarization*, we did not consider the depolarization due to the interaction with particles, which is the term that couples the Cartesian coordinates.

All the above approximations are quite well satisfied in biological tissues, in which the stochastic motion of the scatterers acts as a self-averaging contribution to the quantities considered. The RTE formulation, although quite general, it is not easy to be solved in practical scenarios. Especially for complicated geometries, such as realistic animal or human tissues, it is impossible not only to find an analytical solution, but also to try to approach the problem via numerical calculations: a computational solution of the RTE, in most of the cases, it is not a feasible choice and would require incredible effort at both hardware and software level. Because of all these reasons, often the RTE is further approximated to the more efficient Diffusive Equation (DE).

## I.2.2 DIFFUSION EQUATION

To retrieve a more affordable equation that describes the photon transport within tissues, we should further consider some other aspect of the light propagation. We will not explicitly perform the calculation but we will report the main results and discuss them in the following section.

First of all, it is important to pay attention to the energy conservation implied by the RTE. By integrating it over the whole solid angle we can arrive at [2]:

$$\frac{1}{c_0} \frac{\partial}{\partial t} U(\mathbf{r}) + \mu_a U(\mathbf{r}) + \nabla \cdot \mathbf{J}(\mathbf{r}) = S_0(\mathbf{r}) \tag{60}$$

where $S_0$ is the whole source energy density irradiated in the solid angle. From this result, we notice that the term that accounts for the scattering disappears in the global energy balance. This is expected, since the elastic scattering only contributes for the change of directionality of the photons and does not lead to any loss of energy. Another useful formulation that has to be considered to reach to the final diffusive approximation it is related to the Fick's Law expression. Such formulation expresses the intensity flux for the diffuse light $\mathbf{J}_d(\mathbf{r})$ in the form of:

$$\mathbf{J}_d(\mathbf{r}) = -D \nabla U_d(\mathbf{r}) \tag{61}$$

where we have introduced the *diffusion coefficient D* as:

$$D = \frac{1}{3(\mu_s(1 - g) + \mu_a)} = \frac{1}{3(\mu_s' + \mu_a)} \tag{62}$$

In the previous equation, we introduced a further quantity named *reduced scattering coefficient*, simply expressed by scaling the scattering coefficient via the anisotropic factor:

$$\mu_s' = \mu_s(1 - g). \tag{63}$$

At this point, without proceeding any further, we have to explicitly mention a few considerations. The physical meaning for the diffusion coefficient in the Fick's law formulation, is that it represents the quantity that accounts for the spatio-temporal diffusion of the average intensity within a highly scattering medium. Furthermore, the factor $\mu_s'$ still describes the scattering properties of the media, but it is reduced by taking into account a contribution due to the (mostly) forward directionality of the scattering event. In fact, in typical biological tissue $g$ has been found to have values in the range of: $g = 0.8 - 0.9$ [6] which implies a strong forward directionality of the scattering event. This effectively reduces the overall effect of the scattering, due to the fact that the light has to undergo more changes in the direction before effectively losing its original directionality and becoming diffusive. Together with this consideration we can introduce the concept of two new quantities, the scattering mean free path and the reduced scattering mean free path:

$$l_{sc} = \frac{1}{\mu_s} = MFP \qquad \qquad l_{sc}^* = \frac{1}{\mu_s'} = TMFP. \qquad (64)$$

These two quantities are also called Mean Free Path (MFP) and Transport Mean Free Path (TMFP) and they represent, respectively, the average distance that a particle propagates between two successive scattering events or before loosing completely its original direction.

We are now ready to derive the final expression for the *Diffusion Equation* (DE). By directly inserting Fick's law into the energy conservation equation and considering the diffusion regime for the light propagating through a medium we have:

$$\frac{1}{c_0}\frac{\partial}{\partial t}U_d(\mathbf{r}) + \mu_a U_d(\mathbf{r}) - \nabla \cdot [D\nabla U_d(\mathbf{r})] = S_0(\mathbf{r}) \qquad (65)$$

The DE, compared to RTE, is far easier to afford in terms of the calculation of the final propagating field and numerically much more efficient. It is a common choice, in fact, to model light propagation through diffusive media by using DE rather than RTE, which has been proven to be fast and robust under a broad range of experimental conditions. The above formulation works well for modelling light propagation in biological tissues at optical wavelength, where predominantly scattering overcomes the absorption. In fact, the DE adds a major approximation on top of those included in the RTE formulation, which is that the light diffuses almost isotropically after sufficient scattering events. This imply that the DE is robust when $\mu_s \gg \mu_a$ otherwise the diffusivity is not enough to hold the approximation. It has been proven though [7], that the small variation to the diffusion coefficient via the introduction of a parameter $a$ that ranges from 0.2 to 0.6:

$$D = \frac{1}{3(\mu_s' + a\mu_a)} \qquad (66)$$

where

$$a = 1 - \frac{4}{5}\frac{\mu_s' + a\mu_a}{\mu_s'(1+g) + \mu_a} \qquad (67)$$

extended the validity of the DE also to regimes where $\mu_a \simeq \mu_s'$. Although the modified diffusion coefficient was proven to be useful for pushing further the DE over not-so-highly diffusive regimes, no modifications are feasible to make DE model the light propagation through transparent or quasi-transparent media. In these region with very low scattering compared with absorption, $\mu_s \ll \mu_a$, the diffusion approximation fails and we have to rely on the more accurate RTE or other statistically equivalent approaches. Among other possible choices for modelling light transport in tissue, the most reliable one can be achieved using a Monte Carlo approach to statistically sample the behavior of the photons propagating through the media. We will examine this in detail in the following section, opening the path towards the presentation of one of the main result of our work.

# I.3 Monte Carlo Photon Propagation

As we already discussed in the previous section, the diffusion equation fails to predict the photon diffusion in tissues where the scattering is much smaller compared to the absorption. Although for general biological tissues this approximation holds very well (tissue is highly scattering), there are situations in which the photon transport is far from being diffusive. If, for example, we consider the ocular cavity we can immediately notice that its functionality forces the lens and the vitreous body to be optically transparent, in order to not scramble the light path and make the brain able to process a non-scattered image. We can encounter a certain variety of optically transparent tissue in biology: the vitreous in the eye, the cerebral spinal fluid surrounding the brain and the spine of several animals (humans included), the albumen in the egg, transparent embryos of model organisms (the zebrafish for example) or last but not least optically cleared tissues. In all those scenarios, the DE simply cannot aim at predicting the diffusion of the light which mostly propagates with a ballistic trajectory due to the very low-scattering environment. RTE is requested in all those cases in which the diffusion approximation fails but, as we already discussed, it requires very big efforts for retrieving a solution especially for complex geometries or source extensions.

For this reason, in practice, RTE is not used for light propagation modelling and it is often substituted by its statistical counterpart: the method of Monte Carlo Photon Propagation (MC-PP, or simply MC). This approach is considered a statistically equivalent solution to the RTE, which also implies that it suffers from the same approximations as the RTE and, more importantly, it is able to estimate only ensemble averages. In the following, we will introduce the method by analyzing the approach as treated by Wang and Wu [8] examining the case of the Monte Carlo for Multi-Layered media (MCML) [9]. In the approach that follows, photons are treated as waves at the scattering sites and classical particles elsewhere. Coherence, polarization and nonlinearity are neglected as in the RTE. The structural anisotropy in tissue is neglected as well (elongated fibers in muscles could favor a certain directionality to the scattering) but we will not make any assumption on the magnitude of both the absorption and the scattering coefficient.

### I.3.1 THE SAMPLING OF RANDOM VARIABLES

Every Monte Carlo method relies on the sampling of the random variables used to effectively perform the action that we are interested to simulate by considering their probability distribution. Per each kind of interaction allowed during the photon propagation, then, we have to calculate the *probability density function* (PDF) associated with the event. This function gives the probability that a random variable $x$ assumes a value between $a$ and $b$ by the following relation:

$$P\{a \leq x \leq b\} = \int_a^b p(x)\, dx \tag{68}$$

being the normalization property defined as:

$$\int_{-\infty}^{+\infty} p(x)\, dx = 1. \tag{69}$$

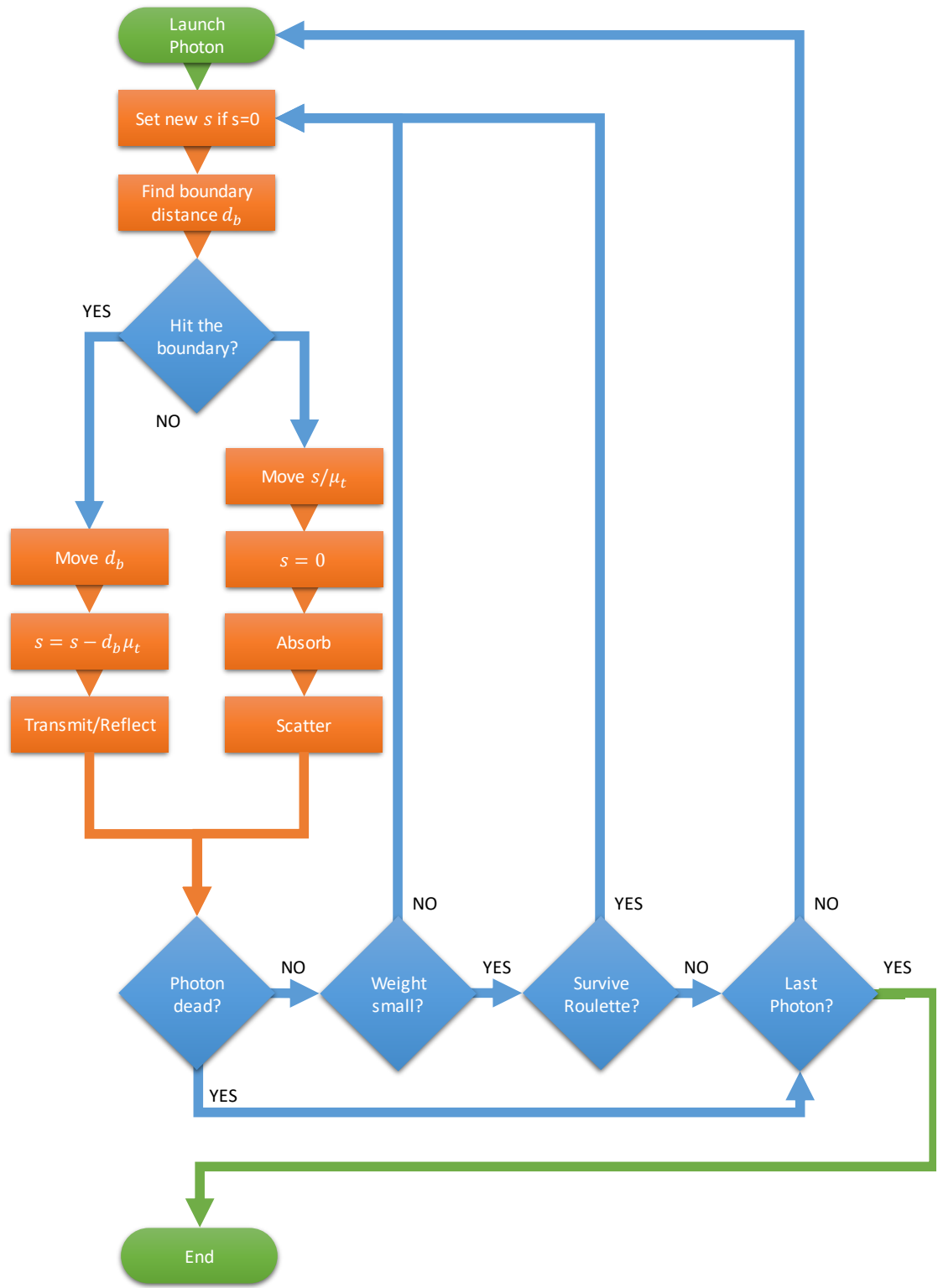

***Figure I-5 | Flow chart description for the Monte Carlo Photon Propagation method.*** *Due to the independence of each run the method is highly suitable for massive parallelization.*

It is important to recall that $p(x)$ is always a non-negative function in its whole interval of definition, otherwise it cannot describe a probabilistic event. It is interesting to notice the relationship between the PDF and its integrated expression:

$$P(x) = \int_{-\infty}^{x} p(x') \, dx' \qquad (70)$$

that is called cumulative distribution function (CDF). The other relation that binds the PDF with the CDF naturally follows their definition and can be found by the inverse operation of the integral, which is the derivative on $x$:

$$p(x) = \frac{d}{dx} P(x). \qquad (71)$$

If we consider a PDF for the event $\chi$ over the interval $(a, b)$ in the general form of $p(\chi)$, the sampling of the variable $\chi$ can be expressed as:

$$\int_{a}^{\chi} p(\chi) \, d\chi = \xi. \qquad (72)$$

In this formulation, $\xi \in [0,1]$ is a pseudorandom number generated by the computer that will sample statistically the variable $\chi$. The expression on the left represents the *cumulative distribution function* (CDF) $P(\chi)$ of the variable $\chi$, which is now related to the pseudorandom variable via the simple equivalence:

$$P(\chi) = \xi. \qquad (73)$$

Moreover, if $P(\chi)$ is sampled biunivocally by $\xi$ in the whole region where it is defined, its inverse transformation correctly samples $\chi$ so that is valid:

$$\chi = P^{-1}(\xi) \qquad (74)$$

This formulation defines the inverse sampling which is commonly referred to as the *inverse distribution method* (IDM).

### I.3.2 MONTE CARLO IN MULTI LAYERED MEDIA

The simulation of the photon propagation that we are going to describe in the following is accomplished in a planar, and multi-layered medium, described with different optical properties. The optical properties commonly used to accomplish the MC-PP are the ones described in the previous section: the absorption coefficient $\mu_a$, the scattering coefficient $\mu_s$, the refractive index $n$ and finally the anisotropy factor $g$. The different layers can have different optical properties. Although the simplistic geometry that we are going to consider, more complex implementations can be treated in exactly the same way. MCML [9] in fact is widely accepted as one of the gold-standard MC-PP methods, to which many other techniques refer to for validation.

*Photon Launching*

The photon is launched perpendicularly to the first layer surface under the form of a *pencil beam*. Such form of the beam implies that every photon is launched exactly at the same position, which can be seen as a delta function of the space, and with the same direction incident perpendicularly to the first layer. The weight of each starting photon is set to $W = 1$. If the external environment has a refractive index of $n_0$ and the first layer $n_1$, then the specular reflectance is equal to the Fresnel formulation:

$$R_{sp} = \left(\frac{n_0 - n_1}{n_0 + n_1}\right)^2. \tag{75}$$

When the photon enters the new medium, the reflectance effect reduces its effective weight by simply taking out the reflected part:

$$W = 1 - R_{sp}. \tag{76}$$

Then, the photon can start its journey throughout the medium.

### Step size

Each moving step can be obtained by sampling the probability distribution for the photon's free path $\in [0, \infty)$. To do so, we firstly consider an infinite turbid medium of which the probability of photon-tissue interaction can be expressed per unit pathlength in the interval $(s', s' + ds')$ as [9]:

$$\mu_t = \frac{-dP\{s \geq s'\}}{P\{s \geq s'\}ds'}, \tag{77}$$

where $P$ is the probability for the condition expressed in the brackets to hold. We have to integrate the equation over $s'$ in the range of $(0, s_1)$ using the condition that $P\{s \geq 0\}$, leading to the result:

$$P\{s \geq s_1\} = e^{-\mu_t s_1}, \tag{78}$$

which is simply the form of the Beer-Lambert law. Equivalently we can say that:

$$P\{s < s_1\} = 1 - e^{-\mu_t s_1} \tag{79}$$

and its corresponding probability density function is given by:

$$p(s_1) = \frac{dP\{s < s_1\}}{ds_1} = \mu_t e^{-\mu_t s_1}. \tag{80}$$

The above PDF can be integrated to find the relation with $\xi$, making explicit the sampling of the step size in the form of:

$$s_1 = -\frac{\ln(1 - \xi)}{\mu_t} = -\frac{\ln(\xi)}{\mu_t} \tag{81}$$

The above expression can be extended, in case the photon experiences free paths $s_i$ through different media, to the more general equation:

$$\sum_i \mu_{t,i} s_i = -\ln(\xi). \tag{82}$$

In this case, the left term defines the dimensionless total step size through the various media encountered. It is worth to notice how the propagation in the clear media does not affect the final value for the step size, due to the fact that the extinction coefficient is equal to zero.

### Moving the photon

After the determination of the substep $s_i$, the position of the photon is updated to the new value by simply adding the translation vector:

$$x \rightarrow x + \mu_x s_i \qquad y \rightarrow y + \mu_y s_i \qquad z \rightarrow z + \mu_z s_i. \tag{83}$$

*Photon absorption*

After the photon is moved to the new position, a fraction of its weight is absorbed by the medium. This, results to be equal to the quantity:

$$\Delta W = \frac{\mu_a}{\mu_t} W \tag{84}$$

that is added as a positive contribution to the local grid element (deposited energy in the medium):

$$A(r,z) \rightarrow A(r,z) + \Delta W. \tag{85}$$

Finally, it is removed from the photon weight, which implies that as soon as the photon moves throughout the media its probability to still be alive is reduced by the quantity:

$$W \rightarrow W + \Delta W. \tag{86}$$

*Photon scattering*

At this stage, the photon has undergone several steps: it has moved with a certain step size, its energy has been deposited and its weight reduced. At this stage, the light particle is ready to be scattered. We can examine the event separating the contribution to the directional deviation into a deflection angle $\theta \in [0, \pi]$, with respect to the original direction, and an azimuthal angle $\psi \in [0, \pi]$, with respect to the plane perpendicular to the original direction.

For the former, the probability distribution can be very efficiently described by the previously introduced Henyey-Greenstein scattering function [10]:

$$p(\cos \theta) = \frac{1 - g^2}{2(1 + g^2 - 2g \cos \theta)^{2/3}}, \tag{87}$$

where the factor $g$ is the anisotropy coefficient and is equal to $g = \langle \cos \theta \rangle$. It has been shown [11] that for biological tissues at optical wavelength that the single scattering event is very well described by this function having values of anisotropy ranging around $g = 0.9$. By using the random number $\xi$ to sample such probability distribution, we can find that the deflection angle can be statistically calculated by:

$$\cos \theta = \frac{1}{2g} \left\{ 1 + g^2 - \left[ \frac{1 - g^2}{1 - g + 2g\xi} \right]^2 \right\} \qquad \text{for } g \neq 0 \tag{88}$$

In case $g = 0$ instead, not relevant for our studies, the angle results equal to $\cos \theta = 2\xi - 1$. For the azimuthal angle, there is no particular request for the directionality, and can be efficiently modelled with a random number, uniformly distributed within the whole span of the round angle:

$$\psi = 2\pi\xi. \tag{89}$$

We have now the rules to scatter the photon, then we can calculate the new direction of the trajectory deflected by the scattering by the new vectors:

$$\begin{cases} \mu'_x = \dfrac{\sin\theta\left(\mu_x\mu_z\cos\psi - \mu_y\sin\psi\right)}{\sqrt{1-\mu_z^2}} + \mu_x\cos\theta \\[2mm] \mu'_y = \dfrac{\sin\theta\left(\mu_y\mu_z\cos\psi + \mu_x\sin\psi\right)}{\sqrt{1-\mu_z^2}} + \mu_y\cos\theta \\[2mm] \mu'_z = -\sqrt{1-\mu_z^2}\,\sin\theta\cos\psi + \mu_z\cos\theta \end{cases} \qquad (90)$$

and the direction updated by substituting those values into the previous ones:

$$\mu'_x \to \mu_x \qquad \mu'_y \to \mu_y \qquad \mu'_z \to \mu_z. \qquad (91)$$

*Photon on a boundary*

During the single propagation, on a certain step size $s$ the photon may hit a boundary, which can be either a boundary with the ambient or with another layer. In this case two situations can happen during the propagation: the photon can be reflected by the boundary or transmitted through it. In both cases, depending on where the photon was, it can continue its journey or be observed as diffuse reflection or transmission if it escaped the external boundaries of the medium.

First of all, we can calculate the distance between the current photon location and the boundary of the first layer that the photon can encounter during its journey having a particular direction:

$$d_b = \begin{cases} (z_0 - z)/\mu_z, & \text{if } \mu_z < 0 \\ \infty, & \text{if } \mu_z = 0 \\ (z_1 - z)/\mu_z, & \text{if } \mu_z > 0 \end{cases} \qquad (92)$$

where $z_0$ and $z_1$ are the coordinates of the upper and lower boundaries with the current layer. Next, we estimate if the step size is greater than $d_b$:

$$d_b\mu_t \le s \qquad (93)$$

Being $\mu_t$ the total interaction coefficient of the current layer. In case the inequality is satisfied, then the photon will hit the boundary and we update the stepsize:

$$s \to s - d_b\mu_t \qquad (94)$$

otherwise the photon movement will simply fit in the current layer and we do not have to do anything else other than move the photon by the quantity $s/\mu_t$, thus the propagation can continue normally. If the photon hits the boundary, we have to calculate the probability of being reflected or transmitted by the layer. To do so, we have to take into account the angle of incidence that we calculate as follows:

$$\alpha_i = \cos^{-1}(|\mu_z|) \qquad (95)$$

and the Snell's law that indicates the relationship between the angle of incidence, reflection and refraction:

$$n_i\sin\alpha_i = n_t\sin\alpha_t \qquad (96)$$

Where the subscript $i$ refers to the medium where the photon is travelling (incidence medium) and $t$ to the medium where it might be transmitted (transmitting medium). In case $\alpha_i$ is larger than the critical angle $a_c = \sin^{-1}(n_t/n_i)$, possible only when $n_i > n_t$, the photon will be reflected and the reflectance will be equal to $R_i(\alpha_i) = 1$. In all the other situations of incidence, the reflectance $R_i(\alpha_i)$ will be estimated by taking into account the Fresnel's formulas [12]:

$$R_\parallel(\alpha_i) = \frac{\tan^2(\alpha_i - \alpha_t)}{\tan^2(\alpha_i + \alpha_t)} \tag{97}$$

$$R_\perp(\alpha_i) = \frac{\sin^2(\alpha_i - \alpha_t)}{\sin^2(\alpha_i + \alpha_t)} \tag{98}$$

that describe the reflectance for the parallel and perpendicularly polarized wave. Since in this approximation the polarization is neglected (thus we assume random polarizations), we will use the average of the above quantities:

$$R_i(\alpha_i) = \frac{1}{2}\left[\frac{\sin^2(\alpha_i - \alpha_t)}{\sin^2(\alpha_i + \alpha_t)} + \frac{\tan^2(\alpha_i - \alpha_t)}{\tan^2(\alpha_i + \alpha_t)}\right]. \tag{99}$$

Now we have to determine whether the photon is reflected or not by statistically sampling with the variable $\xi$. This is quite simple now, because the reflectance already expresses the normalized probability of a photon being reflected by the surface. This implies that reflection occurs if $\xi \leq R_i(\alpha_i)$, otherwise the photon is transmitted. In case the photon is reflected, we have to reverse the last component of the direction cosines:

$$\{\mu_x, \mu_y, \mu_z\} \rightarrow \{\mu_x, \mu_y, -\mu_z\} \tag{100}$$

Otherwise the photon will be refracted and will change direction due to Snell's conditions:

$$\{\mu_x, \mu_y, \mu_z\} \rightarrow \left\{\mu_x \frac{\sin\alpha_t}{\sin\alpha_i}, \mu_y \frac{\sin\alpha_t}{\sin\alpha_i}, \mathrm{sgn}\,(\mu_z)\cos\alpha_t\right\} \tag{101}$$

With this we conclude the description of the MC-PP, leaving the boundary conditions with the external environment for a further reading [8]; we are not interested, in fact, in the calculation of diffuse reflectance and transmittance of the medium.

*Photon termination*

When a photon, during its propagation, reaches very low weight values $W$ it returns very little information about its journey through the media. Unless we want to study effects related to a very last stage of the photon propagation, we need to terminate properly the photon to conserve the energy. Common choice in these scenarios is to use the Russian roulette to "kill" the photon when its weight is lower than a certain threshold level $W_{th}$. In this case, we assign one chance in $m$ ($m = 10$ for example) of surviving with a weight of $mW$. This can be mathematically implemented as:

$$\text{when } W < W_{th} \rightarrow W = \begin{cases} mW & \text{if } \quad \xi \leq \dfrac{1}{m} \\ 0 & \text{if } \quad \xi > \dfrac{1}{m} \end{cases} \tag{102}$$

where $\xi \in [0,1]$ is a uniformly distributed pseudonumber. The reason to perform this choice is that it preserves the total energy of the system: in fact, if the photon would be simply terminated when it falls below the threshold level, we would have removed a small fraction of energy that the photon was still carrying with itself. By adding the surviving condition (in case the photon survives the roulette) of increasing the weight from $W$ to $mW$, the whole system will statistically preserve the total initial energy.

# Chapter II
# THEORY OF CLASSICAL AND AUTOCORRELATION IMAGING

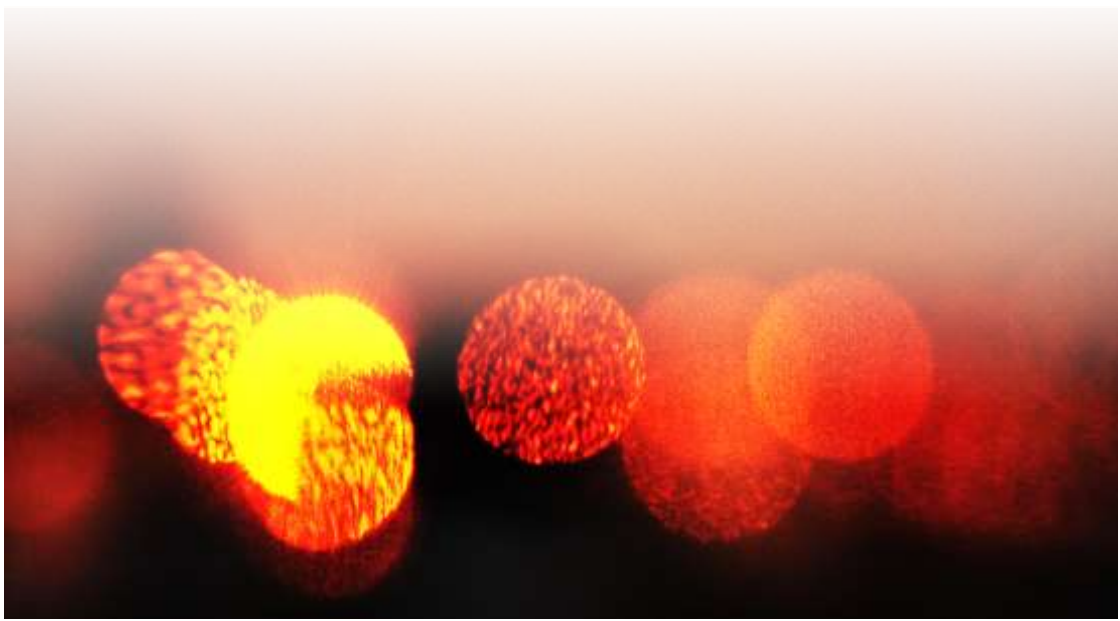

# II.1 General Concept in Optical Imaging

Before to deepen into the details of hidden imaging modalities, it is very important to briefly recap some important concepts of general optical imaging measurements. To do so we will refer to the Fourier formalism as originally treated by Goodman [13]. We will introduce only the general concept useful to understand the classical imaging formalism that will be extended to the hidden imaging modalities discussed in the next sections.

Let us define an object plane of coordinates $(\xi, \eta)$ where an aperture is located in the space. Let z be the distance of this plane with respect to a plane of coordinates $(x, y)$ parallel to it, in which we are interested in calculating the wave-field propagation $U(x, y)$. For distances greater than some wavelengths, it is common to use the *Huygens-Fresnel principle* which states that the field can be written as:

$$U(x, y) = \frac{z}{i\lambda} \iint\limits_{-\infty}^{+\infty} U(\xi, \eta) \frac{e^{ikr_{01}}}{r_{01}^2} d\xi d\eta \tag{103}$$

and where $r_{01} = \sqrt{z^2 + (x - \xi)^2 + (y - \eta)^2}$ is the relative distance between two points located respectively at each of the two planes. Although this formula contains just a few approximations, being the vector field reduced to scalar and the observation distance far away so that $r_{01} \gg \lambda$, the above expression is in general not easy to solve. To reduce the formula to a more usable expression, we need to expand the factor $r_{01}$ to its first order binomial formulation:

$$r_{01} = z\sqrt{1 + \left(\frac{x - \xi}{z}\right)^2 + \left(\frac{y - \eta}{z}\right)^2} \approx z\left[1 + \frac{1}{2}\left(\frac{x - \xi}{z}\right)^2 + \frac{1}{2}\left(\frac{y - \eta}{z}\right)^2\right] \tag{104}$$

and at this stage, we can make some further considerations. For the quadratic term at the denominator, the error committed by dropping off all the factors in the square root is relatively low, so that we can write $r_{01}^2 \approx z^2$. On the other hand, we cannot drop the terms in the exponential, because its value can change drastically for small variations. With these further considerations in mind, we can write the formula as:

$$U(x, y) = \frac{e^{ikz}}{i\lambda z} \iint\limits_{-\infty}^{+\infty} U(\xi, \eta) e^{i\frac{k}{2z}[(x - \xi)^2 + (y - \eta)^2]} d\xi d\eta. \tag{105}$$

The resulting expression can be seen as a convolution product between two function in the form of:

$$U(x, y) = U(x, y) * h(x, y)|_z. \tag{106}$$

where $h$ is the convolution kernel. Explicitly the two functions in the convolution result to be:

$$U(x, y) = \iint\limits_{-\infty}^{+\infty} U(\xi, \eta) h(x - \xi, y - \eta) d\xi d\eta, \tag{107}$$

$$h(x, y) = \frac{e^{ikz}}{i\lambda z} e^{i\frac{k}{2z}(x^2 + y^2)}, \tag{108}$$

and because of its definition, $h$ might be also referred to as the impulse response of the free space propagation. Lastly, we can factorize the quadratic terms leading to the final form of:

$$U(x,y) = \frac{e^{ikz}}{iz} e^{i\frac{k}{2z}(x^2+y^2)} \iint\limits_{-\infty}^{+\infty} \left\{ U(\xi,\eta) e^{i\frac{k}{2z}(\xi^2+\eta^2)} \right\} e^{-i\frac{k}{2z}(x\xi+y\eta)} d\xi d\eta \qquad (109)$$

This is the formulation of the *Fresnel diffraction integral* and the region where this formulation holds is called the *near-field* of the aperture. Interestingly, written in this form, this can be seen to be the Fourier transform of $U$ and $H$, but we will not treat this argument here.

More importantly, we could focus on a very stringent approximation. If, in addition to the Fresnel approximation, we further consider that the distance is long enough to satisfy the following relation

$$z \gg \frac{k}{2}(\xi^2+\eta^2)_{\text{max}}, \qquad (110)$$

then we are in the so called *far-field* region, and the above formula defines the *Fraunhofer approximation*. In this regime, the expression for the field reduces to a very useful expression:

$$U(x,y) = \frac{e^{ikz}}{iz} e^{i\frac{k}{2z}(x^2+y^2)} \iint\limits_{-\infty}^{+\infty} U(\xi,\eta) e^{-i\frac{k}{2z}(x\xi+y\eta)} d\xi d\eta \qquad (111)$$

that is called *Fraunhofer diffraction integral*, which is tightly related to the Fourier transform of the aperture. In fact, if we consider the spatial frequency variables of the form $f_X = x/\lambda z$ and $f_Y = y/\lambda z$, then the far-field the expression results:

$$U(x,y) \propto \mathcal{F}\{U(\xi,\eta)\}_{f_X,f_Y} \qquad (112)$$

where the proportionality ($\propto$) symbol lets us omit a trivial phase factor. Although very simple, this condition is quite strict and in fact, at optical wavelengths, an aperture of 1 cm is found in its far-field at distances approximately equal to 1 km. An equivalent, but less strict, condition can be expressed in the form of:

$$z > \frac{2D^2}{\lambda}, \qquad (113)$$

which is commonly called *antenna designer's formula*. Lastly, it's worth to mention that there is no direct way to express the Fraunhofer diffraction integral in the form of a convolution with an impulse response but, since the far-field formulation is derived as an approximation of the Fresnel integral, even in the far-field the convolution formulation holds.

### II.1.1 FOURIER TRANSFORMING PROPERTY OF THE LENS

Commonly made out of glass of refractive index approximately equal to 1.5, the lens is the basic optical device to perform operations on propagating light fields. Due to the different refractive index compared to that of the air, the light propagates through the lens at different speed and when it emerges on the other side collects a certain delay with respect to its original phase. A propagating wavefront trespassing a lens is delayed in phase and refracted at different direction but, if the lens is thin enough, the refraction can be considered null and the lens acts only as a phase transformation on the input signal. This condition is referred to as the *thin lens approximation* and allow us to write for an input complex field $U_l$ the emerging output as:

$$U_l'(x,y) = t_l(x,y)U_l(x,y). \tag{114}$$

In this case, $t_l$ is the phase delay introduced by the lens, which in exponential form can be written as:

$$t_l(x,y) = e^{ik\Delta_0}e^{ik(n-1)\Delta(x,y)}, \tag{115}$$

where n is the refractive index of the lens, $\Delta$ the function that describes the thickness of the lens and $\Delta_0$ its maximum thickness. We will not go into geometrical details that can be found in [13], but we point out that such phase factor can be found to be tightly related to the geometry of the lens and it results equal to:

$$t_l(x,y) = e^{-i\frac{k}{2f}(x^2+y^2)}. \tag{116}$$

In this formula, we introduced $f$ as the *focal length* of the lens defined by:

$$\frac{1}{f} = (n-1)\left(\frac{1}{R_1} - \frac{1}{R_2}\right), \tag{117}$$

being $R_{1,2}$ the radius of curvature of respectively the input and output face of the lens and having performed the paraxial approximation. Although ideally thin, the lens has a finite spatial size that is important to take into account in the formulation. Because of this it is useful to introduce the *pupil function* $P(x,y)$, which is 1 inside the lens aperture and 0 outside, and it is possible to write the field as:

$$U_l'(x,y) = U_l(x,y)P(x,y)e^{-i\frac{k}{2f}(x^2+y^2)}. \tag{118}$$

We are now interested in considering the field $U_f(x,y)$ at the focal plane (distance $z = f$) of the lens. By using the Fresnel diffraction formula calculated in the focal plane of the lens and taking into account the de-phasing introduced by the optical element, we arrive to write:

$$U(x,y) = \frac{e^{i\frac{k}{2f}(x^2+y^2)}}{i\lambda f}\iint\limits_{-\infty}^{+\infty} U(\xi,\eta)e^{-i\frac{k}{2f}(x\xi+y\eta)}d\xi d\eta, \tag{119}$$

which is equivalent to the *Fraunhofer diffraction intensity* calculated in the focal plane, thus also equal to the Fourier transform of the field entering the input side of the lens. This result holds if the size of the input is fully contained into the pupil function, which basically means that all the field is entering the lens. Moreover, it can be noticed that the frequencies of such Fourier transformation are connected with the geometrical properties of the lens via its focal length characteristics ($f_X = x/\lambda f$, $f_Y = y/\lambda f$).

## II.1.2 GENERAL TREATMENT FOR IMAGING SYSTEMS

So far, we have treated the general case of a system composed by a single lens and analyzed its properties in particular related to its inherent Fourier transform abilities. The question that now we are interested to pose is what are the properties of a more generic ideal *imaging system*, which is a system able to produce a real image in space that can be viewed by an observer.

Let us consider a generic imaging system, composed of an arbitrary number of lenses of different kind and different distances between them. Those set of lenses form what we can think as a black box (an objective lens of a camera for example, in which we do not know its lens assembly), of which we have unique information about the entrance and the exit pupils. In this section, we will refer only to diffraction-limited imaging systems, of which a beautiful description can be found in [13] and that we report in the following citation.

*"An imaging system is said to be diffraction-limited if a diverging spherical wave, emanating from a point source object, is converted by the system into a new wave, again perfectly spherical, that converges toward an ideal point in the image plane, where the location of that ideal image point is related to the location of the original object point through a simple scaling factor (the magnification), a factor that must be the same for all points in the image field of interest if the system is to be ideal."*

*J. W. Goodman*

If the system does not satisfy all those rules it is said to be affected by *aberrations*. With such perfect system, we are interested in imaging the bi-dimensional field $U_O$ generated into the observation plane that we express in function of the coordinates $(\xi, \eta)$. This can be considered the plane where the object of interest is placed and which we want to image with the above combinations of lenses. Because we do not have information about the black box, but we assume it is formed by a sequence of thin lenses, we can consider it as an operator that ultimately acts in a linear way on the object field, transforming it into the final image field $U_i$ which lies in the image plane $(u, v)$. Following the Fresnel near-field approximation, we can in general write its form as:

$$U_i(u,v) = \iint\limits_{-\infty}^{+\infty} U_O(\xi, \eta) h(u, v; \xi, \eta) \, d\xi d\eta, \qquad (120)$$

where $h$ is the amplitude response function of the system to a point source located in the object plane $(\xi, \eta)$. It is basically the function that "transfers" (or maps) a point on the object plane into another point in the image plane. As discussed before, the light amplitude in the image plane is simply the Fraunhofer diffraction pattern of the exit pupil:

$$h(u,v; \xi, \eta) = \frac{A}{\lambda z_i} \iint\limits_{-\infty}^{+\infty} P(x, y) e^{-i\frac{2\pi}{\lambda z_i}[(u-M\xi)x + (v-M\eta)y]} \, dxdy, \qquad (121)$$

centered in the image coordinates ($u = M\xi$, $v = M\eta$), where $(x, y)$ are the coordinates in the pupil plane $P$ and $z_i$ is the distance of the image plane with respect to the pupil. It is now important to introduce a further request for the description of our ideal imaging system: we want, in fact, that this system is ultimately space-invariant, which means that it does not introduce any distortion in the image plane and that the overall image has to be a uniformly magnified version of the object of interest. To achieve such spatial invariance, we have to define the reduced coordinates in the object plane such as:

$$\tilde{\xi} = M\xi, \qquad \tilde{\eta} = M\eta, \tag{122}$$

which allow to rewrite the expression as:

$$h\left(u - \tilde{\xi}, v - \tilde{\eta}\right) = \frac{A}{\lambda z_i} \iint\limits_{-\infty}^{+\infty} P(x,y) e^{-i\frac{2\pi}{\lambda z_i}[(u-\tilde{\xi})x + (v-\tilde{\eta})y]} \, dx dy. \tag{123}$$

The concept of an ideal image can be very well expressed by the following formulation, which takes into account that the image can be magnified but not distorted by the diffraction-limited imaging device:

$$U_g(\tilde{\xi}, \tilde{\eta}) = \frac{1}{M} U_O\left(\frac{\tilde{\xi}}{M}, \frac{\tilde{\eta}}{M}\right), \tag{124}$$

thus, leading to the convolution formulation:

$$U_i(u,v) = \iint\limits_{-\infty}^{+\infty} U_g(\tilde{\xi}, \tilde{\eta}) h\left(u - \tilde{\xi}, v - \tilde{\eta}\right) d\tilde{\xi} d\tilde{\eta} \tag{125}$$

$$h(u,v) = \frac{A}{\lambda z_i} \iint\limits_{-\infty}^{+\infty} P(x,y) e^{-i\frac{2\pi}{\lambda z_i}(ux+vy)} \, dx dy \tag{126}$$

This implies that the image produced in the plane $(u,v)$ is the convolution between the image prediction, given by the geometrical properties of the lenses $U_g(\tilde{\xi}, \tilde{\eta})$, with an impulse response function given by the Fraunhofer diffraction of the exiting pupil.

All these results obtained so far hold for strictly monochromatic systems, in which the field propagating can be described as composed by one single wavelength. Indeed, it is possible to relax the condition of monochromaticity of the wavefront, thus treating the case of polychromatic fields, in a very simple and elegant way. We can consider two kinds of temporal fluctuation of the field's wavelength but we will assume both to be narrowband, which implies that the frequency variation in time is small compared to its central peak. Thus, we reduce our study to two possible kinds of illuminations:

- *Spatially Coherent* illumination, when the frequency variation is unison and the field in all the points at the object plane has the same phase.
- *Spatially Incoherent* illumination, where the variation is absolutely uncorrelated in phase in all the points of the object plane.

In the following, we will examine the frequency response of the system separately for both of the cases, not considering intermediate scenarios.

### *Coherent Imaging systems*
In the case that the object illumination is spatially coherent, the variation of the phase is constant and depends on the time, thus can be written as a complex phase contribution to the impulse response function. Instantaneously then, the narrowband coherent imaging system is equivalent to the monochromatic one, making the results directly extendible by simply adding a time-dependence factor. In this case we can write:

$$U_i(u,v;t) = \iint\limits_{-\infty}^{+\infty} U_g(\tilde{\xi},\tilde{\eta};t-\tau)h(u-\tilde{\xi},v-\tilde{\eta})\,d\tilde{\xi}d\tilde{\eta} \tag{127}$$

where $\tau$ is the time delay of the photon propagation from $(\tilde{\xi},\tilde{\eta})$ to $(u,v)$ and $t$ is, of course, the instantaneous time. As we already mentioned, the narrowband spatial coherence will simply add a global complex phase contribution to the field $U_g$ but such (small) fluctuation, in practice, is not seen by the imaging device. This is due to the fact that the frequency fluctuations are several orders of magnitude faster than the fastest measuring device, thus making the calculation of the time-averaged intensity $|U_i(u,v;t)|^2$ of high importance if we want to proceed further with the study. Mathematically, this leads to:

$$\begin{aligned} I_i(u,v) &= \langle |U_i(u,v;t)|^2 \rangle \\ &= \iint\limits_{-\infty}^{+\infty} d\tilde{\xi}_1 d\tilde{\eta}_1 \iint\limits_{-\infty}^{+\infty} d\tilde{\xi}_2 d\tilde{\eta}_2 \, \langle U_g(\tilde{\xi}_1,\tilde{\eta}_1;t-\tau_1)U_g^*(\tilde{\xi}_2,\tilde{\eta}_2;t \\ &\quad -\tau_2) \rangle \times h(u-\tilde{\xi}_1,v-\tilde{\eta}_1)\,h^*(u-\tilde{\xi}_2,v-\tilde{\eta}_2). \end{aligned} \tag{128}$$

Here we can assume the time difference $\tau_1 - \tau_2 \approx 0$, in fact the multiplication between $hh^*$ is non-zero only when $\tilde{\xi}_1 - \tilde{\xi}_2 \approx 0$ and $\tilde{\eta}_1 - \tilde{\eta}_2 \approx 0$, which means that the two points have to be very close to each other to not lead to null results. If so, we can write:

$$\begin{aligned} I_i(u,v) &= \langle |U_i(u,v;t)|^2 \rangle \\ &= \iint\limits_{-\infty}^{+\infty} d\tilde{\xi}_1 d\tilde{\eta}_1 \iint\limits_{-\infty}^{+\infty} d\tilde{\xi}_2 d\tilde{\eta}_2 \, J_g(\tilde{\xi}_1,\tilde{\eta}_1;\tilde{\xi}_2,\tilde{\eta}_2) \\ &\quad \times h(u-\tilde{\xi}_1,v-\tilde{\eta}_1)\,h^*(u-\tilde{\xi}_2,v-\tilde{\eta}_2), \end{aligned} \tag{129}$$

having defined the *mutual intensity* $J_g$ as:

$$J_g(\tilde{\xi}_1,\tilde{\eta}_1;\tilde{\xi}_2,\tilde{\eta}_2) = \langle U_g(\tilde{\xi}_1,\tilde{\eta}_1;t)U_g^*(\tilde{\xi}_2,\tilde{\eta}_2;t) \rangle \tag{130}$$

as a measure of the spatial coherence of the light emitting from the object plane. Equivalently we can write:

$$U_g(\tilde{\xi}_1,\tilde{\eta}_1) = U_g(\tilde{\xi}_1,\tilde{\eta}_1)\frac{U_g(0,0;t)}{\sqrt{\langle |U_g(0,0;t)|^2 \rangle}} \tag{131}$$

$$U_g(\tilde{\xi}_2,\tilde{\eta}_2) = U_g(\tilde{\xi}_2,\tilde{\eta}_2)\frac{U_g(0,0;t)}{\sqrt{\langle |U_g(0,0;t)|^2 \rangle}} \tag{132}$$

due to the fact that the time varying phasors across the object differ only by a complex constant. By substituting these expressions into the mutual intensity, we have:

$$J_g(\tilde{\xi}_1,\tilde{\eta}_1;\tilde{\xi}_2,\tilde{\eta}_2) = U_g(\tilde{\xi}_1,\tilde{\eta}_1)U_g^*(\tilde{\xi}_2,\tilde{\eta}_2) \tag{133}$$

and then inserting this into the intensity we have that:

$$I_i(u,v) = \left| \iint\limits_{-\infty}^{+\infty} U_g(\tilde{\xi},\tilde{\eta})h(u-\tilde{\xi},v-\tilde{\eta})\,d\tilde{\xi}d\tilde{\eta} \right|^2. \tag{134}$$

We notice that we can define, also in this case, a time invariant phasor amplitude $U_i$ which can be described by the usual convolution equation:

$$U_i(u,v) = \iint\limits_{-\infty}^{+\infty} U_g(\tilde{\xi}, \tilde{\eta}) h(u - \tilde{\xi}, v - \tilde{\eta}) \, d\tilde{\xi} d\tilde{\eta}, \tag{135}$$

leading us to the same result obtained for the monochromatic case. At this stage, it is useful to define the frequency spectra for the input and the output:

$$G_g(f_X, f_Y) = \iint\limits_{-\infty}^{+\infty} U_g(u,v) \, e^{-i2\pi(f_X u + f_Y v)} du \, dv \tag{136}$$

$$G_i(f_X, f_Y) = \iint\limits_{-\infty}^{+\infty} U_i(u,v) \, e^{-i2\pi(f_X u + f_Y v)} du \, dv, \tag{137}$$

and also the *amplitude transfer function*:

$$H(f_X, f_Y) = \iint\limits_{-\infty}^{+\infty} h(u,v) \, e^{-i2\pi(f_X u + f_Y v)} du \, dv \tag{138}$$

as the Fourier transform of the amplitude impulse response function which we found to be space-invariant. If now we focus again on the $U_i(u,v)$, we have that by applying the convolution theorem, the Equation (137) simply reduces to:

$$G_i(f_X, f_Y) = H(f_X, f_Y) G_g(f_X, f_Y). \tag{139}$$

This expresses the result in the formalism in frequency domain and, if we relate this to the physical characteristic of the imaging system, we can point out an interesting feature of the coherent imaging systems. By explicitly writing the expression for the $h$ we have:

$$H(f_X, f_Y) = \mathcal{F} \left\{ \frac{A}{\lambda z_i} \iint\limits_{-\infty}^{+\infty} P(x,y) \, e^{-i\frac{2\pi}{\lambda z_i}(ux + vy)} dx dy \right\} \tag{140}$$

$$= (A\lambda z_i) P(-\lambda z_i f_X, -\lambda z_i f_Y).$$

Finally, regardless of a trivial constant $A\lambda z_i$ and the negative sign in the pupil function (in many applications the pupil is centrosymmetric), we can notice that:

$$H(f_X, f_Y) = P(\lambda z_i f_X, \lambda z_i f_Y). \tag{141}$$

This equation has a very important consequence for the coherent imaging system response in frequency domain. Being the amplitude transfer function equal to the scaled pupil function, that usually is a sharp circular region which is one inside and zero outside, the system effectively behaves as a pass-band in frequency domain, blocking the high-frequency components of the object once the information reaches the image plane. In fact, for the case of circular aperture we have that:

$$P(x,y) = \text{circ}\left(\frac{\sqrt{x^2 + y^2}}{w}\right) \quad \rightarrow \quad H(f_X, f_Y) = \text{circ}\left(\frac{\sqrt{f_X^2 + f_Y^2}}{w/\lambda z_i}\right) \tag{142}$$

that is the mathematical formulation of a band pass filter that blocks the frequencies $f_0 > w/\lambda z_i$.

### Incoherent Imaging systems

When the object illumination is perfectly incoherent in space, the phasor amplitudes across the object vary following statistical laws and ideally they can be expressed as:

$$J_g\big(\tilde{\xi}_1, \tilde{\eta}_1; \tilde{\xi}_2, \tilde{\eta}_2\big) = \langle U_g\big(\tilde{\xi}_1, \tilde{\eta}_1; t\big) U_g^*\big(\tilde{\xi}_2, \tilde{\eta}_2; t\big) \rangle$$

$$= \kappa I_g\big(\tilde{\xi}_1, \tilde{\eta}_1\big) \delta\big(\tilde{\xi}_1 - \tilde{\xi}_2, \tilde{\eta}_1 - \tilde{\eta}_2\big), \tag{143}$$

being $\kappa$ a generic constant. This expression has very interesting implications because we can now write the intensity distribution across the image plane as:

$$I_i(u,v) = \kappa \iint\limits_{-\infty}^{+\infty} I_g\big(\tilde{\xi}, \tilde{\eta}\big) \big| h\big(u - \tilde{\xi}_1, v - \tilde{\eta}_1\big) \big|^2 \, d\tilde{\xi} d\tilde{\eta}. \tag{144}$$

This formulation suggests that for an incoherent illumination, the image intensity is found to be the convolution of the intensity impulse response $|h|^2$ with the ideal image intensity $I_g$ onto the object plane $\big(\tilde{\xi}, \tilde{\eta}\big)$. An incoherent imaging system is, then, linear in the intensity rather than in the amplitude response and it obeys the *intensity convolution integral*. Following the same approach for the coherent imaging system, we can define the normalized spectra of $I_g$ and $I_i$ as:

$$\mathcal{G}_g(f_X, f_Y) = \frac{\iint_{-\infty}^{+\infty} I_g(u,v) e^{-i2\pi(f_X u + f_Y v)} du dv}{\iint_{-\infty}^{+\infty} I_g(u,v) \, du dv} \tag{145}$$

$$\mathcal{G}_i(f_X, f_Y) = \frac{\iint_{-\infty}^{+\infty} I_i(u,v) e^{-i2\pi(f_X u + f_Y v)} du dv}{\iint_{-\infty}^{+\infty} I_i(u,v) \, du dv} \tag{146}$$

and, similarly, the *normalized transfer function:*

$$\mathcal{H}(f_X, f_Y) = \frac{\iint_{-\infty}^{+\infty} |h(u,v)|^2 e^{-i2\pi(f_X u + f_Y v)} du dv}{\iint_{-\infty}^{+\infty} |h(u,v)|^2 \, du dv}. \tag{147}$$

Again, applying the convolution theorem to the intensity distribution at the image plane, we obtain the frequency response of the incoherent imaging system:

$$\mathcal{G}_g(f_X, f_Y) = \mathcal{H}(f_X, f_Y) \mathcal{G}_i(f_X, f_Y), \tag{148}$$

in analogy with the previous section. In this case $\mathcal{H}$ is called the *optical transfer function* (OTF) and its modulus, $|\mathcal{H}|$, the *modulation transfer function* (MTF). Let us now consider some explicit properties of such formalism. First of all, we know that $H(f_X, f_Y) = \mathcal{F}\{h\}$ and also:

$$\mathcal{H}(f_X, f_Y) = \frac{\mathcal{F}\{|h|^2\}}{\iint_{-\infty}^{+\infty} |h(u,v)|^2 \, du dv}. \tag{149}$$

By taking advantage of the Rayleigh's theorem we can then arrive to the formulation:

$$\mathcal{H}(f_X, f_Y) = \frac{\iint_{-\infty}^{+\infty} H(p', q') H^*(p' - f_X, q' - f_Y) dp' dq'}{\iint_{-\infty}^{+\infty} |H(p', q')|^2 dp' dq'} \tag{150}$$

and by changing the variables to:

$$p = p' - \frac{f_X}{2} \qquad q = q' - \frac{f_Y}{2} \tag{151}$$

we end up having a very interesting formulation for the OTF:

$$\mathcal{H}(f_X, f_Y) = \frac{\iint_{-\infty}^{+\infty} H\left(p + \frac{f_X}{2}, q + \frac{f_Y}{2}\right) H^*\left(p - \frac{f_X}{2}, q - \frac{f_Y}{2}\right) dp' dq'}{\iint_{-\infty}^{+\infty} |H(p, q)|^2 dp dq}. \tag{152}$$

The above expression has very important consequences for the OTF, which for the incoherent case results to be the autocorrelation of the amplitude transfer function. If we want to explicitly consider the aberration-free system, for which we have $H(f_X, f_Y) = P(\lambda z_i f_X, \lambda z_i f_Y)$, then it follows:

$$\mathcal{H}(f_X, f_Y) = \frac{\iint_{-\infty}^{+\infty} P\left(x + \frac{\lambda z_i f_X}{2}, y + \frac{\lambda z_i f_Y}{2}\right) H^*\left(x - \frac{\lambda z_i f_X}{2}, y - \frac{\lambda z_i f_Y}{2}\right) dx dy}{\iint_{-\infty}^{+\infty} P(x, y) dx dy}. \tag{153}$$

So, the optical transfer function of a diffraction limited incoherent system is the normalized autocorrelation of the pupil function, which implies that the OTF is always real and positive. Another equivalent method to calculate the OTF is via subsequent Fourier transformations:

$$\mathcal{H}(f_X, f_Y) = \mathcal{F}\{|\mathcal{F}^{-1}\{P(-x, -y)\}|^2\} \tag{154}$$

In which each step represents a Fourier operation on significant optical quantities:

1. Amplitude Point-Spread Function, $APSF = \mathcal{F}^{-1}\{P(-x, -y)\}$
2. Intensity Point-Spread Function, $IPSF = |APSF|^2$
3. Optical Transfer Function, $OTF = \mathcal{F}\{IPSF\}$

With this property of the OTF, we conclude the introductory section on general optical concept applied to imaging systems with this property, which we will encounter in a similar way in the study of incoherent hidden imaging systems (*Chapter IV*). In this special case, in fact, the autocorrelation will play a very important role and the reconstruction ability will be strongly influenced by the clever calculation of such quantity.

## II.2 Speckle Pattern

So far, we have discussed the case of ideal aberration-free imaging systems, which via combination of lenses that refract the light, are able to recreate an object distribution imaged at a certain focal plane into a new imaging plane. In fact, such imaging systems, preserve the phase of the wavefront up to a certain spatial frequency cutoff introduced by the finite dimension of the pupil function. Moreover, as already discussed in the previous section, we noticed that a simple perfect lens introduces a phase delay $t_l(x, y)$ in the wavefront propagating through it, dependent on the local thickness of the lens. Such a delay then is clearly dependent upon the geometry of the lens and defines its Fourier transform ability, making the lens a key tool for the creation of imaging systems.

But what can we say about the completely opposite scenario? Devices such as opaque glasses, ground glass diffusers (or even biological tissues) do not let the light pass through them in a straight way, as illustrated in **Figure II-1**. In fact, the phase of the wavefront emerging from them results scrambled in a complex, random fashion and seems to have lost all the initial information. Considering a simplistic description and similarly to the case of the lens, we can introduce the *thin diffuser* approximation, in which the light exits the layer at exactly the same point where it entered but it experienced a phase delay. The transmittance function of such thin diffuser is simply acting as a phase shift and results [14]:

$$t_s(x, y) = e^{ik\Phi(x,y)}. \tag{155}$$

In this description, the function $\Phi(x, y)$ represents a dephasing introduced by the diffuser, like in the lens, but this time depends upon the complex roughness of the layer rather than on the purely geometrical description of a lens. Since the coarseness of the diffuser is in most of the cases unknown, the phase delay $\Phi$ turns out to be a purely random function. This has a very important implication for the signal transmitted or, equivalently, reflected by such surfaces. The wavefront, scrambled in phase, interferes with itself in a highly intricate fashion, giving rise to a complex intensity pattern (of which an example is shown in **Figure II-2**) that consists in a random alternation of bright and dark spots. The intensity distribution arising from such complicated propagation is called a *speckle pattern* and, although it seems to introduce complexity in the measurements, its statistical properties make it extremely useful in many fields of study [15]. In fact, the problem can be seen as a *random walk* process [16] and all its statistical properties follow naturally from this approach. We are not interested in examining them in detail, for this we refer to more appropriate and exhaustive readings [15] [17], but we would like to point out a few interesting characteristics of such classes of intensity patterns.

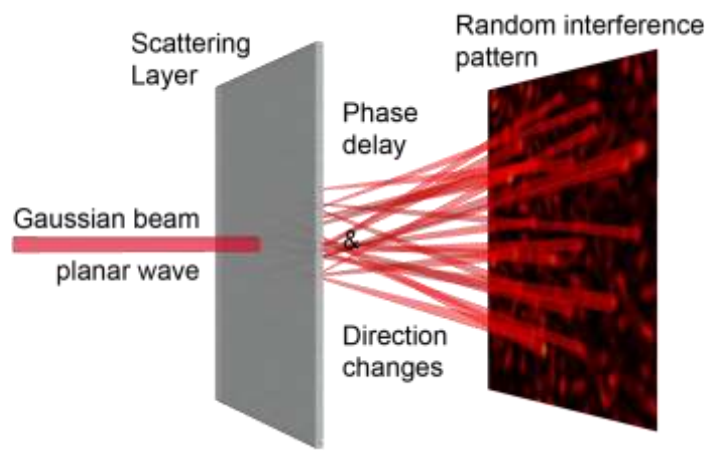

*Figure II-1 | Speckle pattern generation principle. An impinging wavefront encounters a rough surface that scrambles its phase, resulting in a random interference pattern located at its transmission side. Analogous effect is produced also in reflection.*

First of all, a speckle pattern is characterized by its mean value $\langle I \rangle$ around which dark and bright spots fluctuate in a seemingly random fashion. The speckle fluctuation around the mean value is characterized by the standard deviation $\sigma_I$, normally defined as:

$$\sigma_I^2 = \langle I^2 \rangle - \langle I \rangle^2. \qquad (156)$$

This quantity is useful for the definition of the *speckle contrast*:

$$C = \frac{\sigma_I}{\langle I \rangle} = \sqrt{\frac{\langle I^2 \rangle}{\langle I \rangle^2} - 1} \qquad (157)$$

and the *signal-to-noise ratio*:

$$\frac{S}{N} = \frac{1}{C} = \frac{I}{\sigma_I}, \qquad (158)$$

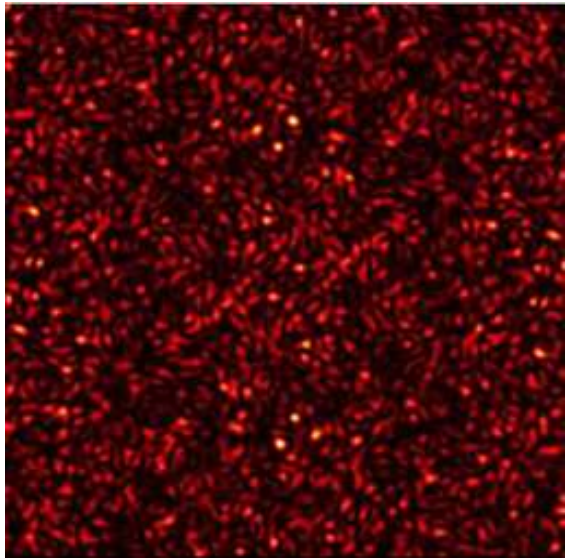

*Figure II-2| Speckle pattern detail. Speckle pattern generated by a (nearly pointless) Gaussian beam propagating through a random scattering layer.*

both of which quantify the visibility of the speckle around their intensity envelope. The maximum contrast condition $C = 1$ is fully satisfied when all the scatterers of the diffuser (or more generically the entities that scramble the phase of the photons) are completely independent one with respect to the other. This means that the phase shift introduced by the layer is uniformly distributed in the interval $(-\pi, \pi)$ and under these condition we have that $\langle I^2 \rangle = 2\langle I \rangle$, which leads to the maximum speckle contrast value of $C = 1$. In this case the probability density function, which describes the intensity distribution of the pattern, follows an exponential decay [17]:

$$p_I(I) = \frac{1}{\langle I \rangle} e^{-\frac{I}{\langle I \rangle}}. \qquad (159)$$

In this regime of light scrambling, the speckle is said to be *fully developed* and we will assume this condition to be satisfied throughout the whole text. Lastly, it is worth to mention that it is possible to define the average speckle grain size, and that this property is related to the autocorrelation of the pattern:

$$C(\delta \mathrm{x}, \delta \mathrm{y}) = \langle I(\mathrm{x}, \mathrm{y}) I(\mathrm{x} + \delta \mathrm{x}, \mathrm{y} + \delta \mathrm{y}) \rangle, \qquad (160)$$

being $(\delta \mathrm{x}, \delta \mathrm{y})$ the relative distance between two different positions. The autocovariance function of the intensity $c_I(\delta \mathrm{x}, \delta \mathrm{y})$ is related to the autocorrelation by the formula:

$$c_I(\delta \mathrm{x}, \delta \mathrm{y}) = \frac{C(\delta \mathrm{x}, \delta \mathrm{y}) - \langle I \rangle^2}{\langle I \rangle^2}. \qquad (161)$$

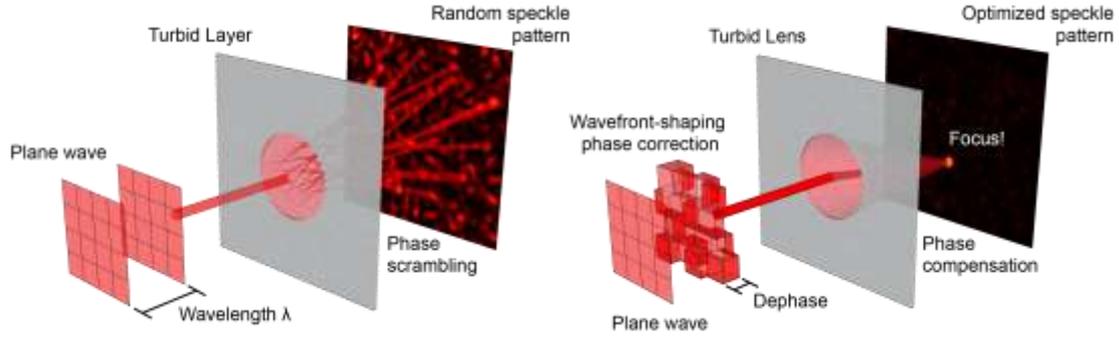

*Figure II-3 | Focusing through scattering layer process. The different phase arising from a complex light path is compensated by a spatial light modulator in order to achieve a highly localized coherent interference at the back of a scattering layer, thus forming a focus spot at user defined position.*

Its full width at half maximum $w$ can be seen as the measure for the typical size of the speckles and we have that for a system composed by spherical scatterers of diameter D this is equal to:

$$w = 1.4\lambda\frac{z}{D} \tag{162}$$

with longitudinal extension of

$$w_l = 6.7\lambda\left(\frac{z}{D}\right)^2. \tag{163}$$

The above results are obtained via statistical calculation in [17] assuming the photon propagation to be approximately equal to a random walk process. A similar result can be calculated in the case of rough surfaces, for which we have that the typical spot size is of the same order of the wavelength:

$$w = \frac{\lambda}{2}. \tag{164}$$

Before concluding the section with a general description of the properties of the speckle patterns, it is worth mentioning that the average width of the spot $w$ is equal to the coherence area of each speckle [18]. This result implies that the underlying phase region behind a speckle spot is slowly varying but, surprisingly, was experimentally proven that the spot sits (in average) on the side of phase-saddle and not directly above the center [18].

As we already mentioned at the beginning of this section, the scattering layer acts as a phase-scrambler for the impinging wavefront, ultimately producing a complex interference pattern. Interestingly, it has been shown however, [19] [20] that it is possible to compensate for such random dephasing and generate a focus even at the back of a highly scattering curtain. The technique is normally referred to as *adaptive optics* and *wavefront shaping* and makes uses of optical phase and/or amplitude manipulators, such as digital micro-mirror devices (DMD) or spatial light modulators (SLM). Interestingly, it has been proven that the autocorrelation function characteristics define the final shape of the focusing spot that we are able to achieve with such compensation. The statistics of the speckle pattern, then, ultimately define the final result of the focusing process. In our complementary works, we provided two different approaches to exploit this important feature, both of them based on the modification of the autocorrelation properties of the pattern. On one side, we tuned the speckle characteristics by cutting off some spatial frequencies of the system. Firstly, we noticed that blocking some Fourier components at low frequencies (with a spatial high-pass filter) unlocked the possibilities to achieve sub-speckle sized focusing [21]. Then we proved that filtering in Fourier domain with a ring (spatial band-pass filter) would in fact lead to a spatial distribution of the speckles that have a Bessel-like autocorrelation [22], rather than the classical Gaussian-like. In fact, by focusing through this systems shown in Figure II-4, we achieved a tunable non-diffractive Bessel focusing [23], which has the inherent ability to be set and moved at user-defined position. On the other hand, we proved that it is possible to tune the autocorrelation properties by designing ad-hoc the diffuser itself, so that the scattering event was probabilistically forced toward particular directions. By laser-ablating parallel rods, randomly distributed within a transparent glass, we created an *anisotropic photonic glass* that exhibited interesting speckle properties. By impinging a normal Gaussian laser onto the glass surface, the scattering due to the jump of the refractive index between glass and air gave rise to a new form of elongated speckle pattern. The resulting intensity distribution of such speckles was exhibiting the same anisotropy of the system, which resulted into the generation of linear regions of coherent interference. The elongated pattern produced in such a way resulted to have an elongated correlation function and, in fact, via adaptive optics compensation we were able to achieve a light sheet focusing through such opaque photonics glass [24]. Interestingly, we found similar structures already present in nature under the form of dentinal tubules, a biological bone-like tissue constituting the teeth (Figure II-5 panel b). The speckle pattern produced by these structures (Figure II-5 panel c-e) exibited the same elongation properties of the anisotropic photonic glasses, thus enabling us to through-biological media light sheet focusing.

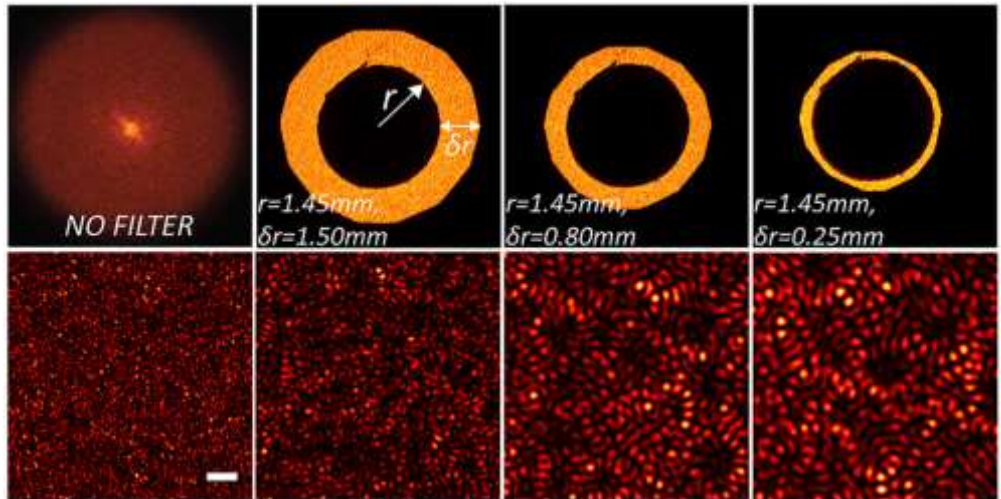

*Figure II-4 | Spatial ring filtering in the Fourier space for of a generic speckle pattern. It is possible to appreciate how the different size of the ring modified the spatial distribution of the speckle interference leading to an amorphous speckle distribution. The figure is adapted from our work [23], which allowed us to obtain a non-diffracting Bessel like focus behind a scattering layer.*

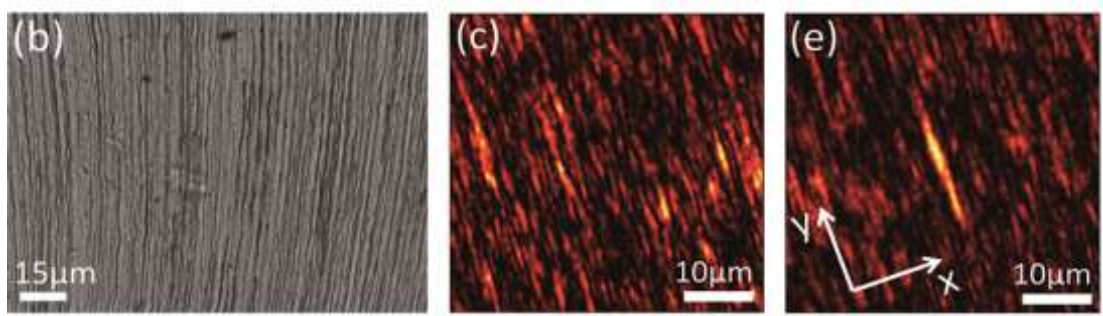

*Figure II-5 | Focusing through dentinal tubules structures. b) On the left it is possible to appreciate the structure of the dentinal tubules under the microscope through which we impinge a laser with a Gaussian profile. C) The microscopic structure of the tubules gave rise to a new kind of elongated speckle pattern. E) by compensating for the scattering through such structure it is possible to achieve a light sheet focusing profile. Figure imported from our work [24].*

As we said throughout this section, the organization and structure of the speckle spots depends upon the geometrical features of the scattering layer that initially scrambled the phase, and upon the frequency components that contributed to the interference pattern. This suggests that some degree of information managed to survive the random scrambling of the light paths and appears to be hidden (or better, encoded) intrinsically within the speckle pattern structure. In the following chapter, we will try to explicitly underline some effects related to the speckle properties, focusing in particular on their not absolute randomness and to the fact that a certain degree of information is still manifested by the structure of the pattern itself.

# II.3 Memory Effect (Isoplanatism)

It has been proven that a carefully developed class of phase retrieval algorithms allow for the recovery of the phase connected with a diffraction intensity pattern. Although the phase information can provide interesting hints about the properties of the measured object, one of the more interesting features of the algorithm is that it allows the possibility to image with lens-free systems. The field of Coherent Diffraction Imaging (CDI) takes large advantage from this capability, mostly due to the fact that for X-ray or electron we do not currently have efficient lens systems. In this case, the computation of the phase unlocks the possibility of imaging by only taking into account the diffraction pattern of the object. In the case of the light-photons instead, such technique does not offer any real advantage over lens imaging, which is usually preferred over lens-less systems. There are, though, some recently opened scenarios in which phase retrieval can play an interesting role in decrypting the information locked into the spatial distribution characteristics of optical speckle patterns.

Let us consider the case of the light speckles produced by photons propagating through a highly scattering curtain and randomly interfering at a certain plane behind such layer. If we do not know any information about the structure of the layer, we could be pushed to think that all the information that the light was carrying with itself has been lost at its back side. The pattern produced, in fact, seems a random distribution of intensity fluctuations crafted by complex dephasing introduced by the layer and no information seems to be retained in this scenario. A hypothetical imaging of a hidden object appears to be impossible without any further considerations.

Firstly predicted [25] and then proven experimentally [26], the memory effect (or isoplanatism theorem) states that while the light is trespassing a scattering layer it remembers much of the wavefront from which it derives. In fact, although the speckle produced seems absolutely random (and it is random, in the sense that the number and position of the particles that scramble the light path is not known) a correlational information is still preserved while passing the layer. For small angle tilting of the incident wave, the speckle distribution does not substantially change its intensity pattern but, instead, the whole pattern translates accordingly to the tilting of the incidence angle $\delta\theta$, as depicted in **Figure II-6**. Although this seems not such an interesting phenomenon, it has some important and not trivial implications. First of all, the direction of the source impinging the layer can be followed, due to the fact that the speckles track this information while the source moves in space. Second, if we consider the layer having optical thickness $L$, the information on a scale finer than the optical thicknesses are lost, but those on larger scales are preserved [26].

Theoretical calculations [25] predict that the pattern correlations-decay in transmission obeys to this function that depends on the optical thickness of the layer and the wavelength:

$$C(\delta\theta) = \left( \frac{k|\delta\theta|L}{\sinh(k|\delta\theta|L)} \right)^2 . \tag{165}$$

Where k is the wave number defined as $k = \frac{2\pi}{\lambda}$. Such phenomenon has been shown also in reflection [26] mode, in which all the characteristics mentioned above have been still proven to work but under a different correlation decay:

$$C(\delta\theta) = \frac{L \sinh(k|\delta\theta|l) \sinh[k|\delta\theta|(L-l)]}{k|\delta\theta|(L-l) \sinh(k|\delta\theta|L)}. \qquad (166)$$

In this case $l$ is the transport mean free path because, in order to be reflected, the light has to firstly become diffusive to have any hope of emerging from the same side where it entered the scattering layer.

For small angular variation, it has been proven [27] that the speckle correlation holds up to angles equal or smaller than:

$$|\delta\theta| \leq \frac{\lambda}{2\pi L} \qquad (167)$$

This implies that for angles larger than this value, the speckles do not only translate but consistently change their pattern distributions.

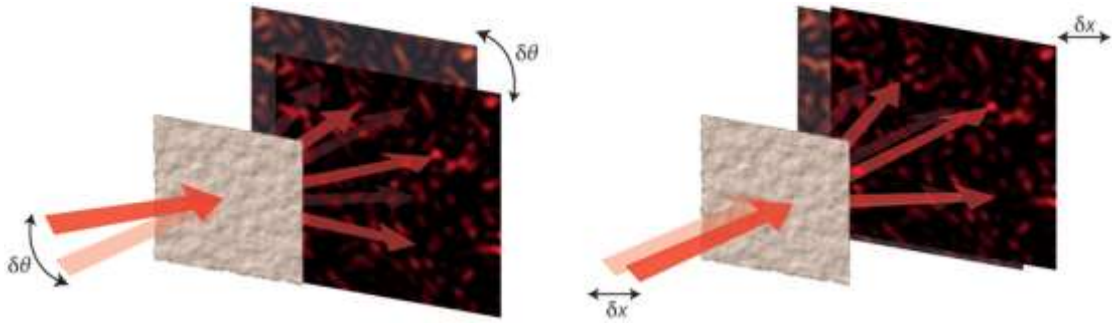

*Figure II-6 | The speckle memory effect. The pattern translates both while tilting the incident wavefront, as in the original formulation by Freund et al. [27], recently also extended to the translation of the input beam [127]. In both cases the position of the input beam can still be tracked by looking at the translation of the transmitted speckle pattern.*

## II.3.1 CORRELATION IMAGING

A very interesting consequence of this phenomenon was foreseen by Freund, that in 1990 mathematically proved [28] that for such small angles (within the memory effect regime) it is possible to define a correlation imaging technique for approaching the imaging through random and multiply-scattering layers. Let us call $I$ and $I'$ two measurement of speckle intensity distribution, for example recorded with a Charge-Coupled Device (CCD), produced by a coherent point source trespassing a planar opaque slab of thickness $L$ and transport mean free path $l^*$. Such layer is assumed to be parallel to the camera recording area in the $xy$-plane. To measure their similarity, we can make use of the normalized cross-correlation function defined as:

$$C(I, I') = \frac{\langle (I - \langle I \rangle)(I' - \langle I' \rangle) \rangle}{\sqrt{(I - \langle I \rangle)^2 (I' - \langle I' \rangle)^2}} \qquad (168)$$

Where the brackets $\langle \dots \rangle$ imply an ensemble average. The numerator of the fraction, under the weak-scattering condition which assumes that scattering wall does not localize the light (Anderson localization effects are excluded), can be expressed with the factorization approximation [29]:

$$\langle (I - \langle I \rangle)(I' - \langle I' \rangle) \rangle = \left| \langle EE'^* \rangle \right|^2 = |\langle E|E' \rangle|^2, \tag{169}$$

where $E$ is the electric field of the wave impinging on it. The transmission of the signal through the slab can be expressed by the complex transfer function $T(\boldsymbol{r}, \boldsymbol{R})$ where $\boldsymbol{r}$ is the position on the back side of the wall and $\boldsymbol{R}$ is the position in front. Considering $U(\boldsymbol{r})$ a field incident on the back surface of the wall, the resulting field on the surface will be given by:

$$V(\boldsymbol{r}) = \int T(\boldsymbol{r}, \boldsymbol{R}) U(\boldsymbol{r}) \, d^2 r. \tag{170}$$

In the following we will assume the Huygens-Fresnel approximation, in which spherical wavefront are treated like parabolas. Denoting the position on a distant screen, the image plane parallel to the wall, $\boldsymbol{x}$ and defining a wake-like vector $\boldsymbol{K} = {kx}/{d_i}$ where $k = {2\pi}/{\lambda}$ and $d_i$ is the distance between the two, we can write for the complex amplitude in the speckle patter:

$$E(\boldsymbol{K}) = \frac{1}{i\lambda d_i} e^{ikd_i} e^{\frac{id_i K^2}{2k}} \int V(\boldsymbol{R}) e^{\frac{ikR^2}{2d_i}} e^{-i\boldsymbol{K} \cdot \boldsymbol{R}} d^2 R. \tag{171}$$

Calculating the cross-correlation function $C(U_1, \boldsymbol{K}; U_2, \boldsymbol{K}')$ between two arbitrary different fields $U_1$ and $U_2$ at respective points $\boldsymbol{K}$ and $\boldsymbol{K}'$ implies the calculation of the quantity $\langle E(U_1, \boldsymbol{K}) | E(U_2, \boldsymbol{K}') \rangle$. To accomplish this, it is useful to notice that we can write:

$$\langle T(\boldsymbol{r}, \boldsymbol{R}) | T(\boldsymbol{r}', \boldsymbol{R}') \rangle = \delta(\boldsymbol{r} - \boldsymbol{r}') \delta(\boldsymbol{R} - \boldsymbol{R}') \langle |T(\boldsymbol{r}, \boldsymbol{R})|^2 \rangle \tag{172}$$

due to the fact that optical fields at different position in the random media are always uncorrelated. The last quantity can be seen as a probability that a photon injected at position $\boldsymbol{r}$ will exit at position $\boldsymbol{R}$ and, in average assuming diffusive propagation, this will not depend upon exact spatial positions but instead on the relative displacement $s = |\boldsymbol{r} - \boldsymbol{R}|$. This implies that we can write:

$$\langle |T(\boldsymbol{r}, \boldsymbol{R})|^2 \rangle = D(s). \tag{173}$$

The diffusion coefficient $D$ can be derived by diffusion theory and will be different between transmission and reflection, depending mainly on the parameters $L$ and $l^*$. We define now a form factor as:

$$F(q) = \int e^{i\boldsymbol{q} \cdot \boldsymbol{s}} D(s) d^2 s, \tag{174}$$

where $\boldsymbol{q} = \boldsymbol{K}' - \boldsymbol{K}$ is the relative displacement between the speckle patterns. We obtain then, a very important result useful for calculating the cross-correlation:

$$\langle E(U_1, \boldsymbol{K}) | E(U_2, \boldsymbol{K} + \boldsymbol{q}) \rangle = AF(q) \int U_1(\boldsymbol{r}) U_2^*(\boldsymbol{r}) e^{i\boldsymbol{q} \cdot \boldsymbol{r}} d^2 r \tag{175}$$

having defined a multiplicative factor:

$$A = \frac{1}{i\lambda d_i} e^{\frac{ikd_i\left(K'^2 - K^2\right)}{2K}}. \tag{176}$$

It is important to notice that the effect of the phase front curvature in the speckle pattern as described by the $K'^2 - K^2$ term in $A$, has no effect in the measured correlation C, which is then only a function of the relative displacement $\boldsymbol{q}$. Moreover, because of the normalization of the cross-correlation, C is also independent of both the other constants appearing in $A$ and on the absolute scale of $U_1$ and $U_2$.

Let us now consider $U_1$ the reference wave of the system, in order to make some considerations upon $C(\boldsymbol{q})$, and let this wave be a diverging spherical wave generated by a point source:

$$U_1 = U_R = \frac{1}{i\lambda d_i}P(x,y)e^{ikd_R}e^{\frac{ikr^2}{2d_R}}. \tag{177}$$

In this description $U_R$ is the reference wave, $d_R$ is the radius of curvature or the spherical wave front and $P(x,y)$ an entrance pupil function. Considering the pupil function is a logical choice, due to the fact that only a limited area of the wall surface can be used during the measurements. Moreover $d_R$ is also the distance of the point source with respect to the scattering layer. Once we have fixed the reference wave, let us consider the response of the system to a source distribution located in a plane which lies parallel to the wall and it is separated from it by a distance $d_o$. The coordinates in this plane, defined as the object plane, are expressed as $(x_o, y_o)$ and its field distribution as $U_o$.

At this point is then possible to calculate the correlation between the object and the speckle pattern [27] [28], obtaining:

$$C(\boldsymbol{q}) = C(x_i, y_i) = |\langle E_R|E_o\rangle|^2 = |U_i(x_i, y_i)|^2 \tag{178}$$

And explicitly:

$$U_i(x_i, y_i) = BF\left(\frac{kx_i}{d_i}, \frac{ky_i}{d_i}\right)\int\int h(x_i, y_i; x_o, y_o)U_o^*(x_o, y_o)dx_o dy_o \tag{179}$$

Here the spatial variables $(x_i, y_i)$ are the relative displacements between the object and the reference speckle pattern in the image plane used in computing the correlation, which also implies $q_x = kx_i/d_i$ and $q_y = ky_i/d_i$.

In the previous expression, we have defined as $B$ a not important phase factor:

$$B = \frac{1}{i\lambda d_i}e^{i\theta_i}e^{-ik(d_0 - d_R)} \tag{180}$$

and the impulse response function of the system:

$$h(x_i, y_i; x_o, y_o) = \frac{1}{\lambda^2 d_i d_o}e^{-\frac{ik(x_i^2+y_i^2)}{2d_i}}e^{-\frac{ik(x_o^2+y_o^2)}{2d_o}} \times$$
$$\times \iint P(x,y)e^{-i\frac{k}{2}\left(\frac{1}{d_0}-\frac{1}{d_R}\right)(x^2+y^2)}e^{ik\left[\left(\frac{x_o}{d_0}+\frac{x_i}{d_i}\right)x+\left(\frac{y_o}{d_0}+\frac{y_i}{d_i}\right)y\right]}dxdy. \tag{181}$$

Although the above expressions seem to be very complicated, they have a very simple and interesting interpretation:

- $h(x_i, y_i; x_o, y_o)$ is exactly the response function of a thin lens [13], with focal length $f_R$ defined by the relation $\frac{1}{f_R} = \frac{1}{d_i} + \frac{1}{d_R}$

- $U_i(x_i, y_i)$ is the exact form for the distribution of the image field located at the focus of such lens [13]. This is true except for the factor B, which is irrelevant and disappears in the normalization process of $C(\boldsymbol{q})$.

The scattering layer, in these terms, acts like a thin lens, in which the phase information is scrambled by the different paths travelled by the photons within the media.

---

*"Upon calculating the measured correlation C in terms of $|U_i(x_i, y_i)|^2$ and normalizing (whereupon B disappears), we obtain exactly the same real image as we would get if the wall were replaced by a thin lens with pupil function P, focal length $f_R$ given above, and a field stop of functional form F placed in the focal plane to limit the observed field of view."*

*J. Freund (1990)*

---

An important feature of such a "turbid lens system" is that the condition for a perfect focus is that $d_o = d_R$ which suggest that for better results the object plane has to be located at the center of the reference wavefront. Interestingly, the magnification is ruled by the ratio between the distances the object and the image plane with respect to the scattering layer:

$$M = -\frac{d_i}{d_o}. \tag{182}$$

The limit in the resolution of the turbid imaging system is dictated by the Gaussian pupil function (with an aperture of $2w$) by $2d_R/kw$. For the depth of field (DOF) it worth a consideration: for a common lens system, the DOF is connected to the distance at which the focus can be displaced with respect to the object before the image quality degrades; with a wall lens, instead, this is connected with the maximum distance at which the object plane can be moved with respect to the center of curvature of the reference wave ($d_R$) before the image loses quality. Due to the fact that mathematically a *wall lens* and a normal *thin lens* are represented by the same equations, the value of $DOF$ for both is equivalent to the classical formulation [30]:

$$DOF = \frac{4}{k}\left(\frac{d_R}{w}\right)^2 = \frac{2\lambda}{\pi}\left(\frac{d_R}{w}\right)^2 \tag{183}$$

Although mathematically equivalent to the corresponding classical lens system, the field of view ($FOV$) of such turbid imaging modality may have an extremely narrow imaging region:

$$FOV = \frac{2d_R}{kL} = \frac{d_R\lambda}{\pi L}, \tag{184}$$

where $L$ has to be replaced with $l^*$ in case we consider such imaging system working in reflection modality. The fact that the $FOV$ is quite limited inexorably depends upon the memory effect. This value is tightly connected to the maximum tilting angle before the speckle pattern produced by a point source start losing the correlation with its originally generated one, leading to $FOV = 2|\delta\theta|d_R$.

Lastly, it is worth to mentioning a few considerations regarding the cross-correlation formulation for $C(I, I')$. By its definition it implicitly requires, an average over different speckle patterns realizations, i.e. the imaging can be correctly outperformed only by averaging uncorrelated speckles obtained by using different wall lenses. This is not desirable if we consider the case of a single wall lens, because it would require the changing of the layer and the repetition of the measurement until sufficient sampling would allow the calculation of a statistically reasonable average. An efficient solution to this problem can be found by formulating an ergodic hypothesis: averaging the correlation function with a single speckle-spot produced over different walls is equivalent to average many different speckles produced by a single wall lens. A single speckle image in fact, presents many different and uncorrelated speckles, which were the results of many independent photon paths within the turbid layer. A sufficiently big single speckle pattern then, contains enough information to allow statistical averaging and can be effectively used as a turbid lens system. For the sake of completeness, by tuning the reference wave is possible to make the turbid wall act also as other optical instruments, opening new paths toward exploration of innovative experimental implementations.

### II.3.2 HIDDEN IMAGING WITH PHASE RETRIEVAL

So far, we have discussed how a combination of opaque wall, correlation properties and the appropriate selection of a reference wave can potentially act as an effective lens system, characterized by interesting properties. But in practice, how can we deal with the possibility of blind imaging, where we do not have control of the reference wave (and so we do not know the speckle response of the system for the reference)?

Bertolotti et al. [31] have shown that this is possible under a certain set of conditions and with the help of a phase retrieval algorithm. In their first implementation, they considered a fluorescent object (a $\pi$ having size of about $50\ \mu m$) hidden behind a diffusive layer. Both the illumination and the detection of the response was performed on the same side, opposite to the position of the sample with respect to the scattering layer. Under this experimental and prohibitive conditions, they made the following considerations. First of all, they scanned the incident angle of the beam (using two galvanic mirrors) in the two directions $\theta = (\theta_x, \theta_y)$. Such scanning procedure guarantees that the whole (hidden) object was illuminated uniformly, because the illuminating speckle translates through the whole object over a distance $\Delta r \approx \theta d$, finally making it fluorescing uniformly. The total amount of the transmitted fluorescence, then, is equal to the convolution (that we denote with the symbol $*$) of the object fluorescent response $O(r)$ and the speckle intensity pattern $S(r)$. Therefore, the total intensity measured as a function of the angle can be seen as:

$$I(\theta) = \int O(r)S(r - \theta d)d^2r = [O * S](\theta). \tag{185}$$

Remarkably such intensity does not resemble at all the image of the object, due to the random nature of the speckles produced by the wall. As we discussed previously, to separate the contribution of the speckle from the object itself, it is fundamental to calculate the cross-correlation of the image $I$ with respect to a known speckle given by a reference wave. In this case, we do not have access to such information, thus we calculate the cross-correlation (that we denote with $\star$) of the detected image with itself, what we will refer to as the autocorrelation of the image:

$$\langle I \star I \rangle (\theta) = \langle O * S \rangle \star \langle O * S \rangle = \langle O \star O \rangle * \langle S \star S \rangle = [O \star O] * \langle S \star S \rangle \qquad (186)$$

In this case, as previously described, the brackets indicate an average over different speckle realizations (different measurements). Since the speckles are assumed to be distributed in a random fashion in space, the speckle autocorrelation is a sharply peaked function, $\langle S \star S \rangle = \delta(r)$ and so disappears from the image autocorrelation. Via the calculation of the image autocorrelation, we are then measuring approximately the autocorrelation of the object itself $\langle I \star I \rangle (\theta) \sim [O \star O]$. Explicitly, for a circular illumination beam of width $w$, we have that:

$$\langle I \star I \rangle (\theta) = \left( \frac{k|\theta|L}{\sinh(k|\theta|L)} \right) \left[ [O \star O] * \left( \frac{2J_1(k|\theta|w)}{k|\theta|w} \right) \right] (\theta). \qquad (187)$$

In this formulation, the first term represents effectively the memory effect, which de-correlates the speckle pattern as we increase the angle. The second term in the convolution representing the average speckle size [32], where $J_1$ is the Bessel function of the first order, and this can be made arbitrary close to the diffraction limit by only increasing $w$. To overcome the need of averaging over different wall lenses, it is possible to start the scanning at sufficiently well separated angles $\theta$, in fact if their separations are larger than the object angular size, the speckles realizations will be effectively independent. This is equivalent to the ergodic assumption, and it has been proven to work well in this regime [31].

So far, we have described and connected the measurement of the autocorrelation of the signal coming from a hidden sample with the autocorrelation of the object itself. This is not enough if we aim at the reconstruction of its spatial intensity distribution and we need to further introduce some concepts to unlock this ability. We already said that we can measure $\langle I \star I \rangle$ and this can be related to the quantity $[O \star O]$, which means that we have the access only to the information about the relative distances within the various parts of the object, but nothing we can say about the object itself. In fact, the autocorrelation operation preserves only the magnitude of the spatial Fourier transform and at the same time lose the information about the phase. By using the convolution theorem, we can write:

$$\mathcal{F}\{\langle I \star I \rangle\} = \mathcal{F}\{O \star O\} = \mathcal{F}\{O\}\mathcal{F}\{O\}^* = |\mathcal{F}\{O\}|^2 \qquad (188)$$

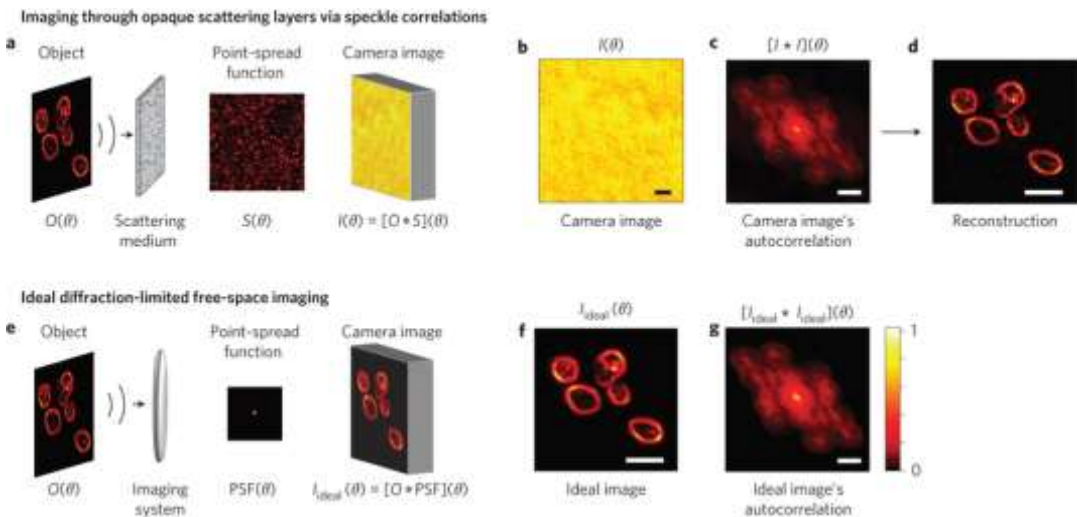

*Figure II-7 | The single shot hidden imaging ability via phase retrieval methods. In this case Katz et al. [33] imaged biological samples hidden behind turbid layer by exploiting the autocorrelation properties of the speckle pattern transmitted through it.*

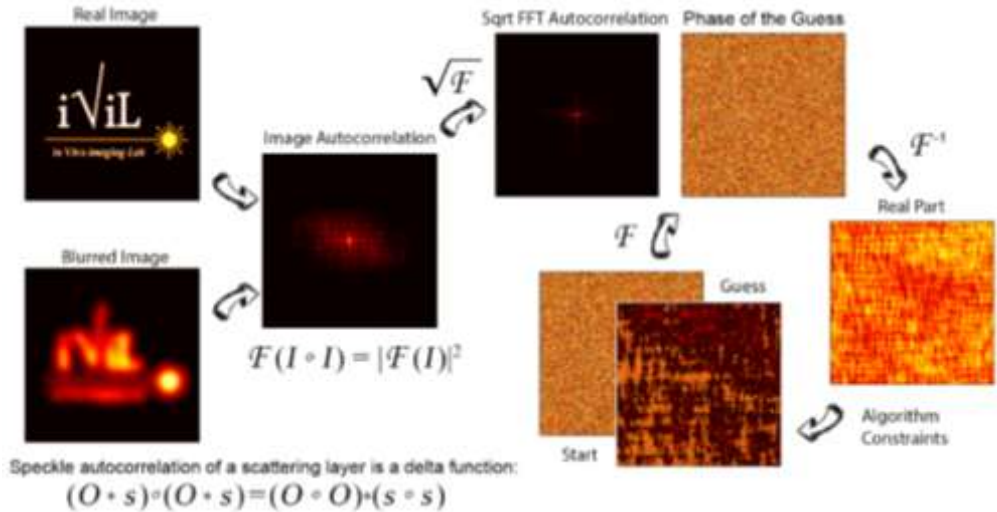

*Figure II-8 |Phase retrieval approach applied for imaging behind scattering layers. The correlation of the object and the hidden object is preserved regardless of the phase information lost while trespassing the scattering layer. Then a phase retrieval can unlock the possibility of imaging behind a scattering layer by simply retrieving the phase connected with the autocorrelation of the object, thus leading to meaningful reconstructions.*

that leaves us with the possibility of the calculation of the modulus of the Fourier transform of the object via the Fourier transformation of the image autocorrelation. We have now access to the modulus information and, if we are interested to image the object, we only miss the phase to connect with such a Fourier modulus. This is formally equivalent to a phase retrieval problem, which we will describe in details in the section that follows. It has been proven, in fact, that the combination of correlation imaging and phase retrieval allows imaging of two dimensional, incoherently emitting, samples hidden by turbid layer or around corners [31] [33].

Remarkably, Katz et al. noticed that such imaging system is equivalent to a classic one but in which the lens is replaced by a turbid layer of which the speckle patter represents its complex Point Spread Function (PSF):

$$I_{ideal} = [O * PSF](\theta) \iff I_{wall} = [O * S](\theta) \tag{189}$$

This combination results in the generation of an aberration-free optical system described in **Figure II-7**, which has incredible potential applications not only for hidden imaging, but also as a microscopic technique as well. By taking advantage of autocorrelation properties and via ad-hoc phase retrieval implementation, we will show how those techniques can be adapted and tuned to serve as a tomographic (and hidden) alignment-free imaging system.

## II.4 Phase Retrieval

The problem of the recovery of a function given its Fourier modulus is called Phase Retrieval and arises in several fields such as electron microscopy, crystallography, astronomy and optical imaging. This is mainly due to the fact that the measurement devices rely on the conversion of a photon flux into an electronic current. This conversion does not allow a direct recording of the phase, mainly due to the fact that the fast oscillation of the electromagnetic field (around $10^{15}$ $Hz$) cannot be followed by any of the currently available electronic devices [34]. Measuring the phase then, requires indirect methods such as interference with another known field in a process that is called *holography*. Although measuring directly the phase is not possible, some features of the electromagnetic field leave rooms for algorithmic phase retrieval. In fact, for a quasi-monochromatic electromagnetic field distribution in a specific plane in space, its far field at large enough distance corresponds to the Fourier transform of its near field. Thus, in this case, the recovery of the phase function associated with the far field would allow the imaging of the object of interest.

Historically the first computational implementation for the phase-retrieving process was proposed in 1978 by Fienup that developed an algorithm able to retrieve the phase connected to a Fourier modulus of a 2D image [35]. This is a very interesting problem, in fact **Figure II-9** interestingly shows the importance of the phase in the image formation. After an initial interest for the algorithmic phase retrieval with potential application for the creation of an optical computer (which was found to be not feasible), it gained interest from the X-Ray imaging community in the rush of increasing imaging resolution [36]. Another field in which phase retrieval is playing a crucial role is Astronomy, where it allows high resolution imaging for adaptive optics aberration correction, mainly due to atmosphere turbulence or imperfection of the optical imaging system [37]. It is also used for speckle interferometry and to overcome the diffraction limit of the imaging systems [38] [39]. Recently, a growing number of interesting works, inspired by a detailed literature in Astronomy, unlocked the possibility to optically image objects hidden behind turbid layers or around corners [31] [33] taking advantage of the information encoded in the speckle memory effect [26].

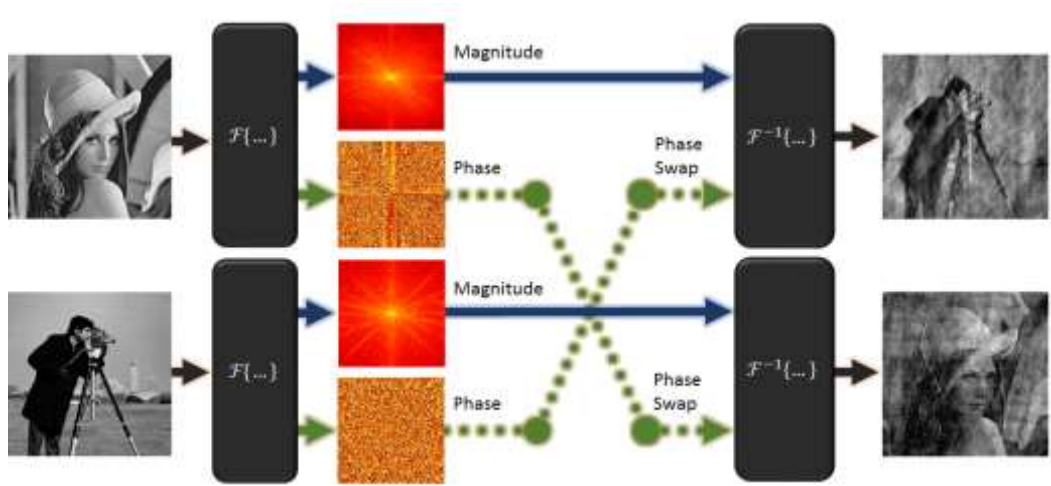

*Figure II-9 | The importance of the phase in the image formation. We can appreciate how the phase mixing between two different images makes the final result look like the original image that contained the phase information. It is clear, then, that the phase has a very high importance in the image formation.*

### II.4.1 MATHEMATICAL FORMULATION

Herewith we offer one of the possible mathematical formulations of the phase retrieval problem. Due to the fact that we will make use of the *Discrete Fourier Transform* throughout the whole work, here we refer to the discretized formulation of the theory. In fact, measuring devices, such as CCD cameras or other sensors, rely to the digitalization process to record the signal that is recorded in a discrete fashion, thus making consistent our assumption of discretized function. Moreover, for the sake of simplicity, we consider the one-dimensional formulation; extension to higher dimensionalities and to the continuous case is straightforward.

Let us consider the real distribution of an object $u(x) \in \mathbb{C}^N$, where $N$ is the number of its discrete sampling in the space-variable $x$. Such object is what we are interested in imaging, so we consider its Discrete Fourier Transform (DFT) $\mathcal{F}$ given by:

$$U(k) \stackrel{\text{def}}{=} \mathcal{F}\{u\} = \sum_{x=0}^{N-1} u(x)e^{-i2\pi\frac{kx}{N}}, \qquad \text{with } k = 0,1,\dots.N-1 \qquad (190)$$

and its oversampled version:

$$U(k) = \sum_{x=0}^{N-1} u(x)e^{-i2\pi\frac{kx}{M}}, \qquad \text{with } k = 0,1,\dots.M-1 \qquad (191)$$

where the term oversampled DFT refers to $M$ points of the DFT of x which have $M > N$. Of course, if we know everything of $U$, it is possible to recover $u$ via its inverse-DFT:

$$u(x) \stackrel{\text{def}}{=} \mathcal{F}^{-1}\{U\} = \sum_{k=0}^{K-1} U(k)e^{i2\pi\frac{kx}{N}}, \qquad \text{with } x = 0,1,\dots.M-1. \qquad (192)$$

Writing the DFT with the exponential formulation $U(k) = |U(k)|e^{i\phi(k)}$, we define the phase retrieval problem as the problem of recovery the function $\phi(k)$ given the associated Fourier modulus $|U(k)|$:

$$\text{given } |U(k)|, \text{ find } \phi(k) \ \mid \ u(x) = \mathcal{F}^{-1}\{U(x)\}. \qquad (193)$$

Let us denote $\hat{u}$ the vector $u$ being padded with $N-1$ zeros. Then the autocorrelation of such a vector is defined such as:

$$c(m) = \sum_{j=\max(1,m+1)}^{N} \hat{u}_j \hat{u}_{j-m}, \quad m = -(N-1),\dots.N-1. \qquad (194)$$

It is well known that the DFT of $c(m)$ is related to the Fourier transform of the signal we were interested in retrieving via the relation:

$$C(k) \stackrel{\text{def}}{=} \mathcal{F}\{c\} = |U(k)|^2 \qquad (195)$$

which has as a major consequence that the phase retrieval problem is also equivalent to the problem of estimating the object given its autocorrelation sequence.

For the sake of completeness, it is worth to mention that the Fourier phase retrieval is a sub-case of a more general class of phase problems wherever a generic measurement can be expressed as:

$$y_k = |\langle a_k, u \rangle|^2, \qquad \text{with } k = 0, 1, \ldots M \qquad (196)$$

In our specific case, the measurement vector $a_k$ can be expressed as a complex exponential function:

$$a_k(x) = e^{-i2\pi\frac{kx}{M}}, \qquad (197)$$

that makes the scalar product equal to the DFT of the signal x that we described in the beginning of the paragraph.

Since our problem starts with a Fourier measurement, in general it does not admit a unique solution; but in this case, lack of uniqueness does not necessarily imply the impossibility to obtain high quality reconstructions. In fact, in such kind of problems there is always a trivial set of ambiguities that does not compromise the reconstruction capabilities of the phase retrieval methods. First of all, the Fourier transform is invariant under three transformations:

1. A global phase shift, $\qquad u(x) \rightarrow u(x)e^{i\phi_0}$
2. Conjugate inversion, $\qquad u(x) \rightarrow \overline{u(-x)}$
3. Spatial translation, $\qquad u(x) \rightarrow u(x + x_0)$.

This implies also that the Fourier modulus is preserved under such transformations and more specifically they imply that the reconstruction of the phase does not lead to its absolute value (1), the object reconstructed can be flipped with respect to any spatial dimension (2) and the absolute object position is lost in the image plane (3). These considerations make the phase retrieval to be always a non-unique inverse problem, but still usable for practical scope. It has been proven that for dimensionality equal or higher than two ($D \geq 2$) the problem admits always a single solution that, except for the above-mentioned transformations, is considered unique [40] [41]. More complicated is the scenario for the one-dimensional problem, in which multiple signal could lead to the same Fourier magnitude (and so to the same autocorrelation sequence) and, even if the support of the image is bounded within a known range, the uniqueness does not exist [42]. In general, in the context of our work where we do not consider 1-D signals, it is important to notice that the uniqueness is guaranteed if the magnitude of the oversampled sequence satisfies $M \geq 2N - 1$.

## II.4.2 ALGORITHMICAL IMPLEMENTATION

So far, we have introduced the problem connected with the retrieval of the phase information for a far field Fourier measurement, but we have not shown in practice how to retrieve such information. The Fourier phase retrieval is an inverse problem that admits unique solutions under some specific conditions, but unfortunately this does not necessarily imply that such solution is always possible to find. In fact, during the years, several different approaches were proposed to solve the problem: most of them rely on alternate Fourier projections, others take advantage of sparsity-based methods and, finally, a third class uses the transport-of-intensity equation (TIE) [43]. In the following, we will focus mainly on the first class and briefly mention sparsity based techniques, which we plan to implement in future works to enhance converging rate and speed.

*Error Reduction (ER)*

The first implementation of the alternating projection method was proposed by Gerchberg and Saxton (GS) [44], in which measurements in different planes (real and Fourier) are alternated to make the final phase determination converging into a meaningful reconstruction. This problem is similar to the phase retrieval problem discussed above, but for the fact that it relies on two modulus measurements: the image (in the real space) and its diffraction in the far-field (Fourier space). A more generalized version of the GS algorithm is called Error Reduction (ER) method and rely only on one measurement in Fourier space. The determination of the phase connected with such modulus is accomplished via the application of some image constraints and allow the reconstruction of the final image.

The ER method [45] consists of a simple four-step algorithm and has been proven to be useful in many of our experimental trials. The iteration starts with an initial guess for the object $g_0(x)$ that can be chosen as a random guess or as a more accurate estimation if we have some prior information. Then at the $j$-th iteration, the algorithm proceeds with this step sequence:

1.  Fourier transform an estimate of the object

$$g_j(x) \rightarrow G_j(k) = \mathcal{F}\{g_j(x)\} \tag{198}$$

2.  Replace the modulus of the transformed object with the originally measured one

$$G_j(k) = |G_j(k)|e^{-i\phi(k)} \rightarrow G_j'(k) = |U(k)|e^{-i\phi(k)} \tag{199}$$

3.  Inverse Fourier transform of such estimation

$$G_j'^{(k)} \rightarrow g_j'(x) = \mathcal{F}^{-1}\{G_j'(k)\} \tag{200}$$

4.  Calculate the estimate of the object eliminating (setting to zero) the values in the region $\Gamma$ where the assumption of the image being real and positive are violated

$$g_{j+1}(x) = \begin{cases} g'_j(x) & \text{for } x \notin \Gamma \\ 0 & \text{for } x \in \Gamma \end{cases} \tag{201}$$

The process is iterated until the modulus of the Fourier transform of the calculated object $|G_j(k)|$ is equal to the one measured $|U(k)|$, or when the computed image fully satisfies the object domain constraints. The convergence of the algorithm can be evaluated by the calculation of the recovery error function, which for the GS has been proven to be monotonically non-increasing in function of the step $j$:

$$E_i = \sum_k \left| |G_j(k)| - |U(k)| \right|^2 \rightarrow 0 \quad when \ g_j(x) \rightarrow u(x) \tag{202}$$

Although the function is monotone, this does not guarantee that the process will converge to the exact reconstruction. In fact, the iteration could stagnate in local minima that might not be close to the true reconstruction, leading to meaningless results. Although the problem of the stagnation is very well known and several approaches were proposed to tackle it [46], the main disadvantage of the ER method is the slow convergence rate. To tackle this many other PR methods were proposed, but in general they follow the same general scheme proposed in **Figure II-10**.

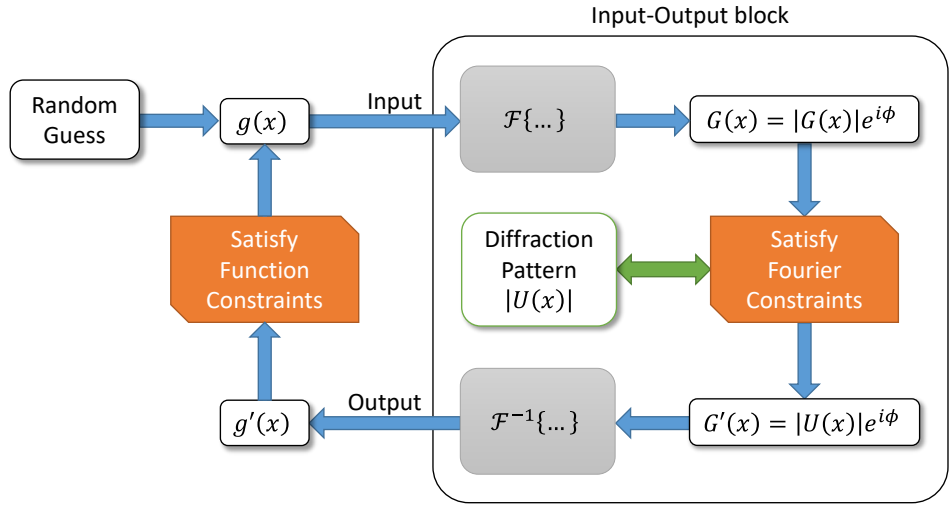

*Figure II-10 | Block diagram of the phase retrieval methods. The algorithm is equal for all of the methods ER, HIO and OSS and only changes which they differ only in their fourth step, where a different decision for the input of the next iteration is taken accordingly with their characteristics. Graphics adapted from [45].*

### Hybrid Input-Output (HIO)

A more efficiently converging method, with respect to the ER, is the Hybrid Input-Output algorithm, which has been proven to retrieve phase solutions faster either for the case of two intensity measurements and for the single plane measurement [45]. Although it is substantially similar to the ER algorithm, there is no mathematical proof that HIO converges to a solution. The only difference with respect to the ER is in the fourth step of the algorithm, while setting the operation accomplished due to object constraints:

4.  Applying a correction to the object estimate that retains the information of the previous step:

$$g_{j+1}(x) = \begin{cases} g'_j(x) & \text{for } x \notin \Gamma \\ g_j(x) - \beta g'_j(x) & \text{for } x \in \Gamma \end{cases} \tag{203}$$

In this way, the chosen function $g_{j+1}(x)$ is no longer the best estimate of the object, like in the ER, but it can be seen as a driving function for the next $g'_{j+1}(x)$. The parameter $\beta$ then, is the variable that controls the modification for the estimation and can push the retrieval of the phase toward faster paths.

The HIO forms a subclass of the more general Input-Output algorithms, where the operation from 1. to 3. described in the ER method can be grouped into a unique non-linear operation with an input $g$ and output $g'$ (rounded block in **Figure II-10**). A useful property of such formulation, is that the output always satisfies the Fourier-domain constraints, therefore if such output also satisfies the object-domain constraints it is a solution of the phase problem. In general, we want that a small change in the input is retained in the output and a logical choice for such variation can be:

$$\Delta g_j(x) = \begin{cases} 0 & \text{for } x \notin \Gamma \\ -g'_j(x) & \text{for } x \in \Gamma. \end{cases} \tag{204}$$

This implies that when the constraints are satisfied there are no variations to the object, while if changes are needed then they must lead to zero changes via a negative variation in the region Γ. This also implies that the most logical choice for a basic *Input-Output* approach is

$$g_{j+1}(x) = g_j(x) + \beta \Delta g_j(x) = \begin{cases} g_j(x) & \text{for } x \notin \Gamma \\ g_j(x) - \beta g'_j(x) & \text{for } x \in \Gamma \end{cases} \quad (205)$$

An interesting property of such kind of non-linear systems, is that if an output g' is used as an input, its output will be itself. Since the Fourier transform of g' already satisfies the Fourier-domain constraints, g' will be unaffected and it goes through the system. Therefore, irrespective of what input resulted in the output g', the output g' can be considered to have resulted from itself as an input. From this point of view, another logical choice for a next input would be:

$$g_{j+1}(x) = g'_j(x) + \beta \Delta g_j(x) = \begin{cases} g'_j(x) & \text{for } x \notin \Gamma \\ g'_j(x) - \beta g'_j(x) & \text{for } x \in \Gamma \end{cases} \quad (206)$$

that can be defined as an *Output-Output* algorithm. It is worth noticing that if we set $\beta = 1$, such formula reduces to the ER algorithm and due to the fact that typically the optimal value for $\beta$ is not one, the ER can be seen as a non-optimal choice within a broader class of algorithms. It is clear then that the HIO is a hybrid combination between the above-mentioned *Input-Output* and *Output-Output* algorithms, that tries to combine the advantages of both, attempting to avoid stagnation problems [46].

In conclusion, HIO can be seen as an iterative algorithm that jumps back and forth in real and reciprocal space applying some operations which push towards the retrieval of the phase associated with the measurement. The no-density region in the real space due to the oversampling of the Fourier transform (assumed to be estimated via the calculation of the autocorrelation) and the assumption of non-negativity are used as constraints in the real space, together with the imposing of the Fourier modulus in the reciprocal space.

### *Oversampling Smoothness (OSS)*

The previously mentioned methods for the solution of the phase retrieval problem work reasonably well with noise-free measurements. High frequency noise in fact, introduces perturbations in the Fourier domain and can lead to instability in the path towards the convergence or, even worse, compromise the possibility of finalizing the phase retrieval problem. In those cases, oversampling in the ratio of 2 or more in frequency domain helps the reconstructions, but in general noisy measurements will compromise quality and convergence. In particular, the oversampling condition assumes that the region outside the reconstruction area has to be set to zero (or converge to), but in many applications the noise reflects its presence in this region creating persisting patterns that mislead the reconstruction.

A new approach was proposed to tackle this effect, forcing a smooth intensity profile in this region via the application of a variable filter outside the image support and finding a balance between ER and HIO toward the search of a global minimum in the space of the solutions [47]. Due to the properties described above, the method is called Oversampling Smoothness (OSS)

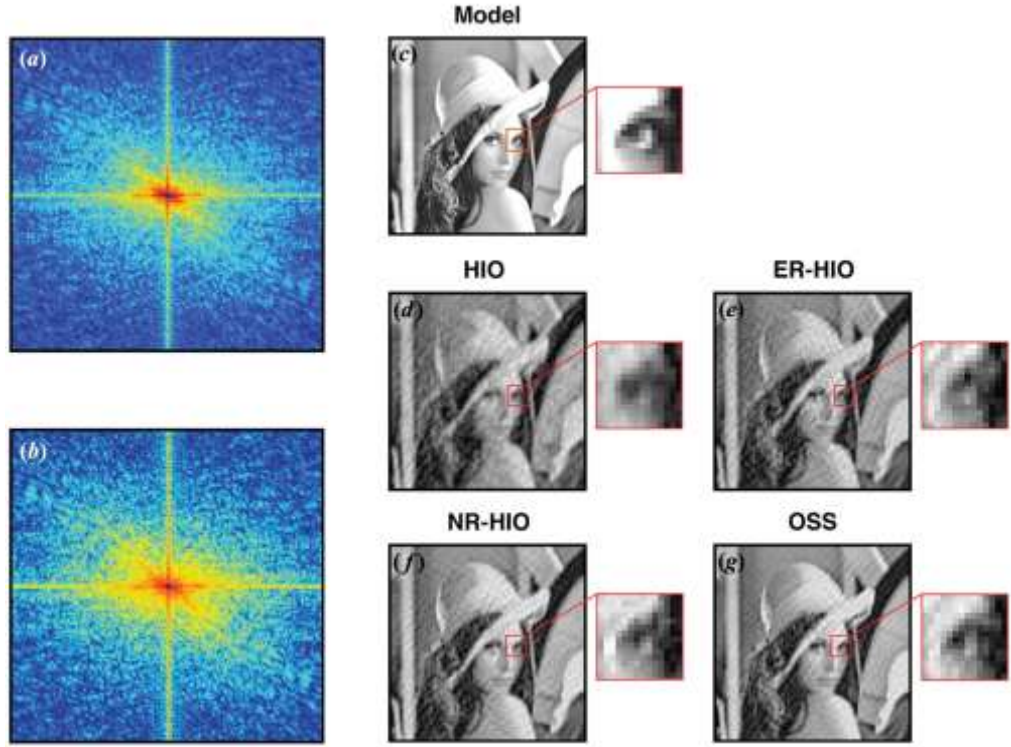

*Figure II-11 | Comparison of phase retrieval approaches based on the alternating projections. a) The noise-free oversampled diffraction pattern of Lena and b) its corresponding pattern perturbed with Poisson noise. C) is the original image, d-g) the reconstructions in presence of noise by using different approaches. We can appreciate the good result obtained by the OSS method presented in the work [47].*

and it is again based on the alternating projections approach of the ER, but with a modified last step:

4. Calculate the image with the HIO criteria and then applying Gaussian smoothing in the region outside the support:

$$g''_j(x) = \begin{cases} g'_j(x) & \text{for } x \notin \Gamma \\ g_j(x) - \beta g'_j(x) & \text{for } x \in \Gamma \end{cases} \tag{207}$$

$$g_{j+1}(x) = \begin{cases} g''_j(x) & \text{for } x \notin \Gamma \\ \mathcal{F}\{G''_j(k)W(k,\alpha_j)\} & \text{for } x \in \Gamma \end{cases} \tag{208}$$

Where $W(k,\alpha_j)$ is a Gaussian smoothing function in the reciprocal space, with a width $\alpha_j$ linearly decreasing in function of the iteration step:

$$W(k,\alpha_j) = e^{-\frac{1}{2}\left(\frac{k}{\alpha_j}\right)^2} \tag{209}$$

Typically, the value for $\alpha$ ranges from values equal to N and it is decreased during the iterations until $1/N$, where N is the minimum size of the image. It is worth noticing that at the first steps, where $\alpha = N$, the algorithm behaves like the HIO because the filter allows almost all the frequencies to pass to the next step. Instead, when the filter size is decreased down to $\alpha = 1/N$, the algorithm suppresses almost all the frequencies from outside the support region and behave like the ER.

*Other methods for Phase Retrieval*

Although we described three different algorithms useful to find the solution of the phase problem, they all belong to the same class of alternate projections algorithms. New approaches have been proposed along the years to find different ways to approach the problem of phase retrieval. Some new classes of algorithm are currently proposed and investigated, such as the semidefinite programming approach, the transport of intensity calculation and sparsity based methods that include further prior information useful for the resolutions of the problem in some specific scenarios. Although a comprehensive review [34] describes them in detail, we are not interested in those even if we do not exclude possible future implementations, in particular in hidden imaging.

*Comparison of the PR-methods*

All of the proposed methods are able to recover the phase of a diffraction intensity pattern but the difference between them is their relative sensibility to noise, efficiency in the exploration of the solution-space and stagnation. Among the others, the ER method is the slowest in the converging rate and it is very sensitive to noisy measurements, although it is the only method which can be mathematically proven to possess a solution for the phase problem. HIO is the most widespread, due to its simplicity and efficiency, but more often is found in combination with final correcting ER iterations to further push down the recovery error function. OSS instead is a promising new trend, which could allow higher resolution retrieval in highly noisy measurements. **Figure II-11** qualitatively shows an example of different results in presence of noisy measurements, in which it is possible to appreciate that the combination of ER-HIO returns better results than pure HIO implementations and interesting results reconstruction with OSS, which seems to obtain the sharpest results among the others. In general, though [34] [48], there is not a strict rule on what kind of algorithm to prefer with respect to the others, but the choice is left to the final user, that is left with the hurdle to find the correct balance between quality and execution speed.

# Chapter III
# CLEAR LAYERS

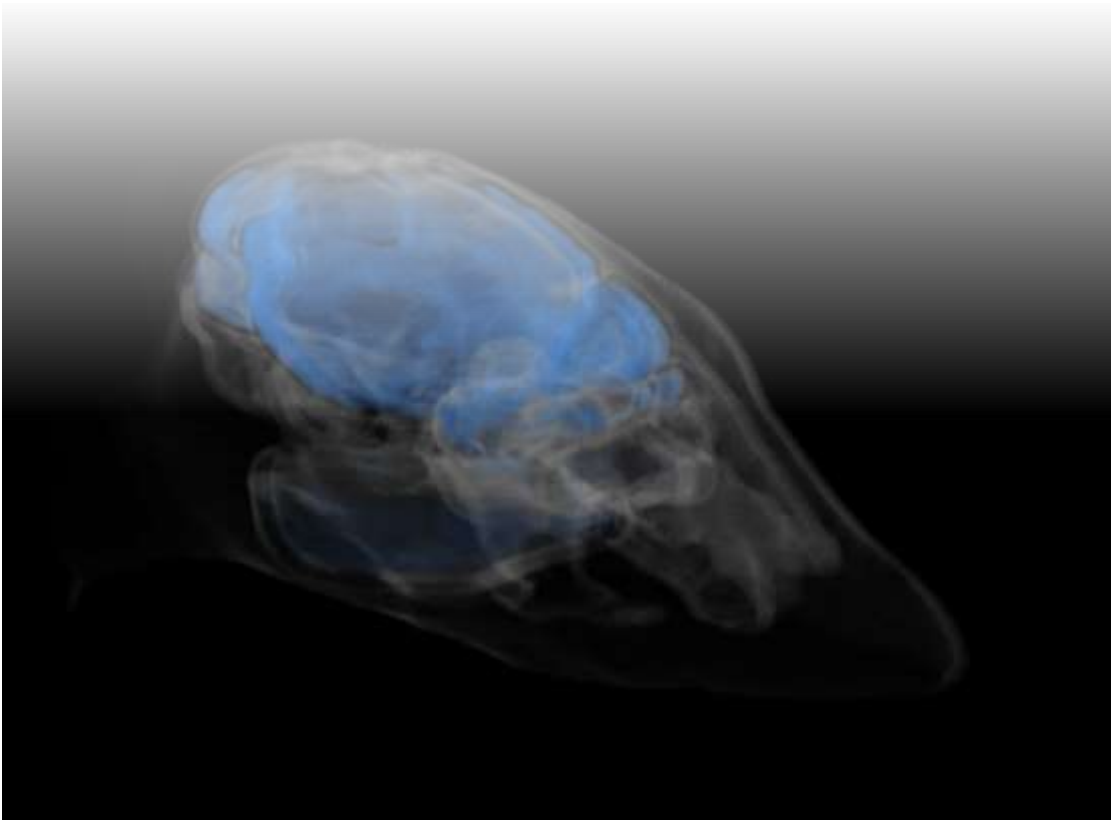

# III.1 Small Animal Imaging

In the first theoretical chapter, we have introduced the concept related to the Radiative Transfer Equation (RTE), the Diffusive Equation (DE) and the Monte Carlo Photon Propagation (MC-PP). So far, we have discussed details about their efficiency and the approximations which we are forced to take into account whenever we prefer one approach over the others. For tomographic imaging purposes, especially for what concerns the small animal imaging, we need such techniques to model the light diffusion within the tissue (or the whole body) to be able to infer the distribution of the quantity that we are interested to image. In fact, when the dimensions of the specimen are bigger than the transport mean free path of the light through tissue (typically in the order of millimeters), it is no longer possible to obtain any kind of reconstruction by using direct imaging approaches. Among the others, one of the most important problems in biomedical optical imaging is the location and quantification of the fluorescence emitted by a certain class of molecules, the *fluorophores*, which are specifically designed to tag various organic structures of interest, such as tumors, plaques, necrosis and many others. If we are, for example, interested to locate a tumor mass in the brain of a mouse, one possible approach could be labeling such tissue with a fluorophore and, by exciting it with the proper wavelength, study its fluorescence emission. The challenge, in this case, is that the visible light (both the excitation and emission) is strongly scattered by the tissue, not leaving any chance of direct observation of the tumor's fluorescence distribution. To overcome this limit, some techniques were proposed to be able to reconstruct the signal of interest by having some a-priori information, such as anatomical shapes of the sample (external and/or internal), its optical properties and the knowledge of the characteristics of the excitation sources. In fact, linking this information with the experimental observation (with detectors or cameras) of how the light propagated through the body, we can approach the fluorescence reconstruction problem. These techniques are commonly referred to as Fluorescent Diffuse Optical Tomography (fDOT) or Fluorescence Molecular Tomography (FMT), and fall into the class of *inverse imaging problems*. The solution of such inverse problems relies on efficient modeling of the photon propagation through the tissue, which is variated to minimize the distance between experimental measurement and numerical model. Here RTE, DE and MC-PP come into play and the results strongly depend upon their accurateness. In general, the DE approach is preferred over the others, due to the fact that biological tissues are highly scattering (the light propagates under diffusive regimes) and because it is very fast and computationally efficient. The fDOT, or more in general Diffuse Optical Tomography, commonly neglects or assumes as insignificant the presence of optically clear regions in biological tissues, estimating their contribution as a small perturbation to light transport. Since the whole imaging community rushes for reaching higher resolution, we examined in detail the inaccuracy introduced by this practice in the context of a complete, based on realistic geometry, *virtual fluorescence Diffuse Optical Tomography* experiment, where a mouse head is imaged in the presence of *cerebral spinal fluid*. Despite the small thickness of such layer, we point out that an error is introduced when neglecting it from the model with possible reduction in the accuracy of the reconstruction and localization of the fluorescence distribution within the brain. The results, obtained throughout the extensive study presented in the following [49], suggest that fluorescence diffuse neuroimaging studies can be improved in terms of quantitative and qualitative reconstruction by accurately taking into account optically transparent regions, especially in the cases where the reconstruction is aided by the prior knowledge of the structural geometry of the specimen. Thus, this has only recently become an affordable choice, thanks to novel computation paradigms that allow to run Monte

Carlo photon propagation on a simple graphic card, hence speeding up the process a thousand folds compared to CPU-based solutions.

### III.1.1 STATE OF THE ART AND PROBLEM DESCRIPTION

Accounting for the effect of light scattering through biological tissue is the major challenge of non-invasive biomedical imaging at optical wavelengths [50] [51]. Despite the fact that light scatters mainly in the forward direction, after a few transport mean free paths ($TMFP$), i.e. a few millimeters of propagation, light becomes completely diffusive losing any information on the initial directionality. Many approaches are being currently developed to enable imaging in different scattering regimes defined as Microscopy, Mesoscopy and Macroscopy [51] depending on the specimen or the functionality considered in the study. The general rule of thumb is that deeper imaging corresponds to lower resolution ability. Enhancement on this side has been feasible in part thanks to the increasing contribution of complex computational methods that deal with data acquisition and post processing, registration, light diffusion simulation and inverse problem based reconstruction. In particular, optical tomographic methods such as Optical Projection Tomography (OPT) [52], Optical Coherence Tomography (OCT) [53] as well as fluorescence-based methods like Selective Plane Illumination Microscopy (SPIM) [54] and Fluorescence Molecular Tomography (FMT) [55] or in the more general form fluorescence diffuse optical tomography (fDOT) used in small animal studies [56] offer the capabilities to reconstruct quantitatively three dimensional models with resolutions down to the sub-cellular level. In this context, the development of novel computational techniques, such as automatic image segmentation and registration [57] [58], image reconstruction [59] [60], phase retrieval [31] [33], computational light diffusion models [61] [62], and the creation of accurate virtual biological phantoms with specific optical properties are playing a crucial role in terms of quantitative and accurate reconstruction of the measured specimen. The development of the biomedical imaging field is therefore tightly bound to the increasing complexity of the computer architectures, which nowadays enables fast parallel computation, approaching quasi real time data processing for most of the applications. One of the most highly time consuming methods, the Monte Carlo Photon Propagation (MC-PP) that we introduced in the first theoretical chapter, can be trivially parallelized [63] [64], increasing the speed up to thousand-fold by its implementation in modern GPU programming paradigms (such as CUDA and OpenCL). This allows simulations of light diffusion in a feasible time scale even in low cost desktop personal computer solutions. Nowadays, as we have already discussed, the most common simulation involving models with complex realistic geometries is performed by solving the Diffusive Equation (DE) to obtain the photon flux distribution within the tissue, which represents an approximate solution of the Radiative Transfer Equation (RTE) [2]. We have seen that the DE however, is accurate only in the diffusive regime, when scattering predominates absorption. Consequently, optically clear layers embedded in scattering tissues constitute a challenge for modelling the photon propagation and therefore they are either commonly neglected [65], treated using radiosity theory [66] [67] [68], included taking into account special boundary conditions [69] or used in mixed DE-MC interfaces [70] and coupled RTE-DE [71] approaches, introducing further approximations or limiting their application to planar slab or simple geometries. On the other hand, MC-PP, being directly based on RTE, is in general a more accurate choice to model propagation through non-scattering regions, despite the fact that it requires intense computational efforts to reach a good signal-to-noise ratio.

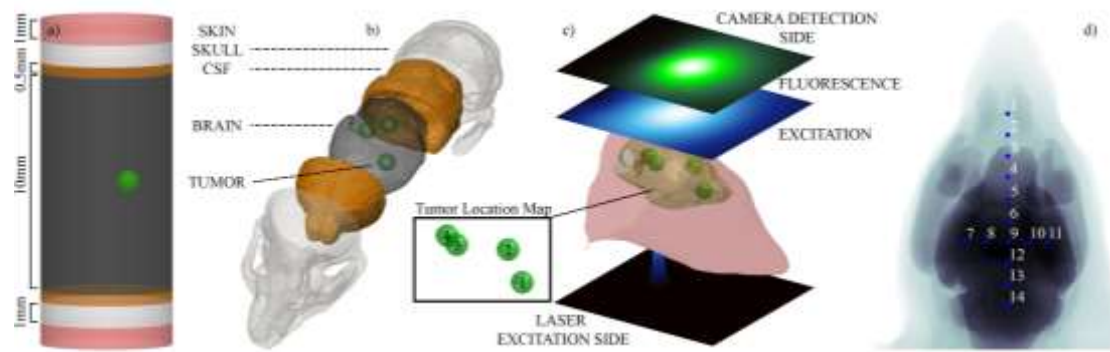

*Figure III-1 |Description of the cylindrical and mouse head models used for the simulations. a) Cylindrical phantom (*cyl_CSF model*) used to validate the calculation of the fluorescent field using MC photon propagation. The sandwich of layers maps the complex mouse head geometry displayed in panel b) in a simple and useful representation. b) Exploded view of the realistic mouse head model used in the simulations. It is possible to notice the brain (depicted in light black) encapsulated by the thin Cerebral Spinal Fluid layer (matte orange). The two tissues are embedded in the skull (light gray) and the skin (pink contour shown in panel c). The other structures (eyes, muscles, skin and glands) are not shown in the graph but were included in the model. The synthetic FMT experiment is presented schematically in c-d). The laser excitation side at the lower surface of the mouse head propagates through the mouse's head reaching the camera side where is detected and stored. The field generated by the laser diffusion excites the fluorescence of the tumor masses (green spheres) which emit photons detected by the camera. Assuming GFP as fluorophore, in this picture the blue represents the excitation and the green the emission. ID numbers in the inner box map the different tumor position examined in our work. Last, in panel d), the raster scan path of the laser at the lower surface of the mouse. The ID numbers identify the position of the source in the plane at the bottom of panel c).*

Even if, in general, biological tissues strongly scatter light due to numerous refractive index discontinuities at the cellular level, in nature it is quite common to find either almost transparent model specimens (such as *Danio rerio* embryos and *C. elegans*) or optically clear biological tissues (fluids, ocular or other empty cavities etc.) enclosed in highly scattering regions. Moreover, of great interest in Neuroimaging, is the fact that the brain and the spine of complex animals is completely submerged into the cerebral spinal fluid (CSF), an optically clear liquid which provides functional regulation of the blood flow as well as mechanical and immunological protection [72] [73] [74] [75]. Such fluid, that circulates in the subarachnoid space all around the brain, is composed by 99% of water which makes it absolutely optically transparent and therefore not possible to consider in the diffusive regime of photon propagation, the commonly used choice especially for FMT and fDOT neuroimaging reconstructions. In this context, the aim of this work is to present a complete study of the effect introduced by the inclusion of the cerebral spinal fluid in the synthetic model of a mouse head. A complete virtual FMT approach based on realistic experiments [76] [77] is used to characterize extensively the complexity of the CSF geometry contribution at the detection level both for the excitation and the fluorescence photon propagation. The forward modeling of both the excitation and the fluorescent emission coming from tumor inclusions in the brain are accomplished with the novel MCX code [63], which enables fast photon propagation in voxelized models based on realistic geometries and perfectly suits the requirements for a straightforward implementation of a camera detection scheme.

## III.1.2 MATERIALS AND METHODS

To perform all the simulations and data analysis reported in this work we used a desktop personal computer based on the CPU Intel i7-4930K equipped with 32 GB of RAM and a GPU nVidia GeForce GTX 780Ti with 2880 CUDA cores. All the data were processed and analyzed in MATLAB environment where for the Monte Carlo simulations we used the MC eXtreme [63] package (release version 0.9.7-2) to exploit the speed of the GPU computation. Tests, evaluations and cross-validations of the simulations using DE were implemented using the Finite Elements Method (FEM) solver from the Toast [61] package. With this low-cost configuration, we reached a speed of about $10^6$ photons per second, computing demanding tasks of about $10^{10}$ photon propagating through the volume in less than 3 hours. Since the most important parts of the calculation are easily parallelized we expect that this time can be further reduced dramatically with the use of more up to date hardware.

*Synthetic Mouse Head Model*

Cylindrical phantom geometries are interesting models to understand the behavior or to validate a methodology, but lack in accuracy when compared to realistic models due to their idealistic geometries. The present study is based on the utilization of a realistic mouse head model with the inclusion of the cerebral spinal fluid layer surrounding the brain. In order to include such layer we performed a new segmentation based on the data freely available from the Digimouse project [78] extending the approach used in our previous works [79] [80], coating the brain with a clear layer structure while keeping the CT measurement for the skull as a fixed reference. Although we are aware of the presence of other ventricular structures containing CSF, we decided not to include them in the segmentation. The Digimouse dataset did not show accurate details on their structure, thus making inaccurate their location within the brain. We considered the brain from the Digimouse atlas as a whole structure constituted of the same tissue's optical properties. We allowed slight brain surface modification (contraction or expansion +0.1mm) in respect to the original segmentation, coating it with CSF layer at various thicknesses. The obtained CSF area was adjusted onto the cryosection dataset (**Figure III-2** a). The area defined in such a way overlaps the dark region in the cryosection data enclosed between the skull and the brain (**Figure III-2** panel b), exactly where the CSF is expected to be located. Since the cryosection images are grayscale intensity measurements, a good segmentation of the new layer is the one that better includes the dark-pixels in the region. We can search for that by looking at the average pixel intensity value in the new layer, where the best segmentation corresponds to the one having the minimum average intensity. From the position of the minima in plot of **Figure III-2**, it is possible to notice that no modification of the brain is needed to obtain a good segmentation for the CSF. We are simply filling the space left between brain and skull in the Digimouse atlas that, in its original representation, had the same value of the skin. The result of the segmentation process is depicted in **Figure III-1** panel b) where it is possible to notice the thin clear layer obtained from the data segmentation that encapsulates the brain. Despite the very thin average thickness of the layer being 0.20+0.05mm, the final CSF's volume inserted is 62.5mm3 which represent quite a wide 15% of the total brain volume of 407.4mm3. Those values are in agreement with what expected from the anatomy, since the cerebral spinal fluid commonly extracted from an adult mouse is in ratio of ~10% in respect of its brain volume [74] [75]. We decided to take into account irregularities of the subarachnoid space to mimic folds on the brain surface by randomly removing clear voxels at the brain boundary with a probability of 50%. From now on in the following text, we will refer to this virtual realistic mouse phantom as the *CSF model*. We name, instead, *noCSF model* the one in which the clear layer is substituted with the adjacent brain tissue. Contracting the surrounding structures to preserve the brain size would have falsified the model, because in realistic FMT experiments the overall shape of the specimen can be accurately measured.

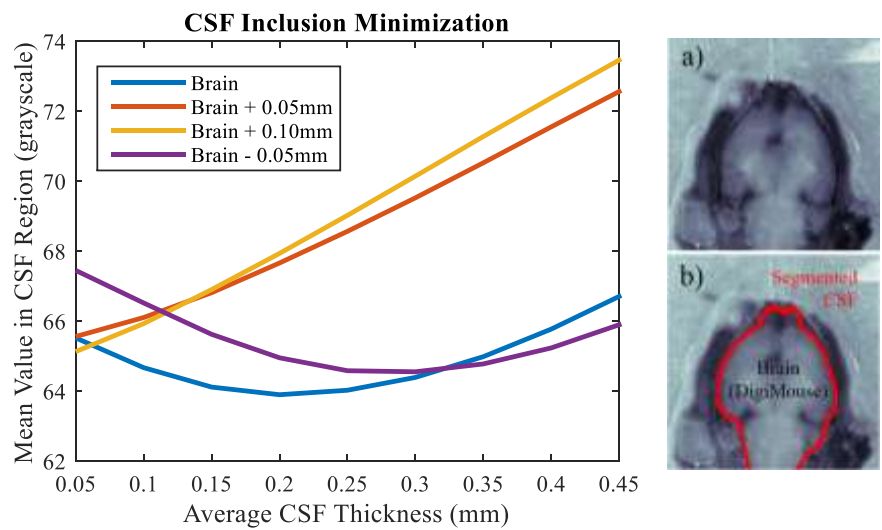

*Figure III-2 | Minimization of the grayscale intensity value for the included CSF region registered onto the cryosection data. A unique minimum is found for the blue curve which represents the original brain segmentation, while its modifications do not give better results thus confirming the quality of the original Digimouse brain segmentation. a) Original data and b) the final layer projected to the corresponding cryosection intensity in grayscale. The skull was kept always fixed assuming its shape known from the CT scan.*

### *Virtual FMT Experiment*

Inverse problem solutions, towards accurate and unique tomographic reconstructions based on diffuse optical measurements, are strongly depended on computational methods to model the forward problem [50]. FMT, in particular, is a modern in-vivo imaging technique [55] that grounds the quality of the fluorescence reconstructions on the calculation of the light diffusion field in biological specimens, thus efforts both at experimental [77] and computational [76] [81] level have been proven to result in apparent tomographic improvement.

In this scenario, we are interested in studying the benefits of the inclusion of clear layers in realistic mouse head models. To quantify the improvement obtained we have carefully designed a complete virtual Monte Carlo FMT experiment as follows:

1. A Gaussian beam is impinged at the lower surface of the mouse head, opposite to the camera detection side, obtaining the excitation field.
2. The emitted fluorescence field is propagated from different spherical fluorescent regions within the brain excited by the previously calculated excitation field.
3. The process is repeated, for all the different beam scanning position collecting the camera detections from the top side of the head (**Figure III-1** panel c and d).

The calculation of the excited fluorescent emission follows the same criteria of the previously validated MC-fluorescence method in paragraph A, except the fact that now; we use a realistic mouse head geometry. We performed two independent set of simulations:

1. Gaussian laser excitation and fluorescence emission for the CSF model, this data will be used as ground truth and considered as the virtual laboratory measurement.
2. Gaussian laser excitation and fluorescence emission with the *noCSF model*, this is expected to give results compatible with DE because diffusive approximation holds in every region of this model.

Every simulation was performed using always the same beam size of FWHM=0.7mm for the Gaussian laser source. As depicted in **Figure III-1** d) 14 scans at different beam positions where performed in transmission for each set of measurements using $10^{10}$ photons, impinging the laser on the lower surface of the mouse head and collecting the camera detection on the opposite side. Each propagation excites 4 different fluorescent spherical regions of D=2mm diameter at different positions, composed by equally spaced fluorophores at the center of each voxel included in the tumor volume. Such spheres are composed of 33371 cubic voxels with 0.05mm resolution. From the center of each voxel a set of 106 photons were launched isotropically to mimic the emission of excited fluorescence and the resulting diffusion is stored in the fluorescence camera detection matrix. Since the problem is linear, by simply summing the appropriate sources, and weighting their contribution with the forward excitation probability, it is possible to build the normalized fluorescent emission as described in part II section A.

*Validation of the MC-Fluorescence method*

The aim of the present study is to analyze the fluorescence contribution excited by laser diffusion through the body via MC-PP. We decided to carefully design a MCX-based algorithm, which starts by calculating the excitation field due to the propagation of a laser source through the body and then weights the emission field for each fluorophore with its local excitation field probability. A cylindrical phantom made by a stack of different tissue layers one on top of the other has been used for the validation of the model. The phantom geometry is composed by either 5 or 7 layers with different optical properties, listed in Table I, that mimic an ideal mouse head. **Figure III-1** panel a) shows the 7-layer phantom constituted of skin, skull, CSF and brain tissues to which we will refer to as the *cyl_CSF model*, while the 5 layer phantom (*cyl_noCSF model*) is obtained by replacing the clear layer with brain tissue to preserve its overall shape. At the center of the cylindrical phantom, and displaced 1 cm from its vertical axis, a spherical fluorescent source ($S_{tumor}$) of diameter *D=2mm* is implanted, which represents a brain tumor mass. The layered phantom together with the position of the green fluorescence source are shown in **Figure III-1** a). A laser excitation field with a Gaussian profile coming from the bottom ($S_{exc}$) with full width at half maximum of *FWHM = 2mm* is propagated through the body leading to the excitation field ($\Phi_{exc}$) computed by solving DE and MC propagation. At this step the excitation field that reaches the spherical region will emit fluorescence with a distribution of $S_{fluo} = S_{tumor}\Phi_{exc}P$, where P is the probability of a photon being absorbed and emitted by the fluorophore and is set to be constant $P = 1$.

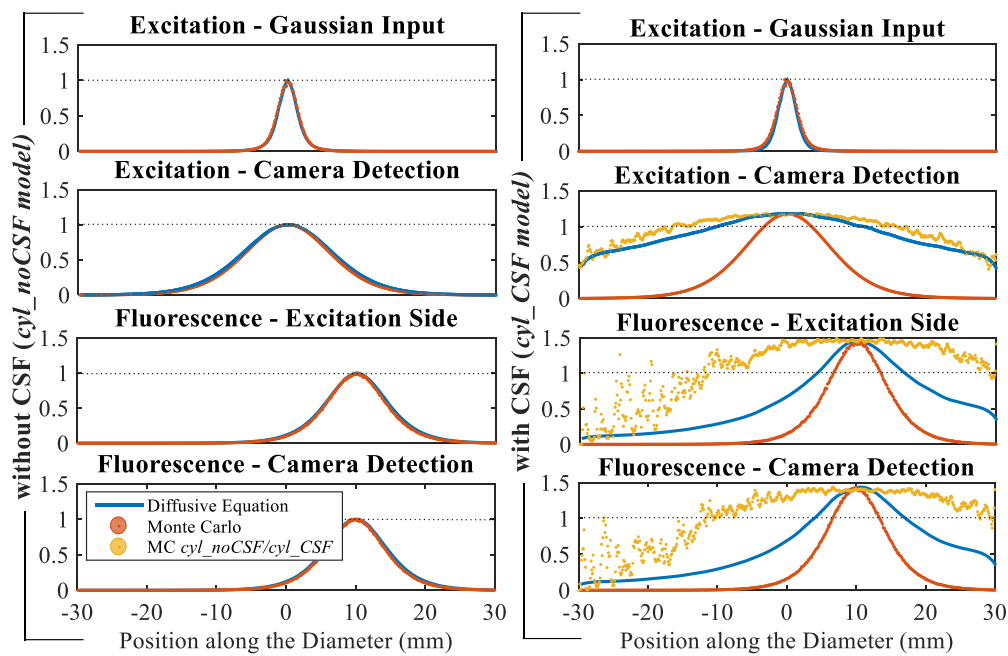

***Figure III-3 | Intensity profiles on the top and bottom side of the cylindrical phantom.*** *On the left, results neglecting the CSF tissue used to validate the fluorescence model with MCX. On the right, results of the solution of DE and MC including the CSF layer. The inclusion of a thin clear layer in the model solved with DE produces diversion from the MC solution. The results for cyl_CSF are peak normalized using the corresponding peak values from cyl_noCSF as reference.*

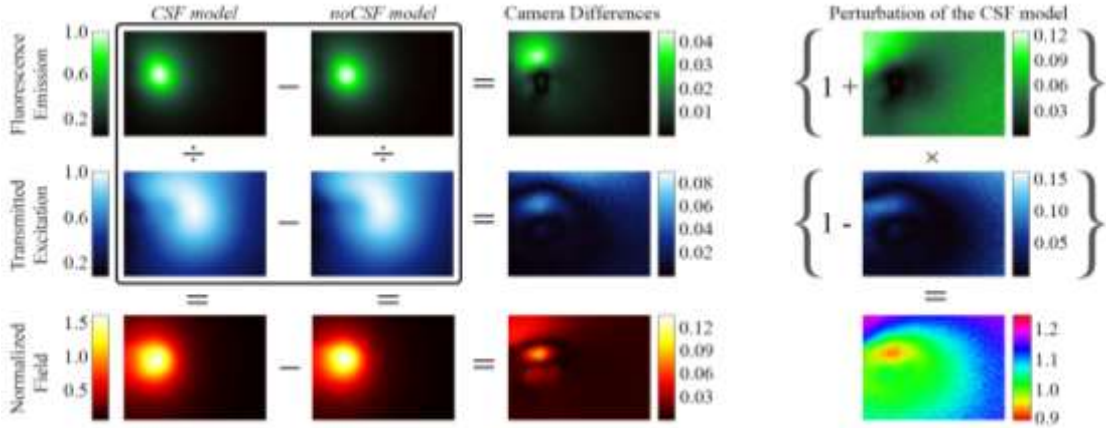

*Figure III-4 | Schematics of the analysis performed with the final results for the CSF and noCSF realistic models. A black box delimits the camera acquisitions of the simulated data. Moving from left to right we analyze the difference of the two, from top to bottom we calculate the normalized field. The color code is green for fluorescence data, blue for laser excitation and red for normalized data. By looking at the difference in the camera detections results that the CSF inclusion modifies the output, resulting in a visible circular region located exactly on top of the brain. On the right side the calculation of the perturbation factor Π. The green region (Π~1) is where the approximation of small perturbation for the noCSF compared with the CSF holds, while it does not in all the surrounding environment.*

For a more flexible MC implementation we decided to calculate the total fluorescent field ($\Phi_{fluo}$) by simply summing up the fields produced by each individual fluorophore which constitute the tumor. Assuming them as isotropic sources ($S_{iso}^i$) the total field can be written as $\Phi_{fluo} = \sum S_{iso}^i \Phi_{exc}^i P$, where $\Phi_{exc}^i$ is the value of the excitation field at the fluorophore location $i \in S_{tumor}$. In this way, it is possible to store every propagation set and modify the shape of the tumor region by simply changing the list of fluorophores included in the summation. After the calculation of the total field for both MC and DE solution, each field as well as the camera detections were normalized to allow direct comparison of the results.

## III.1.3 RESULTS

Thanks to the speed offered by the combination of MC and GPU, an extensive amount of simulations were performed to validate and statistically examine the data we obtained. We found out that to reach a very good signal to noise ratio (SNR) in transmission, a minimum amount of $10^9$ photons had to be propagated within the typical dimensions for the volumes treated. Typically, in each simulation we run 100 sets of $10^8$ to check statistics of the simulations, keeping the MC SNR within a threshold lower than 1%. For the fluorescence emission, the average number of photons simulated ranged from a minimum of $10^9$ to $10^{11}$ depending on the volume of the emitting sphere considered in each particular experiment.

TABLE I
OPTICAL PROPERTIES OF MOUSE HEAD TISSUES

| ID | Tissue Type | $\mu_a$ $(mm^{-1})$ | $\mu_s$ $(mm^{-1})$ |
|----|-------------|---------------------|---------------------|
| 1 | Skin | 0.0191 | 6.6 |
| 2 | Skull | 0.0136 | 8.6 |
| 3 | Eye | 0.0026 | 0.01 |
| 4 | Masseter Muscles | 0.0240 | 8.9 |
| 5 | Lachrymal Glands | 0.0240 | 8.9 |
| 6 | Brain | 0.0186 | 11.1 |
| 7 | Cerebral Spinal Fluid | 0.0026 | 0.01 |

For all the tissues simulated in our model we assumed the same anisotropy coefficient which for biological materials points in the forward direction (g = 0.9) and the same refractive index (n = 1.37). Values for the optical properties were taken from literature [37]. We report the effective reduced scattering coefficient $\mu_s' = \mu_s(1-g)$ to refer directly at the transport mean free path defined as $TMFP = 1/\mu_s'$ in the final discussion.

TABLE II
LOCATIONS OF SOURCES AND FLUORESCENT TARGETS

| ID | x | y | z | ID | x | y | z |
|----|-----|-------|---|----|------|-------|-------|
| 1 | 7.0 | 9.25 | 0 | 11 | 16.0 | 12.25 | 0 |
| 2 | 8.5 | 9.25 | 0 | 12 | 17.5 | 9.25 | 0 |
| 3 | 10.0 | 9.25 | 0 | 13 | 19.0 | 9.25 | 0 |
| 4 | 11.5 | 9.25 | 0 | 14 | 20.5 | 9.25 | 0 |
| 5 | 13.0 | 9.25 | 0 | | | | |
| 6 | 14.5 | 9.25 | 0 | ID | *Fluorescent Targets* | | |
| 7 | 16.0 | 6.25 | 0 | 1 | 12.0 | 9.0 | 10.15 |
| 8 | 16.0 | 7.75 | 0 | 2 | 15.0 | 11.2 | 12.0 |
| 9 | 16.0 | 9.25 | 0 | 3 | 18.0 | 7.5 | 13.0 |
| 10 | 16.0 | 10.75 | 0 | 4 | 20.0 | 9.0 | 13.0 |

The main table lists the locations of the 14 laser sources impinging the skin on the lower side of the mouse head (in mm unit). In the sub-table on the right, the central location of the 4 spherical targets inserted within the brain during this study.

*Monte Carlo Fluorescence in a Cylindrical Phantom*

At the first step, the validation for the fluorescence MC-PP is needed in order to proceed further in the study. Considering the two *cyl_noCSF* and *cyl_CSF models* with the 5 and 7 layers cylindrical geometry, the first one without the clear layer, we propagated the laser excitation and we collected the output at the camera detection plane at the lower surface of the phantom. We used the same Gaussian input for both the MC simulations. The validation of the method is performed in the 5 layers phantom without CSF regions (the *cyl_noCSF model*, leftmost column in **Figure III-3**), where both DE and MC returned comparable results at any stage of the simulation always within less than 1% of error. We conclude therefore, that our method for modelling the fluorescence with MCX within the brain is consistent and it can be used in the following sections of this work. With the *cyl_CSF model*, instead, we immediately notice that, in presence of a thin (*0.5mm*) clear layer, solving the DE returns an inconsistent solution at the camera detection side, as depicted in the right column of **Figure III-3**. The laser diffusion is corrupted already at the first crossing with the clear region, resulting to an incorrect calculation for the photon flux in more than 86% of the remaining volume. The corrupted total flux that reaches the fluorescent target in the brain region leads to an incorrect propagation for the fluorescence emission as well, while MC calculations are robust in every step of the simulation preserving consistent diffusive solutions for the Gaussian input. Excitation and camera detection sides return indistinguishable profiles due to the z-symmetric location of the emitting fluorescing target. It is important to report that, the output profiles of the laser beam excitation are not strongly affected by the presence of the clear layer itself, making DE/MC *cyl_noCSF* and MC *cyl_CSF* results almost indistinguishable except of a scaling factor due to different total intensity detected, as already shown in our preliminary work [79]. By plotting the ratio function *cyl_CSF/cyl_noCSF* for the MC simulations (orange plots in **Figure III-3** right column, translated along y from the peak of *cyl_CSF* down to the origin for better visibility) we can notice the effect of the CSF inclusion onto the output profiles. The ratio is quite flat and symmetrical around the peak and increases everywhere else, implying that the signal is negligibly broadened when the CSF is inserted but no substantial changes in shape of the signal are found. This is mostly due to the simple geometry of the phantom used, in which the approximation introduced by removing clear layers does not affect significantly the light propagation. Neglecting the CSF layer, in this case, could be considered as a small perturbation (around 2% in terms of output shape, around 20% in terms of total intensity transmitted) compared to the ground truth solution given by the *cyl_CSF* model.

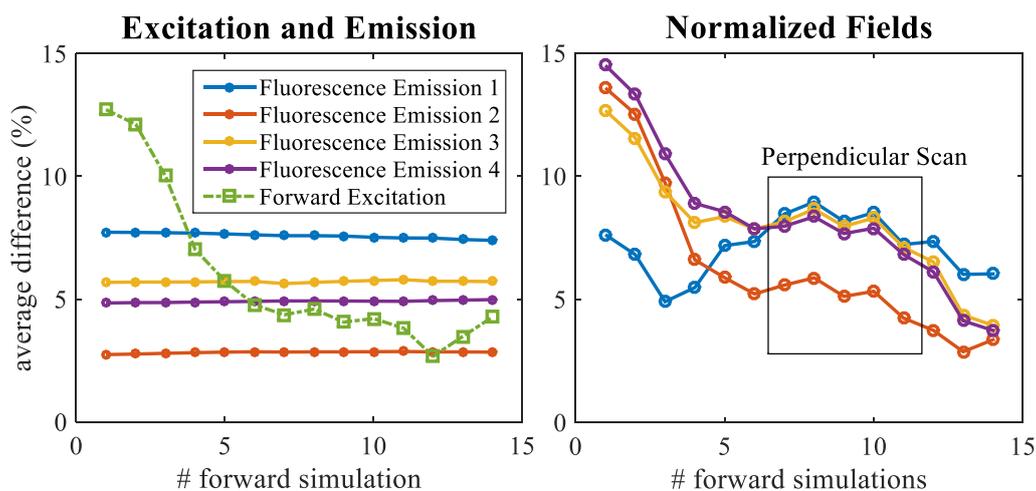

*Figure III-5 | Error committed during the simulations without CSF.* The left panel shows the average difference at the camera detection both for excitation and fluorescence emission at each step of the scan, where the forward simulation number identifies the position of the beam given at Figure III-1 panel d). The right plot represents the difference after the calculation of the normalized data taking into account excitation and emission. The box high-lights the scan perpendicular to the mouse head direction.

*Virtual MC-FMT experiment for Mouse Head*

Following the validation of the MC simulation for the fluorescence within an inhomogeneous cylindrical phantom of the brain, we proceed on to examine the results of the simulations in the more realistic case of the mouse head geometry. Impinging a Gaussian laser beam at the lower surface of the head phantom and collecting the light transmitted at the top, we compared the results with and without the inclusion of the CSF layer, raster scanning the mouse head as showed in **Figure III-1** panel d). Compared to the cylinder geometry a completely different situation arises when considering such complex geometries. The analysis we have performed running our *CSF* and *noCSF model* provides insight on the underlined phenomena. **Figure III-4** presents schematically the calculations performed in the virtual FMT experiment with and without the CSF together with the difference between the two situations. Raw camera images for both the excitation and fluorescence fields are shown and it is possible to notice, by subtracting one to the other $\Phi_{\text{diff}} = \Phi^{CSF} - \Phi^{noCSF}$, the structural differences between the two models in column 3.

The presence of the CSF then, becomes significantly visible and inevitably introduces inaccuracies in the normalized field. Calculating the average percent variation for each of the simulations performed (image-average of the variation field defined as $\Phi_{var} = |\Phi^{CSF} - \Phi^{noCSF}|/\Phi^{CSF}$), shown in the plots of Figure III-5, it is possible to notice another interesting phenomenon. By looking at the average variation between the models, it turns out that the excitation output error strongly depends upon the source positioning, while all the different fluorescent target regions, once excited by the laser at every scanning point, are affected by an almost constant average error of the camera intensity distribution. Normalizing the fluorescence field with the excitation, i.e. calculating the normalized field $\Phi_N = \Phi_{fluo}/\Phi_{exc}$ [2] [76], complicates even further the situation and spreads the error along each simulation as shown in Figure III-5 b). Performing a statistical analysis of the obtained data, we pointed out that the error committed is all but uniform at the camera level. We present thus, the distribution of the errors in the percent variation field($\Phi_{var}$) that help us to evaluate the probability of committing an error by using the *noCSF model*. The histogram in Figure III-6 clearly shows this for fluorescence at position 3, although an identical trend was found for all the other source positions.

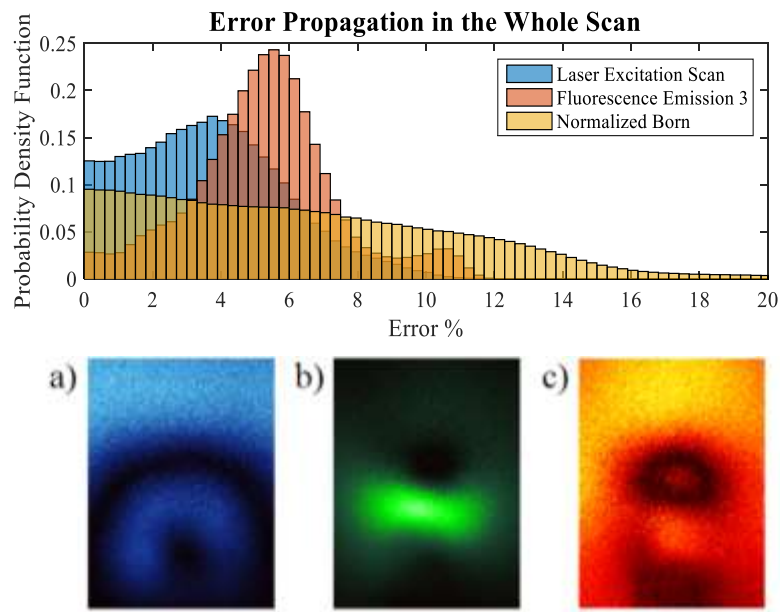

**Figure III-6 | Histogram showing the pixel error at the camera level during the whole scanning process of fluorescence at position 3.** *The distribution of the errors in the presence of the CSF is not predictable and it is possible to notice how such inaccuracy is spread up to the normalized measurements. The lower panels show the differences at camera level for the: a) laser excitation, b) fluorescent emission and c) normalized measurements at the scanning step 9.*

The error committed in the excitation affects the fluorescence emission, and is spread around by the normalization procedure. Although absolute pixel error in normalized camera detection typically ranges around 5–10% by examining the absolute difference of camera detections in the lower part of Figure III-6 we can easily visualize the real effect caused by the introduction of the clear layer in the mouse model. It is worth noticing that already at the excitation level the presence of the CSF layer, is strongly visible resulting to the dark region in Figure III-6 a), and significantly influences the fluorescence as shown in Figure III-6 b). The combination of those two errors strongly affects the normalized measurements used for the fluorescence distribution reconstruction, as presented in Figure III-6 c). We found a wide range of unpredictable differences at every stage of the synthetic experiments performed. In particular, we noticed that the dark ring appears at every beam propagation after transmission through the brain, making it a characteristic fingerprint of the presence of the clear layer. Even if a relatively very thin CSF layer is included, the data always presents a complexly shaped function and not a simple rescaling as in the case of the cylindrical phantom, making the correction of *noCSF* impossible. Considering the *noCSF* data as a small perturbation in comparison to the *CSF model* seems to be not uniformly valid. To check this we calculated the perturbation of the normalized field introduced by considering the following formula:

$$\frac{\Phi_{fluo}^{CSF}}{\Phi_{exc}^{CSF}} = \frac{\Phi_{fluo}^{noCSF} + dE_{fluo}}{\Phi_{exc}^{noCSF} + dE_{exc}} \sim \Pi \cdot \frac{\Phi_{fluo}^{CSF}}{\Phi_{exc}^{CSF}} \quad (210)$$

where $\Phi$ are the fields indicated by their superscripts and subscripts, $\Pi$ is a perturbation term and $dE$ are the errors with respect to the *CSF model*, such as $dE = \Phi^{CSF} - \Phi^{noCSF}$. In this way, the perturbation term can simply be written as:

$$\Pi = \left(1 + \frac{dE_{fluo}}{\Phi_{fluo}^{noCSF}}\right)\left(1 - \frac{dE_{exc}}{\Phi_{exc}^{noCSF}}\right) \quad (211)$$

and as we can see in Figure III-4 on the right column such error is absolutely not constant on the field detected by the camera. We found out that regions where the perturbation assumption ($\Pi \sim 1$) still holds (green) are present together with regions where the approximation is no longer valid (blue to red) in a complex variety of configurations. Remarkably in every normalized measurement we found the red region located exactly on top of were the fluorescence was emitting.

Output profiles at camera detection, then, inevitably contain non-trivial information about the presence of the clear CSF region, yielding its description very important to include if we aim to increase accuracy of forward-based inverse reconstruction models. It is clear that the presence of the cerebral spinal fluid influences the shape of the output profile, i.e. the camera detected intensity, therefore cannot be uniformly considered as a small perturbation to the photon propagation.

### III.1.4 DISCUSSION AND CONCLUSIONS

With this extensive study, we carefully characterized the importance of accurately considering clear layers for neuroimaging applications. Although the mouse head is very small and the clear cerebral spinal fluid layer is quite thin compared to the surrounding structures, the geometrical complexity of the surface makes its inclusion a not negligible perturbation for the photon propagation, while aiming at accurate simulations. We have initially shown that, in planar geometries, the Diffusive Equation returns inaccurate field calculations in the presence of a transparent layer. Nevertheless, not considering such layer seems to be acceptable in the case of ideal geometries where the presence of CSF acts only as a scaling factor for the output profiles. The shape of the transmitted field is affected only by a slight symmetric broadening possibly due to the straight propagation of the photons in the transparent tissues, but can still be ignored making the *cyl_noCSF model* still usable while solving DE. This was verified comparing MC-PP in the presence of clear layers (*cyl_CSF model*) with Diffusive Equation solutions using cylindrical phantoms (*cyl_noCSF model*), a widely-used tool for preliminary tests and studies but, indeed, not suitable for accurate modeling nor for more complex FMT reconstructions of biological specimens. However, the assumption that in simple planar structures, thin transparent layers can be considered as small perturbations to the photon propagation is completely rejected when complex realistic geometries are investigated. This is the case when complex and high resolution atlases [78] [82] have to be used to create geometrically detailed synthetic models in combination with accurate simulation tools, aiming to increase resolution abilities. We found that structural geometry plays an important role in photon propagation. In several ways, throughout this text, we extensively quantified the error introduced when the CSF is neglected. We studied the differences in the output signal of the system at the detection plane (i.e. the measurements in a real FMT experiment) by plotting the average camera differences (Figure III-5). We also drew histograms to show the statistical error distribution per pixel difference (Figure III-6) and the variation of the *noCSF model* as a function of the CSF thickness (Figure III-7). In addition, we directly showed the camera detection fields for each case (Figure III-4, Figure III-6 panels a-c), the transversal fields differences (Figure III-7) and finally the effect of considering the *noCSF model* as a perturbation compared to that of *CSF* (Figure III-4, right column). A broad range of differences in the output profiles at detection level emerged in every stage of the synthetic FMT experiment, possibly leading to addition of inaccuracy in the inverse reconstruction process. The main message emerging from our study is that the measured signal difference between the *noCSF* and the *CSF* starts as an almost uniform scaling for simplistic geometries, such as layered cylinders, but becomes a significant structural difference when a more realistic atlas is used.

The CSF layer in the mouse head model then leaves trace of its presence not only by its optical properties, but most probably due to its intrinsic geometrical features. Indeed, it is worth noticing that the perturbation to the correct field solution of the optically transparent layer becomes apparent when replaced with an optically scattering layer. In fact, the CSF is embedded between different scattering regions and this region need to be reconfigured to accurately solve the Diffusive Equation; either expanding the skull or the brain by 10% of its volume or, even worse, shrinking all the remaining tissues surrounding the brain to cover the gap left in the atlas. It is this replacement that introduces an unpredictable error, reducing the reconstruction ability of diffused optical tomography in the presence of transparent regions. Even by performing a brief study of the effects of a CSF with varying thickness, we found interesting results. The left graph in Figure III-7 presents the percent variation of the camera detection for the *CSF model*, increasing the thickness of the segmented clear layer, compared to that of *noCSF*. It is visible that in such a case the excitation field is affected by a slowly increasing perturbation, while the fluorescent field is strongly influenced by the CSF thickness. Such thickness related effect then is suggesting that a correct segmentation of the clear region is crucial to correctly calculate the excitation and emission fields.

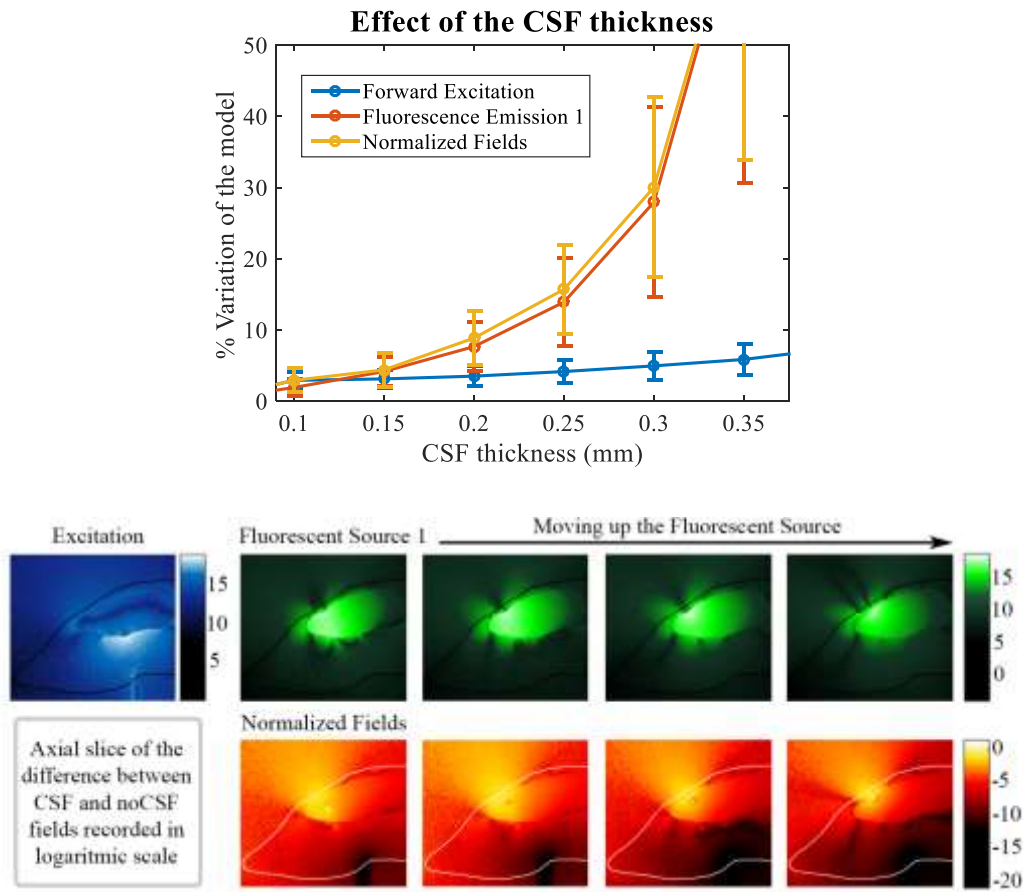

***Figure III-7 | Light diffusion at different regimes within the mouse sample for excitation source at position 9 and fluorescence 1.*** *On the left plot we report the percent variation as a function of the thickness of the CSF layer compared to the noCSF model. Thickening the clear layer results in an increased variation at the detection level, proving the importance of a correct segmentation of such tissue. On the right, the color code indicates the forward laser propagation (blue), the excited fluorescence (green) and the normalized excitation-emission field (red). The drawings show the axial slice of the mouse's head in which we display the difference of the CSF in respect to the noCSF model. We use the logarithmic scale to reduce the effect of the high contrast in the camera differences in order to improve their visualization. It is worth noticing, by looking at the field's difference, how the whole brain field become visible although the only perturbation introduced is a thin clear layer surrounding it. By moving the source slightly up towards the CSF layer, the difference does not remain constant modifying the shape and features.*

At last, it is worth to look at what happens inside when inserting a very thin clear layer surrounding the brain. Figure III-7 shows the difference of the fields inside the head as a function of the depth of the fluorescence source 1. It is quite evident to notice how the whole brain results are affected by the clear layer insertion, both at the excitation and emission, and how it is not possible to correct with the normalization procedure. Moreover, moving slightly up towards the skull the fluorescent tumor results in a modification of the shapes and features of the fields, complicating even further the scenario. This extended study would not have been possible if fast parallel Monte Carlo GPU [62] implementations have not become an affordable tool in terms of time and hardware resources. Until now only expensive CPU clusters could have allowed the same study for such complex realistic models, making the usage of MC-PP for tomographic reconstructions a highly costly tool. We believe that the need for accurate and quantitative high resolution diffuse optics measurements [51] [55] [77] can now fully exploit the speed offered by GPU Monte Carlo techniques opening room for new forward and inverse approaches. In this context, the introduction of clear layers in the computational models is a natural step towards modelling realistic photon diffusion within biological tissues and indeed of great importance to accurately outcome diffuse optical tomography shortcomings. Parallel computational techniques are entering convincingly in every step of Computed Tomography processes, from fast image segmentation [58], inverse reconstruction algorithms [60] to forward simulation methods [63] [64] unveiling rapid and advanced enhancement at every step of the imaging process. In this scenario, we believe that GPU-MC methods have the potential to be used for diffuse optics tomography reconstruction because of their superior accuracy in handling at the same time optically clear and scattering regions. Especially brain, spinal cord and eye imaging can be directly taken into account without neglecting the clear layer in which they are embedded or constituted. In our future and currently ongoing studies we are planning to implement accurate reconstruction techniques such as the matrix free approach [83] [84] currently used only with diffusive approximations, offering further room for improvement and direct application to in-vivo laboratory measurement. Moreover, inspired by our current results, we are beginning a novel study on the effects introduced by thickness changing of the cerebral spinal fluid layer, as it typically happens while Alzheimer's disease progresses in human brain. We strongly believe that this will open room for new discussion, not only for the increased resolution abilities, but more interestingly for the definition of new kind of biomarkers that at the moment can only be taken into account with accurate Monte Carlo simulations. Further simulations together with experimental studies will be performed in the near future to understand more deeply the subtle role played by the cerebral spinal fluid for Optical Neuroimaging.

# Chapter IV
# SCATTERING LAYERS & AUTOCORRELATION IMAGING

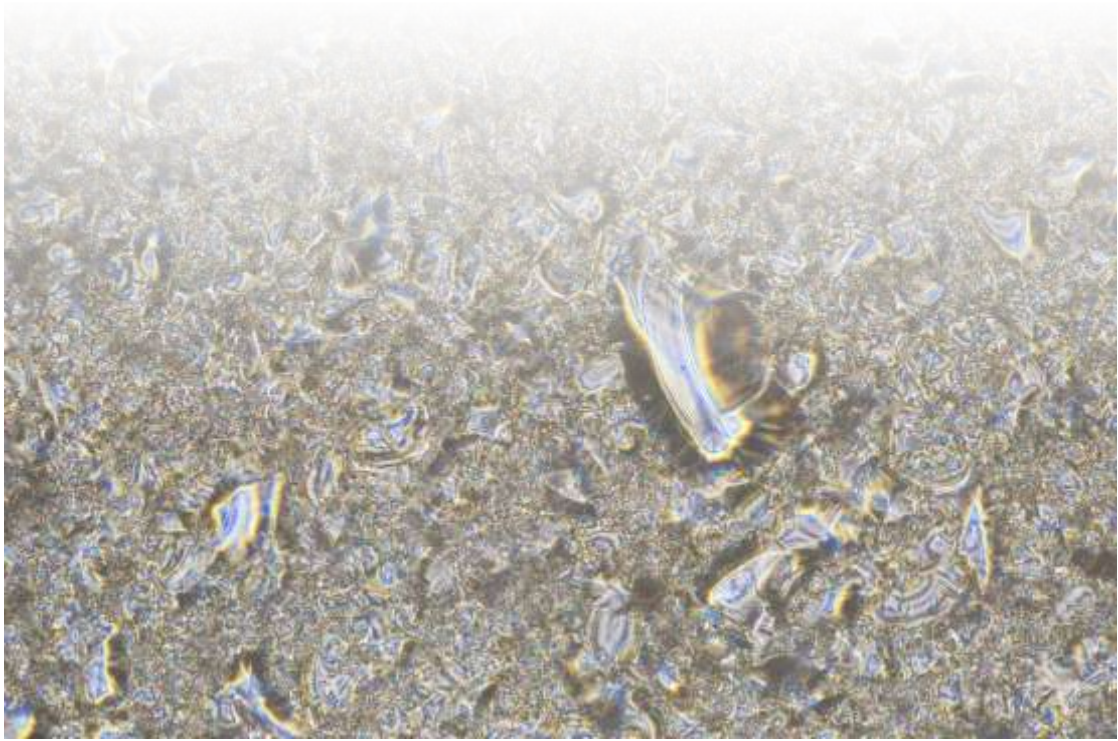

# IV.1 Hidden 3D Imaging

We have analyzed so far how tissue scattering, in particular biological, being often particularly intense at optical wavelengths precludes the possibility to perform direct measurements. In fact, only those ideal cases in which the specimen is transparent or thinner than $1 \, TMFP$ allow us to use microscopic techniques to image it either in two- or three-dimensions. Inverse reconstruction approaches have to be taken under consideration for all those cases of dealing with diffusive measurements. To complicate further the situation, the presence of non-scattering regions surrounded by highly scattering tissues introduces complexity in the reconstruction. Even if it might sound counterintuitive (i.e. we could be pushed to think that less scattering would facilitate measurement), the reason for this is that the theoretical model commonly used as a starting point for the reconstructions fails at regimes of low diffusivity. Different approaches have to be taken into account for such extreme cases, in which higher resolution is requested and thus clear layers cannot be anymore considered as an anatomically negligible structure.

On the other hand, purely scattering layers could absolutely hide the object of interest. In regimes of high light scrambling, in which the phase of the photons also plays an important role, we might not be able to use any of those theoretical approaches examined so far. When we examined the formulation of the RTE, DE and MC-PP, in fact, we have always implicitly (and voluntarily) forgotten the phase contribution, which so far we considered as a non-wanted effect that reduces the resolution ability of the imaging system. Phase-related effects express themselves as an apparition of random speckles in the measurements that we might be pushed to reduce to obtain sharper images. But, in some cases, the scattering is so much that what we observe is just a, seemingly random and information-less, speckle pattern. In the same way as the glass shower hides the person behind it, scattering layers could hide interesting biological structures such as cocoons, eggs, or the skull itself hiding the brain. This is the other side of the coin in light propagation, in which an object of interest is totally hidden due to random phase scrambling of its envelope. In this scenario of seemingly impossible condition to produce and retrieve any image of the hidden object we will focus the following chapter, analyzing what could be done in optical imaging for tomographic application for imaging objects through scattering layers.

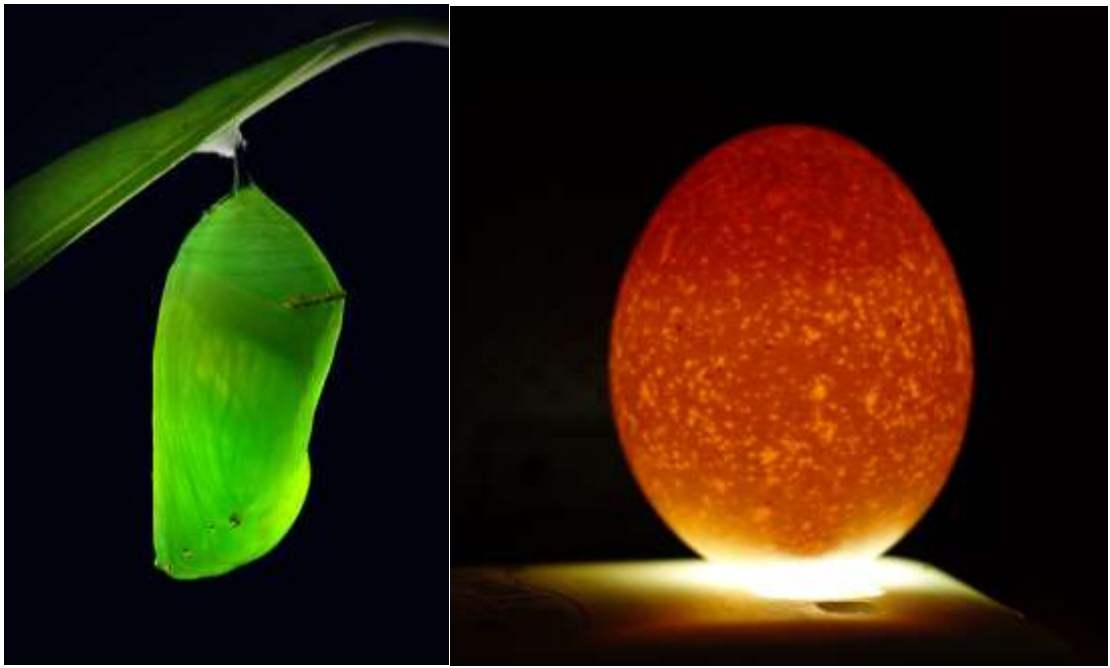

*Figure IV-1 | Scattering layers enclosing interesting biological entities. On the left the Monarch butterfly cocoon (source: the internet). On the right, a chicken egg illuminated with a bright light LED from its bottom side (Nikon D3, Sigma 105 mm f2.8 objective lens, Daniele Ancora).*

As we already discussed, in fact, optical tomography in biomedical imaging is a highly dynamic field in which non-invasive optical and computational techniques are combined to obtain a three-dimensional representation of the specimen we are interested to image. Although at optical wavelengths scattering is the main obstacle to reach diffraction limited resolution, recently several studies have shown the possibility to image even objects fully hidden behind a turbid layer exploiting the information contained in the speckle autocorrelation via an iterative phase retrieval algorithm. In the present chapter, we explore the possibility of blind three-dimensional reconstruction approach based on the Optical Projection Tomography principles, a widely used tool to image almost transparent model organism such as *C. elegans* and *D. rerio*. By using autocorrelation information rather than projections at each angle we prove, both numerically and experimentally, the possibility to perform exact three-dimensional reconstructions via a specifically designed phase retrieval algorithm, extending the capability of the projection-based tomographic methods to image behind scattering curtains. The reconstruction scheme we propose is simple to implement, does not require post-processing data alignment and moreover can be trivially implemented in parallel to fully exploit the computing power offered by modern GPUs, further reducing the need for costly computational resources. We would like to mention that, the present work has been published under the form of two conference proceedings [85] [86] and is currently under consideration in Scientific Reports. Furthermore, we have been invited to present this project in "Methods", a high impact and invited-only journal published by Elsevier.

## IV.1.1 HIDDEN IMAGING MODALITIES – STATE OF THE ART

One of the most common trends nowadays is to combine novel optical biomedical imaging modalities with advanced computational tools to perform high-resolution and accurate three-dimensional tomography [51]. In fact, depending on the specimen of interest, the depth and the resolution at which we want to image, a broad choice of techniques currently can cover mostly every measurement scenario. However, the continuous need for new tools and novel analysis techniques operating in non-usual conditions is pushing the scientific community to import approaches from diverse fields of research [51]. One of the most theoretically and experimentally established technique is the Optical Projection Tomography (OPT) [52] that allows investigation of optically transparent samples. While performing an OPT measurement, the common assumption is that the specimen is only absorbing and the scattering is considered negligible. In this case strong theoretical background supports the image reconstruction, for example via inverse Radon Transform or using other iterative reconstruction techniques [87] [59] [60]. Although several techniques are specifically designed to aid the reconstruction under difficult conditions [88] [89], when the scattering dramatically increases or, even worse, in case the sample is embedded into an opaque curtain (such as egg, shell, or cocoon) the technique immediately results inefficient. At the moment, then, we are optically blind in this imaging regime, although it has been shown that imaging behind turbid biological samples and around corners [31] [33] is possible under specific conditions, opening possibilities to image some interesting biological entities such as the Drosophila pupae enclosed in its cocoon, as well as shells or even eggs of different animals. In fact, recently, phase retrieval techniques are experiencing an increase of interest in the optical (and biomedical) imaging community due to the possibility that they offer to reconstruct objects hidden by opaque curtains. The basic principle that lies behind the use of such technique in order to enable hidden imaging is that, under the memory effect conditions [26], the autocorrelation properties of the object measured are preserved even in the case that the phase information is lost.

In the case that an object of interest is hidden by a scattering layer of optical thickness L, placed at distance $u$ and emitting at wavelength $\lambda$, the chances of retrieving its phase are related to the dimension of such object. The memory effect principle states that two points located at distance D will produce the same, but shifted, speckle pattern if their distance is shorter than:

$$D < \frac{u\,\lambda}{\pi\,L}\,,\tag{212}$$

which intrinsically implies that the two speckles produced are highly correlated. On the other hand, if the points are located at greater distance, such correlation decays rapidly resulting to the generation of a different speckle pattern. The superposition of the two shifted speckle patterns will still retain the information of such displacement in its autocorrelation, allowing us to assume it to be identical to the autocorrelation of the two points located at distance $D$. This principle will hold for any point distribution confined into a circular window which has a diameter equal to the field of view ($FOV$):

$$FOV = \frac{u\,\lambda}{\pi\,L}\,.\tag{213}$$

For any object confined within this $FOV$ the autocorrelation will be preserved even if the phase is scrambled by the scattering layer, giving us useful information to image in such difficult scenario. In this description, the scattering layer acts as a lens which results to a magnification $M$ given by the ratio between the distance of the camera focal plane in respect to the turbid layer $v$ and the distance of the object hidden behind the layer $u$:

$$M = \frac{v}{u}.$$ (214)

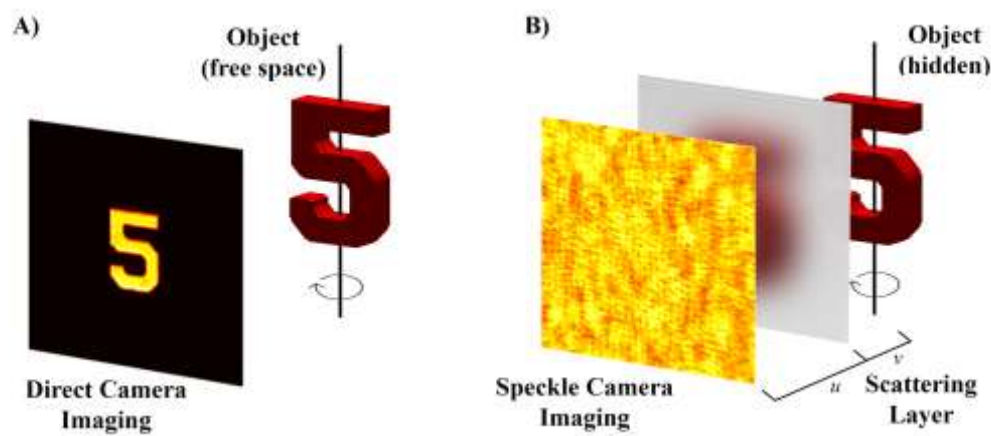

**Figure IV-2 | Schematics of an ideal OPT hidden experiment.** *On the left panel A), the direct imaging of the object selected from the USAF target. Consequently, we numerically created a volume object (the thick red 5) to perform the OPT synthetic experiment. In panel B) the same object hidden by the scattering layer produces a speckle pattern that retains the autocorrelation information. Both the camera imaging shown in this scheme were acquired experimentally.*

Knowing that the autocorrelation of an object is related to the magnitude squared of its Fourier transform we only miss the phase information to correctly retrieve the object. The class of phase retrieval algorithms [45] exactly do so by an iterative process which assumes some spatial and object constraints. For hidden imaging purposes, the object constraints are simple and practically they assume the object to be real, positive and confined within the $FOV$ of the scattering lens. It has been proven to be robust and efficient to retrieve objects hidden even behind biological tissues but at the moment lacks of studies for three-dimensional imaging. In this scenario, we examine the possibility and the feasibility of a three-dimensional implementation for performing tomography of objects hidden by scattering curtains, importing the concepts from the well-studied Optical Projection Tomography but using instead autocorrelations rather than projections in direct space.

The work presented here naturally follows a preliminary study [85] and, at the current stage, focuses on experimental proofs and numerical study for the feasibility of hidden three-dimensional imaging even though we are currently dealing with experimental measurements. We have tuned the technique to image behind a scattering curtain and then we performed computational simulations of an ideal OPT experiments. In these experiments, we perturbed the measurements by drifting the object during the rotation and we scrambled the phase of its measurement to replicate the effect of hiding the sample behind a scattering curtain. In these difficult measurement conditions, the calculation of the autocorrelation sinogram appeared to be an interesting choice to realign or retrieve the information of the specimen we are interested to measure.

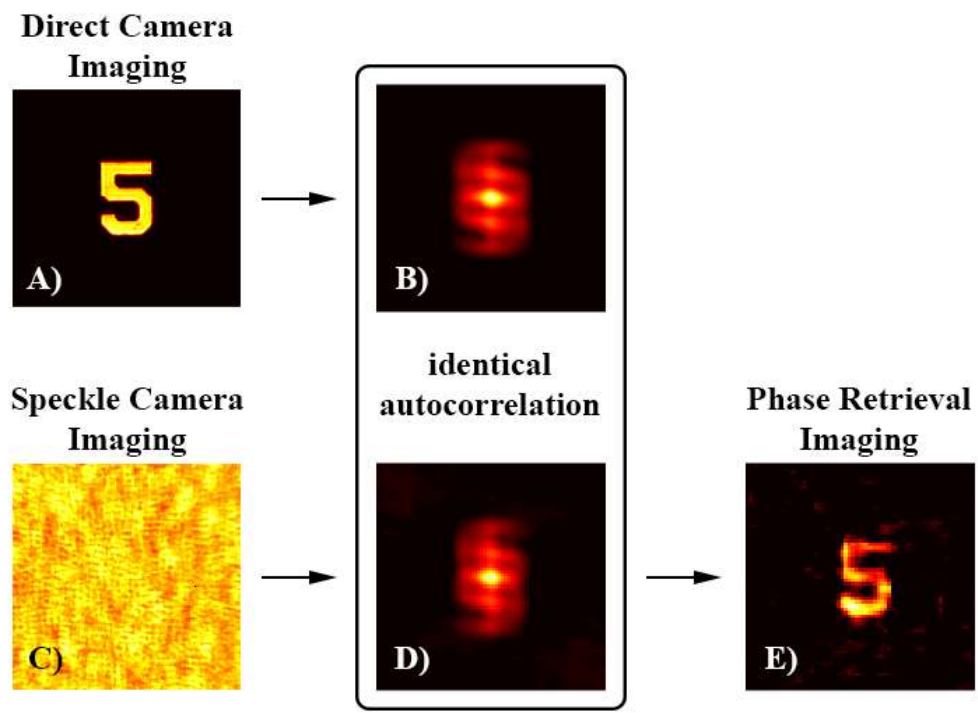

*Figure IV-3 | Schematics showing the autocorrelation properties. A) Direct imaging of the sample 5 in the USAF target and B) its autocorrelation calculated with xcorr2. C) the same sample hidden by a scattering layer produces a speckle pattern that seems to be information-less but as we can see in D) retains the same autocorrelation properties of the object itself. E) the autocorrelation can be exploited in order to obtain a reconstruction of the hidden object by using a PR algorithm to retrieve the correct phase.*

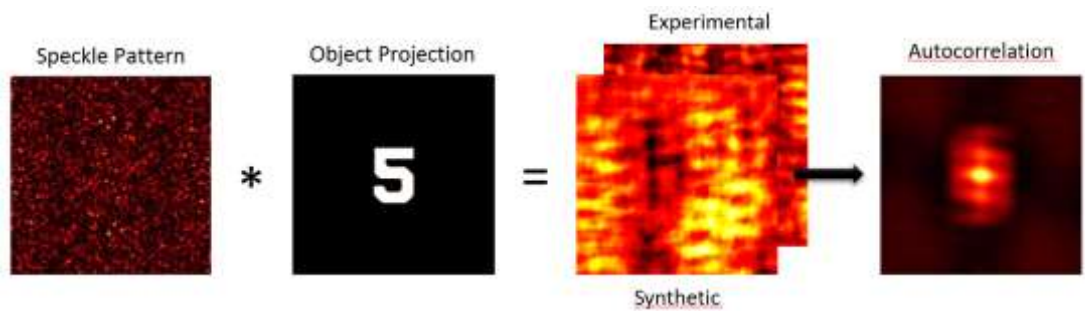

**Figure IV-4 | Numerical method to hide the sample behind a scattering layer.** *The convolution between a point-source generated speckle pattern and the object at a certain projection returned a speckle distribution that totally resembles the one we measured experimentally.*

## IV.1.2 MATERIALS AND METHODS

Initially, we tested experimentally the phase retrieval for imaging behind a scattering layer. We used a PCO Pixelfly camera with a resolution of 1.4 MP coupled to a band pass filter working in the range of $635 \pm 10 \ nm$. A 5x objective lens was used to image the object and the speckle pattern in front of the layer. We isolated an object representing a 5 in the Thorlabs USAF target (from the group 5) and which we illuminated from the backside using a nearly collimated white Light Emitting Diode (LED) placed behind a 120-grit ground glass diffuser. Another 120-grit ground glass diffuser from Thorlabs was placed in front of the object as already been done by several works presented in the literature [31] [27]. The object then results to be absolutely hidden between two opaque layers and was previously imaged removing the scattering layer located in front of the camera in order to compare the autocorrelations of the direct imaging and with the speckle produced by the turbid layer.

The data acquisition, simulation and processing was entirely performed using in-house developed software under MATLAB R2016a. After the acquisition of the direct image of the object (the 5) we created a volumetric object representing a thick "5" that can be seen in **Figure IV-2** panels A-B. We used this object for the simulation of an ideal OPT experiment designed in three different ways. Firstly, we used a three-dimensional radon transform to reproduce a full rotation of the object in steps of 1 degree. Then we perturbed the projections introducing a constant drift of 0.1 pixels per degree in both directions in the camera plane to reproduce a measurement performed in a badly aligned setup. Once we have everything ready for the reproduction of a classical tomographic experiment we hided the sample at the back of a hypothetical scattering layer. To do so we operated accordingly to what we have studied in details in the chapter dedicated to the theoretical analysis of the memory effect. We have in fact, explained how by tilting or shifting the impinging wavefront the speckle pattern obtained would simply translate without modifying its distribution within a certain range (that is the field of view). Now, if we imagine to "draw" a five starting from its upright corner, its corresponding single-point speckle pattern would be "smeared" following the same movement. This corresponds to the mathematical operation of the convolution, thus we can obtain the effect of hiding the object behind a scattering layer by just convolving its projection at angle $\theta$ with a generic speckle pattern (**Figure IV-3** and **Figure IV-4**). In fact, we found a great agreement with the experimental results, thus confirming the correct modelling of the speckle measurement. The repeating of this operation for every projection angle would, then, mimic a three-dimensional measurement performed in hidden modalities. At last, for each of those synthetic experiments, we calculated the autocorrelation of each single projection, stacking them one after the other and obtaining an autocorrelation sinogram. This produced 4 different kind of sinograms, which we report in **Figure IV-5** for completeness: the ideal measurement with its aligned sinogram, a sinogram affected by a constant drift, the speckle pattern sequence at different angle and their respective autocorrelation sinogram. We would like to point out that since what we have discussed so far throughout the text, all the three kind of measurement (ideal, drift and hidden) produced *exactly* the same autocorrelation sinogram. It is possible to notice that the operation of the autocorrelation has some interesting consequences to the misaligned and hidden dataset. We can comment this by looking at **Figure IV-6**, where we report the difference of camera detection at 0° and 360°. Clearly, the ideal measurement gives exactly the same image and returned zero differences between the two while the measurement affected by drift have non null differences. Even worse the case of the speckled measurement, in which the randomness is still visible by looking at their difference. But, instead, for all of the above measurements, the autocorrelation always returned a zero difference, due to the fact that it is centrosymmetric by definition. Drifts and speckles are then not seen by the autocorrelation, which implies an always centered autocorrelation sinogram in case we rotate the object in any of the way described.

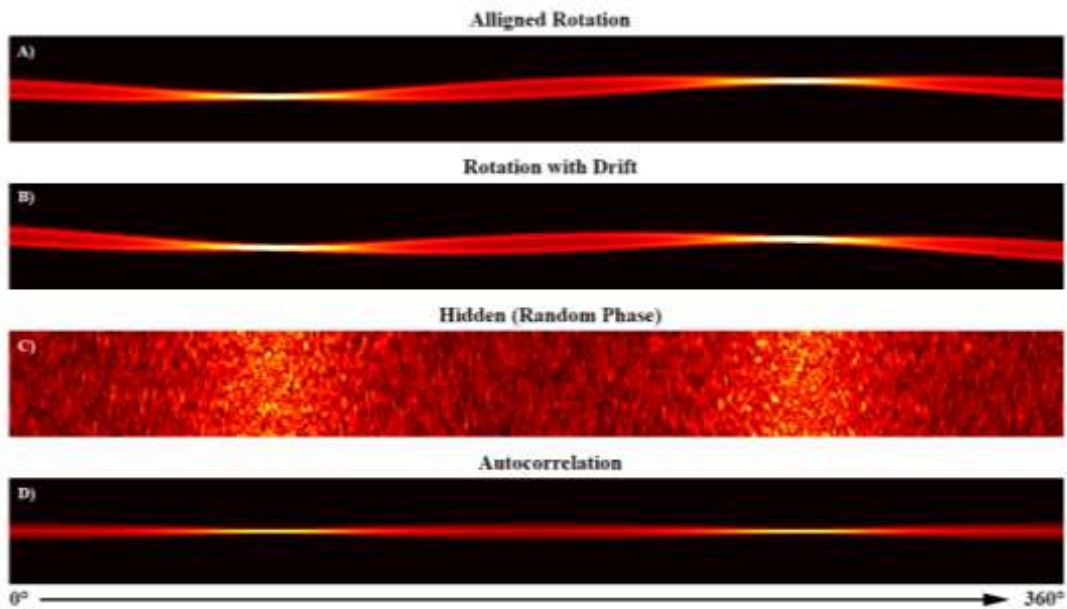

*Figure IV-5 | Various sinograms obtained during the numerical OPT experiment. A) perfectly aligned measurements lead to a uniformly oscillating sinogram. B) diagonal drift bends the direction of the sinogram, producing a dataset that needs post-processing alignment. C) recording different speckles at different projections produces a dataset that seems not containing any information regarding the object rotation. D) The autocorrelation sinogram obtained by calculating each single autocorrelation of every projection results to be always aligned regardless of the object position.*

The autocorrelation was calculated per each of the projections by using the xcorr2 function in MATLAB and related to the Fourier transform of the image by its square root. Calling the image projection $I$, its autocorrelation $R$ can be expressed by an inverse Fourier transform of its energy spectrum:

$$R(\theta) = I(\theta) \star I(\theta) = FT^{-1}\{|FT\{I(\theta)\}|^2\} = \text{xcorr2}\big(I(\theta)\big). \tag{215}$$

$R(\theta)$ then is exactly the autocorrelation sinogram that results to be the same for the ideal measurement, the one affected by the drift and the hidden imaging. Thus, at this stage, we will focus only on the autocorrelation sinogram, which we are going to unlock with the use of the phase retrieval algorithm by repeating the PR for every projection. The PR was implemented with the classical approach proposed by Fienup [45]. We executed 2000 iterations of the Hybrid Input-Output (HIO) keeping the constant $\beta = 0.9$ and after that we reduced the noise with a run of 1000 iteration of the Error-Reduction (ER). Everything was implemented using the MATLAB CUDA extension for a fast execution of the PR using a normal desktop with a GPU nVidia GeForce 780Ti. The images of the speckles were normalized with respect to their intensity envelope by dividing them with a low pass filtered version of the pattern itself.

The starting point of the PR, then, is the modulus of the Fourier transform $A$ of the image, estimated by the calculated autocorrelation:

$$A(\theta) = \sqrt{\mathcal{F}\{R(\theta)\}} = \sqrt{\mathcal{F}\{I(\theta) \star I(\theta)\}} = |\mathcal{F}\{I(\theta)\}|. \tag{216}$$

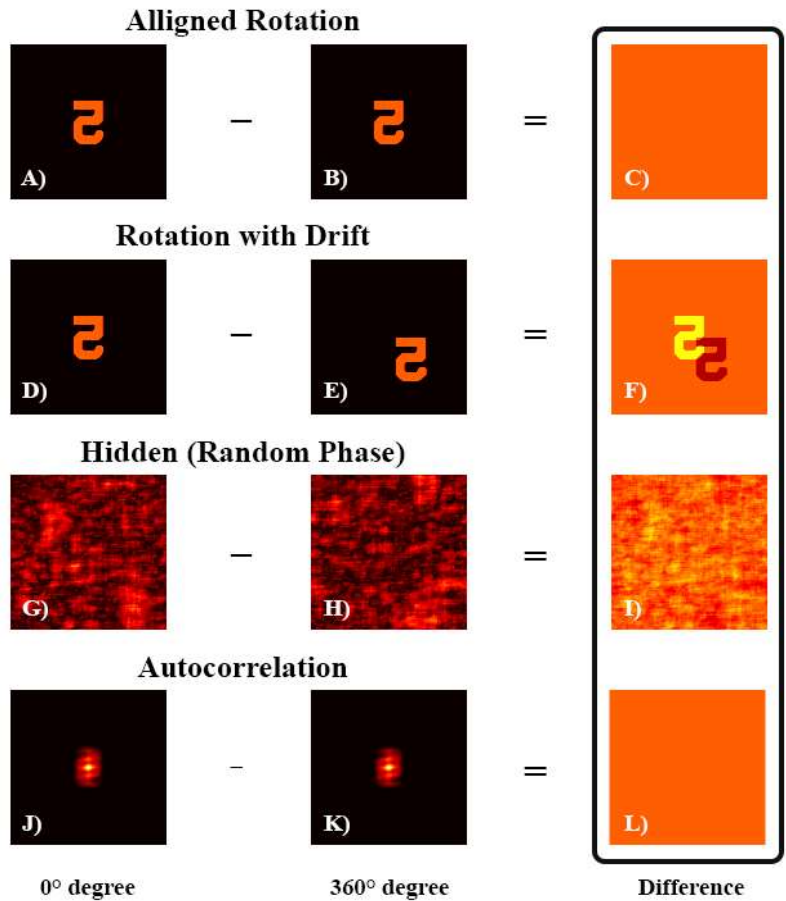

**Figure IV-6 | Diagram of the initial and the last projection after different synthetic OPT experiments.** *A-C) for a perfect rotation there is no difference between the first and the last projection since the object returns to the initial position after a full rotation. D-F) shows the misalignment due to a diagonal drift while rotating the object. To obtain a reconstruction in this case the dataset needs to be re-aligned. G-I) difference between speckle patterns recorded after a full rotation, the dataset cannot be aligned. J-L) autocorrelations, instead, are intrinsically aligned even if the sample or the speckle shifted in any direction perpendicular to the camera detection plane.*

Expressed in this way the phase retrieval problem consists of finding the solution for the phase Φ that returns the estimate of the object G which has minimized distance between its Fourier modulus and $A$:

$$G = |\mathcal{F}\{G\}|e^{i\Phi} \quad \text{is solution when } |A - \mathcal{F}\{G\}| \to 0 \qquad (217)$$

In general, the PR starts typically guessing the phase Φ and via iterative Fourier transformation, applying the object constraints of being real and positive, returns the estimation of the object $G$. We tested the algorithm both with experimental data, coming from hiding the USAF target behind a diffuser and with numerical simulation, comparing and discussing the results obtained.

### IV.1.3 RESULTS AND DISCUSSIONS

Since the procedure adopted so far might result slightly confusing we will briefly recap the current stage of the work: in fact, we are interested in setting up a clear understanding on how to proceed in order to face a hidden three-dimensional reconstruction using OPT and Phase Retrieval approaches. First of all, we acquired a direct imaging of a sample (a 5 in a USAF target) in order to calculate its autocorrelation. Successively we hided it enclosing the sample in between two diffusers, making direct imaging impossible. We acquired the speckle pattern produced in front of the scattering layer placing the sample at different distances, checking their autocorrelation properties. Within the memory effect regime, every speckle pattern retained the same autocorrelation properties of the original object, allowing us to retrieve its reconstruction by making use of the phase retrieval algorithm (**Figure IV-3**).

Since the sample used in the experiment was a flat object (imaged in **Figure IV-3**), we decided to create a virtual "thick" 5 (red sample in **Figure IV-2**) object to reproduce numerically an OPT experiment. Firstly, we simulated a perfectly aligned rotation and then we introduced a constant diagonal drift to reproduce the effect of a misaligned experiment. Successively we scrambled the phase of each of the projection, obtaining the speckle pattern that resemble the ones obtained in a hidden OPT experiment. At last, for each of the three datasets, we calculated their autocorrelation in order to compare them. Analyzing the results obtained after this numerical experiment, noticed some interesting properties. By comparing the first and the last projection of each of the dataset (**Figure IV-6**) we can notice how both misaligned and a hidden imaging are affected by strong differences that will lead to wrong reconstruction or no reconstruction at all. In both cases, if we consider their inverse Radon transform, we will either obtain a reconstruction full of artifacts (for the misaligned measurements) or a reconstruction of just a random noise for the hidden measurement. Interestingly, in each of the previous cases, (**Figure IV-6** panels J-L) the autocorrelations are identical and always intrinsically aligned to the center even if the data were affected by a constant drift or if the measurements where performed in hidden modality. This fact is very promising for hidden reconstructions because we would have not access to its rotation trajectory behind the scattering layer. In this case, we can always assume that the rotation of the autocorrelation is aligned regardless of the position of the object. In fact, this lead us to naturally obtain an autocorrelation sinogram (**Figure IV-5** panel D) which always rotates exactly in the center of the image plane, where the autocorrelation has its highest peak. Moreover, this sinogram can be always obtained starting from the aligned sinogram (A), the sinogram with a constant drift (B) and the seemingly information-less sequence of speckle pattern (C) offering us the possibility to perform aligned imaging even in such difficult conditions.

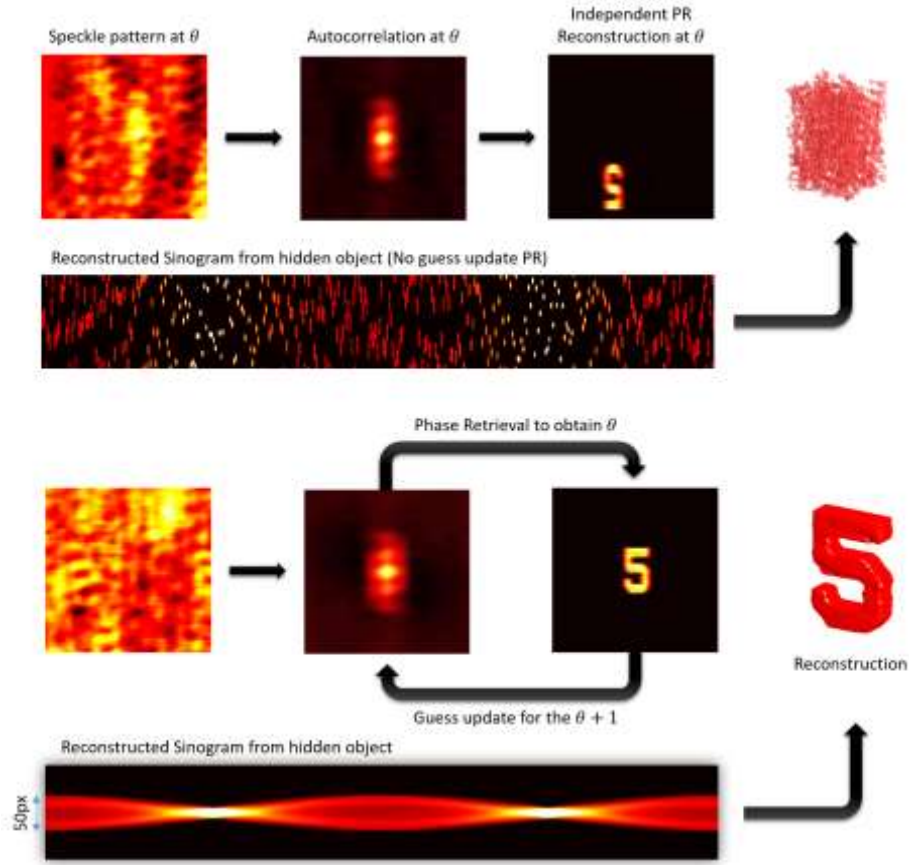

*Figure IV-7 | Different approaches for hidden three-dimensional imaging. In the upper case, the run of independent PR at different angle projection returned an absolutely meaningless projection sequence. By using, instead, the reconstruction of a previously recovered projection iteratively we managed to obtain an aligned sinogram even starting from its random speckle patterns (lower part). The resulting sinogram was possible to invert with an inverse Radon transform that finalized the three-dimensional hidden reconstruction.*

With the autocorrelation sinogram we have lost an absolute positional information of the object, but we have recovered an intrinsic alignment that we have to efficiently exploit to obtain a meaningful reconstruction. As first approach, we have run independent phase retrieval algorithms per each projection and examined the final result obtained. In the upper part of **Figure IV-7** we can notice that independent run of PR randomly reconstructed the object in the image space, which is a classical problematic connected with the phase ambiguity. We have already discussed this fact when talking about the uniqueness of the PR problem, it is clear to conclude then that independent runs are not a feasible reconstruction choice. If we however run several, independent PR reconstructions at the first projection $\theta = 0°$ starting each time with a different random guess, we will obtain reconstructions randomly distributed in space and, with 50% probability, affected by a centrosymmetric flipping. Now let us imagine that we can select one which has the correct flipping: this is always possible to shift in the center of the image plane by moving its center of mass. We have now a reconstruction centered in the image plane but we cannot repeat this for all of the independent projections, because in some cases the center of mass might be moved due to inaccuracies of the PR algorithm in general. Under certain conditions, we can connect this reconstruction with the next one. If the sample after the rotation at $\theta + 1$ is still in the memory effect regime and if the rotation step is smaller than the angular memory effect, we could start the new PR with the retrieved object at $\theta$. In fact, the retrieved object at the previous rotation step is a good estimate for the following, due to the fact that the projection had undergone a very slight change. By repeating this "feeding" operation iteratively we can go through the whole autocorrelation sinogram, thus unveiling the hidden object sinogram. The results of this iterative operation are shown in the lower part of **Figure IV-7**, in which it is possible to notice that the resulting sinogram obtained was perfectly aligned. The alignment of this new dataset is so good that applying the inverse Radon transformation to this sinogram lead us with a very good three-dimensional reconstruction of the hidden object.

In the early stage of this work we analyzed and tested the properties of a novel three-dimensional autocorrelation-based imaging, in order to exploit them for hidden three-dimensional imaging. Although very interesting works have been done exploiting the phase retrieval algorithm for two-dimensional imaging behind scattering layers or to perform three dimensional aligned reconstructions, none of them studied the possibility of imaging three dimensional objects behind scattering layers. So far, we have shown that it is possible to retrieve a meaningful sinogram even in the case of speckle projections, which will open the possibility to perform a hidden reconstruction. Currently in fact we are implementing our results in order to image biological entities in hidden mode to further advance the state-of-art of the optical imaging modalities. Not only hidden imaging modalities could gain a boost by this technique, but also misaligned measurements could be automatically registered by the intrinsic alignment of the object autocorrelation. In fact, in the following section we will focus on this peculiar point that emerged from this study, exploiting the autocorrelation sinogram and phase retrieval methods in a new fashion, to overcome misalignment in the every projection-based three-dimensional reconstruction.

# IV.2 Phase Retrieved Tomography

We have explained, so far, that the retrieval of the phase connected with the speckle's autocorrelation, obtained in the presence of a highly diffusive layer, unlocked the possibility of hidden or behind the corner three-dimensional reconstructions. From the above study, several interesting features emerged while solving the hidden reconstruction problem, mostly related to the combination of the autocorrelation imaging together with backprojection-like approaches used for the tomographic reconstruction. Among others, the fact that the autocorrelation is an always centrosymmetric function, regardless of where the object was in the image space, made us think some important implications. Some times in fact, while aiming at three-dimensional imaging of non-transparent and turbid specimen, the sample itself can hide its own part that is found to be on the other side with respect the camera detection. Thus the specimen could act as the turbid layer of itself, deteriorating the resolution in function of the depth at which we image. To make a more concrete example let us think about a hypothetic spherical tissue distribution (we will present a concrete case in the following), which has some interesting fluorescent structures sparse in its bulk. Due to the environmental tissue scattering, some fluorophores located on its farthest side might be blurred or, even worse, invisible in case such specimen would be bigger than $1\,TMFP$ in diameter. To overcome this limitation, we can think of going directly to image this "dark side" but, of course, mechanical movements, different camera positioning geometries and other external perturbations can limit the registration of the two (or more, if needed) different views. Several ways have been approached to solve this reconstruction problem but no one has introduced a general rule to perform such kind of reconstruction. Even if it might seem that autocorrelation imaging would behave even worse in this scenario, we proved an interesting way to exploit its mathematical properties. Recalling from the theoretical background that a single phase retrieval problem has an infinite space of possible solutions, which they all differ by trivial translations and centrosymmetric flipping, we have solved this randomness in the reconstruction location by an iterative feeding technique for the sinogram reconstruction. There is a possibility, then, to anchor the reconstructions connected with different angular views, which led us to unlock robust hidden tomography. Furthermore, we can prove in fact that the autocorrelation operation "commutes" with the inverse Radon transformation, giving us the possibility to calculate the three-dimensional autocorrelation of the object free of misalignment problematics.

Thus in the following, we present the new Phase-Retrieved Tomography (PRT) method to radically improve mesoscopic imaging at regimes beyond one transport mean-free-path and achieve high resolution, uniformly throughout the volume of opaque samples. The method exploits multi-view acquisition in a hybrid Selective Plane Illumination Microscope (SPIM) and Optical Projection Tomography (OPT) setup and a three dimensional Gerchberg-Saxton phase-retrieval algorithm applied in three dimensions through the autocorrelation sinogram. We have successfully applied this innovative protocol to image optically dense three-dimensional cell cultures in the form of tumor spheroids, highly versatile models to study cancer behavior and response to chemotherapy. We have thus achieved a significant improvement of resolution in depths not yet accessible with the currently used methods in SPIM/OPT, while overcoming all registration and alignment problems inherent to these techniques. Due to its formulation, PRT can be of course also implemented in imaging of objects hidden behind scattering curtains by multi-angle acquisition of the emitted speckle and calculation of autocorrelation sinograms.

## IV.2.1 STATE OF THE ART AND PROBLEM INTRODUCTION

The use of light for visualizing biological processes in living organisms has been a foundational paradigm for biology for almost four centuries, since the invention of the microscope. A variety of methods have been invented and applied in imaging across the scales from macroscopic to mesoscopic to microscopic levels [51]. Optical techniques allow *in-vivo* three dimensional imaging and interrogation of biological samples, ranging from measurements in the macroscopic regime with fluorescence tomography [90], light sheet microscopy [54] and optical projection tomography [52] approaching the mesoscopic regime, down to the microscopic level with confocal and multiphoton microscopy [91] [92]. In recent years, new advancements of imaging technology have achieved in going beyond the limits of conventional microscopy with super-resolution microscopy [93] [94], awarded a Nobel Prize in 2014. These methods heavily rely on the ability of both the hardware and the software to produce meaningful images in terms of resolution, quantification and accuracy. Especially, the tomographic imaging methods cannot abstract from the use of fast and efficient computational techniques, with experiment and algorithms being interlinked in such strong way that every improvement on one side rapidly affects the other. Furthermore, computational methods are nowadays becoming so important that they offer opportunities to develop better imaging methodologies or unlock possibilities to image under prohibitive conditions dictated by different regimes of light scattering in tissues [51]. Although tomographic principles are conceptually simple and well described and characterized, optically sectioning the sample for internal functional and structural inspection often requires a tremendous effort on both computational and experimental aspects in order to achieve a satisfying image quality. This becomes even more relevant and demanding, in regimes at around one transport mean free path and higher, where strong light scattering restricts high-resolution imaging in the macroscopic scales. Despite the advancement in optical technologies that have pushed biological imaging beyond fundamental limits, the depth to resolution ratio still does not allow deep tissue high-resolution imaging. One solution to that problem has been provided by various and invasive chemical methods, referred to as optical clearing [95], that chemically alter the optical properties of tissues. The price to pay in these cases is severe, since the investigated tissue needs to be fixed; thus exchanging the possibility of live *in-vivo* imaging for high resolution.

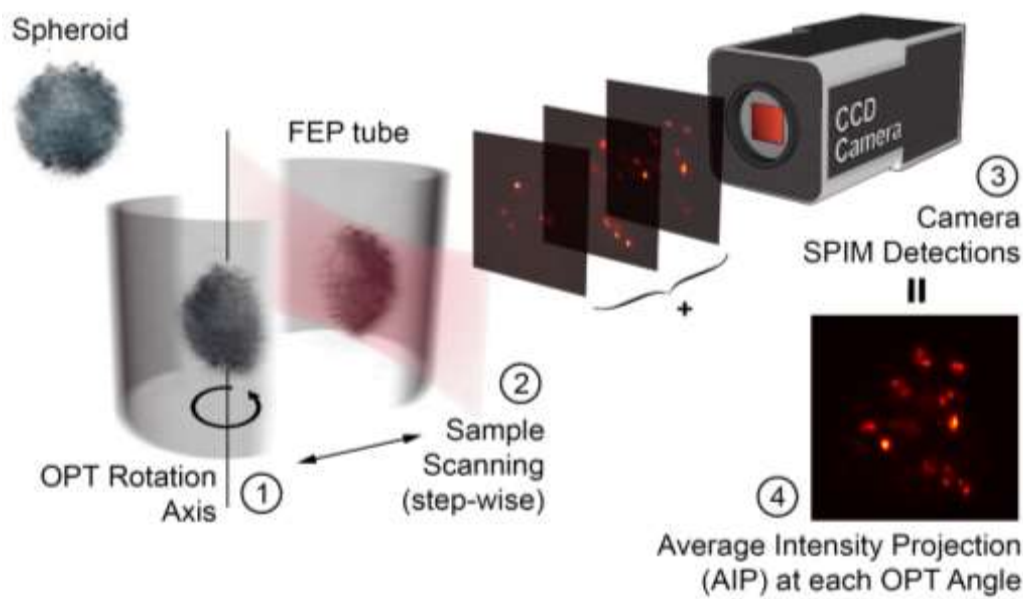

*Figure IV-8 | Schematic depiction of the data acquisition for the SPIM-OPT setup. Starting from the left a tumor spheroid, stained with the DRAQ7™ fluorescent dye, is inserted into a FEP tube and mounted to the rotation stage. (1) The specimen is rotated at a known angle and then (2) is scanned through the light sheet along the detection axis. The illumination light sheet is established orthogonally to the detection axis and it fulfills the objective focal plane, in order to reduce out of focus contribution during the camera acquisition. While translating the specimen a SPIM detection (2D images stack) is stored (3). Finally (4), the data acquired is used to calculate the Average Intensity Projection (AIP) as a function of the angle. This procedure is repeated by rotating the sample (1) in steps of 2° until performing a complete rotation.*

As we already discussed and studied in some of our works [23] [21], a radical new approach in overcoming the limitations imposed by multiple light scattering in biological media, is based on the use of active optical elements for the accurate control of the impinging wavefront, which provides a compensation for the random refractive index variations in tissue. Contrary to popular belief, multiple scattering in the optical paths can be exploited to predetermine light propagation, to reverse image distortion, to focus through or inside turbid media, to improve image resolution, to retrieve otherwise hidden features and transmit images through optically opaque media. These pioneering approaches [96] [20] [31] are based on the utilization of adaptive dynamic wavefront shaping and phase retrieval applied for controlling and inversing light diffusion. Interestingly, autocorrelation-based imaging is currently being employed in novel promising applications that allow, under certain conditions dictated by the memory effect range (intrinsic isoplanatism) [26], the visualization through turbid media and even behind corners [31] [33].

As mentioned above, Light Sheet Fluorescence Microscopy (LSFM) or Selective Plane Illumination Microscopy (SPIM) as we are going to refer to the technique throughout the text [54] [97], is one of the most widely used method for direct *in vivo,* real time [98] visualization of internal structures and functions in model organisms. Of high scientific interest are, among others, *Caenorhabditis elegans* [99], *Danio rerio* (specifically in heart imaging) [100]*,* and multicellular tumor spheroids (MCTS) [101]. In this case, the illumination of the specimen is accomplished with a light sheet illuminating perpendicularly to the detection axis. Different camera images are recorded, either by scanning the sample by changing the position of the light sheet or translating the specimen throughout a fixed single plane illumination. In such a way, the tomographic volume is built with very little computational effort, particularly in cases of optically transparent or chemically cleared specimens. Although the technique works reasonably well in the microscopic regime, challenges arise in mesoscopy where higher scattering and absorption impede uniform and localized illumination or emission, resulting in shadows and blurring at the camera detection level. Many approaches have been proposed to tackle these complications, such as combining multiple projections at different angles [54], pivoting the light sheet for double illumination [102], using multi-view geometry [103], or mixed approaches based on forward light modelling such as Mesoscopic Fluorescence Tomography (MFT) [104]. All of them require dedicated care for alignment and co-registration processes during both experimental and post-processing stages. Even the very robust and easy to implement Optical Projection Tomography (OPT) [52], a technique based on the acquisition of projections at different angles and reconstructing the final image using back-projection algorithms, needs to take into account possible misalignments of the measuring scheme. Among other challenges [105] [106] [107], the possibility that the sample could potentially exit the field of view along with translational and rotational misalignment drastically deteriorate the quality of the inverse reconstruction. Computational techniques for finding the Center of Rotation (CoR) [108], correcting for sample movements [109] [110] and selecting an appropriate Region of Interest (RoI) [111], are often used in order to overcome these shortcomings, although they are complicated and sensitive to fluctuations between measurements.

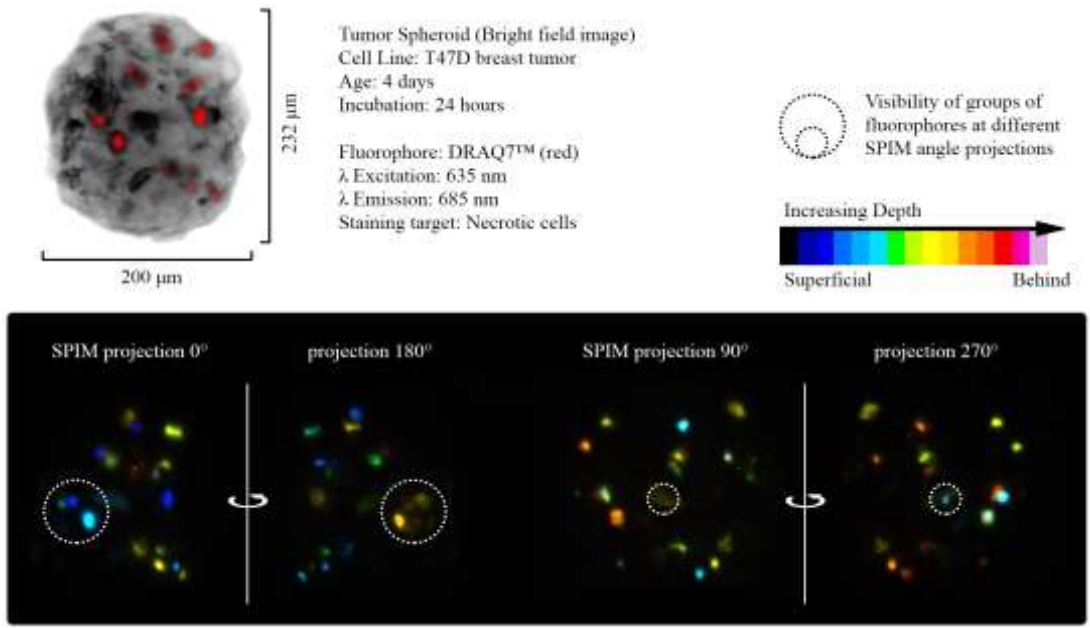

*Figure IV-9 | Imaging the spheroid in Mesoscopic regime. **A**, Bright field image of the spheroid imaged at 0°. The non-uniform structure of the tumor mass is clearly noticeable. **B-E**, SPIM-AIP at perpendicular angles with respect to one another. Dashed circles point out groups of fluorophores not visible at all measured angles, due to significant light scattering from the spheroid (mesoscopic regime).*

In the following section, we present a new tomographic approach based on the reconstruction of the sample's three-dimensional autocorrelation rather than on direct imaging of the specimen itself, that radically improves imaging beyond the one transport mean-free-path limit of mesoscopy. By combining the functional information from SPIM and the structural information from OPT in a complementary fashion, our new computational technique uniquely aligns the dataset by exploiting the mathematical properties of the image autocorrelation. Because of the novel use of the algorithm combined with SPIM/OPT, we found appropriate referring to this approach with the name Phase-Retrieved Tomography (PRT). We have applied PRT for three-dimensional image reconstruction of early stage necrosis distribution in a human-breast tumor spheroid, which well represents a mesoscopic regime proof of concept scenario. MCTS is the best-characterized and most widely used scaffold-free 3D culture system that takes advantage of the inherent ability of many cancer cells to self-organize into spherical clusters [112] [113]. MCTS is gaining huge attention as a pre-clinical drug-testing model and a large body of literature over the past few decades highlights the usefulness of this model system in translational cancer research and drug discovery. In our study, we successfully reconstructed the fluorescence emitted by the spheroid's necrotic cells in the entire volume and in resolution not accessible with SPIM/OPT alone, due to scattering, eliminating at the same time the need for data alignment and registration. By solving the phase-retrieval problem related to the three-dimensional autocorrelation, the PRT reconstruction method extends the applicability of widely used microscopy techniques towards the mesoscopic regime, providing unprecedented insights within so far turbid biological models.

### IV.2.2 THE PHASE-RETRIEVED TOMOGRAPHY METHOD

We are now ready to discuss the principles behind the PRT method. To do so, we will firstly analyze the numerical results obtained with a virtual phantom, then proving its principles by a theoretical approach. Then, we will propose a stepwise method to perform imaging in a lab measurement, which will be presented in the next section. We anticipate that the present method is based on the preliminary observation that the autocorrelation sinogram is always centered, an important feature that already have been exploited for hidden three-dimensional imaging. In this part of the study we will step further, backprojecting for the first time the autocorrelation sinogram in order to obtain what we believe to be the three-dimensional autocorrelation of the object. In fact we will prove this theoretically, thus making us ready for the final experimental imaging application.

#### *Phase-Retrieved Tomography – Numerical Validation*

To test the validity of the proposed method, we performed a numerical validation using a three dimensional Shepp-Logan phantom, a commonly used model for testing the performance of reconstruction algorithms. We used a freely available tool for the generation of such phantom in the Matlab environment [114], creating a cubic volume of 128 pixels per side. The whole process is shown in **Figure IV-10**, in which each image is the Average Intensity Projection (AIP) along the axis of rotation. The 3D phantom was Radon transformed in the domain of $\theta \in [0°, 180°]$ with steps separated by $1°$ using the corresponding Matlab function. This results in the direct projections sinogram, which is perfectly aligned since the axis of rotation around which we performed the Radon transform was fixed to the center of the image. We verified that the reconstruction obtained inverting the sinogram with the filtered inverse Radon transform gives correct results (**Figure IV-10** 3C). We then simulated a vibrational perturbation in the sample rotation by displacing it randomly in three directions at each $\theta$ of the radon transform, resulting in a misaligned projections' sinogram shown in panel D. The random shift introduced was in the range of [-5, +5] pixels in every spatial direction. In this case, the calculation of the inverse Radon transform results to a noisy reconstruction (panel E), in which features are no longer visible. Vibrational misalignment and the subsequent reconstruction artifacts represent an unsolved problem in real experiments and appropriate reconstruction algorithms have not been demonstrated to date. Here we introduce a solution based on the calculation of the autocorrelation for each of the projections, in such a way that we create an autocorrelation sinogram which is perfectly aligned, as we have recently presented in one of our preliminary studies [85]. We noticed a perfect agreement between the three-dimensional autocorrelation calculated from the direct object and the one reconstructed from the sinogram (**Figure IV-10** panels F-H). In practice, the quality of the 3D autocorrelation is never affected by the vibrational-noise level in the original sinogram, in fact any displacement is simply not propagated in the autocorrelation space. Since we have a better estimation for the three-dimensional object's autocorrelation rather than the object itself, we can exploit this information with a three-dimensional implementation of a phase retrieval algorithm (which will be described later) in order to accomplish the reconstruction and retrieve the real object. As can be seen in panel I, the reconstruction obtained with Phase-Retrieved Tomography (PRT) matches the original object (panel A) even if the dataset was perturbed with strong vibrational noise.

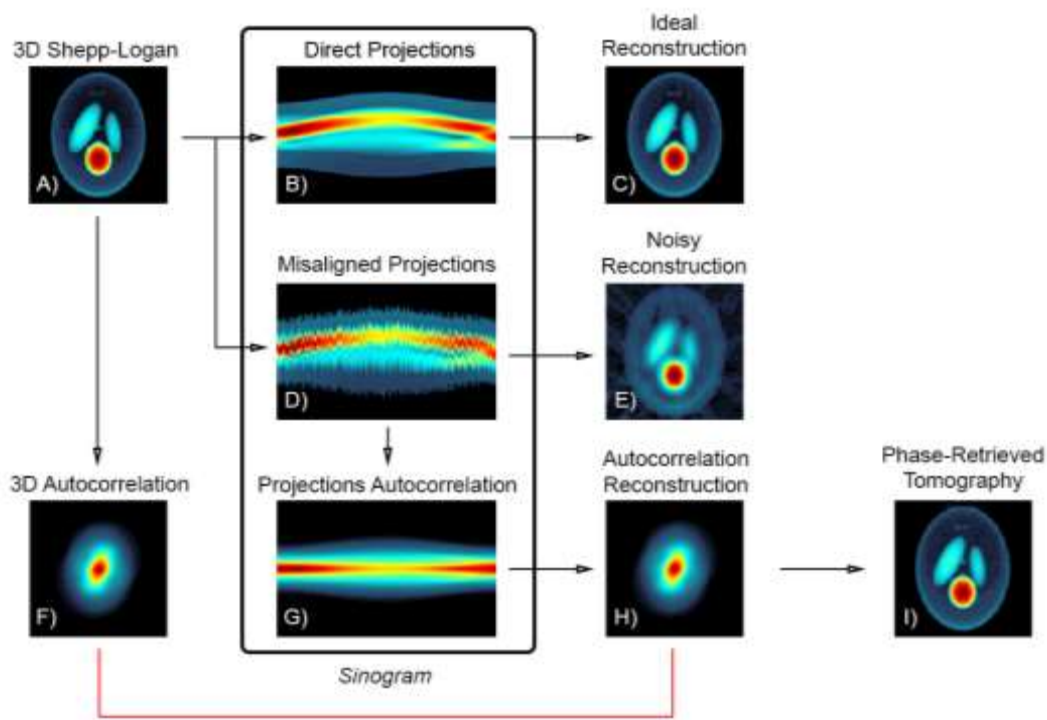

**Figure IV-10 |** *A) 3D Shepp-Logan phantom used to test the reconstruction algorithm in comparison with classical backprojection methods. B) Sinogram of the projections at various angles, which is obtained as the Radon transform of the object in the range θ ∈ [0°, 180°] and C) its inverse radon transform. D) Misaligned measurements due to random shift of the object while rotating, that perturbs the sinogram and return wrong reconstruction at E). F) Original object's autocorrelation directly calculated. G) Naturally aligned sinogram of the autocorrelation of each single projection from D) and its inverse radon transform H). It is worth noticing the agreement between F) and H). Finally, I) is the Phase-Retrieved Tomography reconstruction of the object using only the backprojected H).*

### *Phase-Retrieved Tomography – Theoretical Support*

Our proposed method is based on the calculation of the three-dimensional autocorrelation of the investigated specimen and the use of a three-dimensional phase retrieval to form the final reconstruction, as shown in previous works for two dimensions [115]. Calculating a 3D autocorrelation by the Radon transform of 2D camera autocorrelations requires that the autocorrelation of the projection of the specimen is equal to the projection of its three-dimensional autocorrelation at each angle. In the following section, we prove that the two quantities are identical. For simplicity, we treat the 2D case considering a two-dimensional object to be reconstructed starting from its 1D projections.

Considering $(x, y)$ as the spatial coordinates in which the specimen exists and $(\eta, \xi)$ as their respective translational coordinates, let us define the following quantities:

Object of interest: $\qquad O(x, y)$

Object autocorrelation: $\qquad A(\eta, \xi) = O(x, y) \star O(x, y)$

$$= \int O(x, y)\, O(x + \eta, y + \xi)\, dx\, dy$$

Projection of the Object (at angle $\theta = 0$): $\quad P_O(y) = \int O(x, y)\, dx$

Projection of the Autocorrelation ($\theta = 0$): $\quad P_A(\xi) = \int A(\eta, \xi)\, d\eta$

The object $O$ is finite and limited in space, defined in a closed region $[x, y] \in \mathrm{R}$, and every integral considered is defined up to the region boundaries. In this representation $P_O(y)$ is the 1D camera detection at angle $\theta = 0$.

To be able to reconstruct the object autocorrelation by calculating the autocorrelations of the object projections literally means that the following relation must be satisfied:

$$P_O(y) \star P_O(y) = P_A(\xi). \tag{218}$$

Explicitly we can write:

$$\int P_O(y) \, P_O(y + \xi) dy = \int A(\eta, \xi) \, d\eta \tag{219}$$

$$\int P_O(y) \, P_O(y + \xi) \, dy = \int O(x, y) \, O(x + \eta, y + \xi) \, dx \, dy \, d\eta. \tag{220}$$

Both terms are integrated along y, so we can compare the arguments of the integrals:

$$P_O(y) \, P_O(y + \xi) = \int O(x, y) \, O(x + \eta, y + \xi) \, dx \, d\eta. \tag{221}$$

Now we focus on the integration along the translation $\eta$:

$$P_O(y) \, P_O(y + \xi) = \int O(x, y) \left\{ \int O(x + \eta, y + \xi) \, d\eta \right\} dx. \tag{222}$$

It is possible however, to make the following consideration regarding the argument in brackets; for finite objects in space, a linear translation does not influence their definite integral, i.e. the projection is preserved for translations along the axis perpendicular to the detection axis. This means that integrating in the translation $\eta$ projects the object into its perpendicular axis, eliminating the dependence in $x$ (one coordinate could be seen as the translation of the other) and hence:

$$\int O(x + \eta, y + \xi) \, d\eta = \int O(x + \eta, y + \xi) \, dx = P_O(y + \xi). \tag{223}$$

This implies that:

$$
\begin{aligned}
P_O(y) \, P_O(y + \xi) &= \int O(x, y) \, P_O(y + \xi) \, dx \\
&= P_O(y + \xi) \int O(x, y) \, dx \\
&= P_O(y + \xi) \, P_O(y).
\end{aligned} \tag{224}
$$

Which exactly proves our initial claim. Theoretically then, it is possible to calculate the autocorrelation of the object by the Inverse Radon transformation of the autocorrelation of the projections.

### *Three-Dimensional Phase Retrieval algorithm*
We have always implicitly considered the imaging process presented so far being based on the calculation of autocorrelations for each (full resolution) camera detection at every angle of rotation. In practice however, the raw camera image was always cropped with a squared window of 300 pixels around the object to reduce the size of the matrices considered. Then the autocorrelation $A(x, y, \theta)$ of the cropped camera image $I(x, y, \theta)$ at each step of the rotation $\theta$ is calculated according to:

$$A(x', y', \theta) = I(x, y, \theta) * I(x, y, \theta) = \sum_{x,y} I(x, y, \theta) I(x - x', y - y', \theta). \tag{225}$$

The $A(x', y', \theta)$ is the autocorrelation sinogram that we back-project with the inverse Radon transform operator ($\mathcal{R}^{-1}$), obtaining the three-dimensional object autocorrelation $A_{3D}$:

$$A_{3D}(x, y, z) = \mathcal{R}^{-1}\{A(x', y', \theta)\}. \tag{226}$$

The autocorrelation can be related to the power spectrum $P(x, y, z)$ by the Wiener-Khinchin theorem:

$$P(k_x, k_y, k_z) = |\mathcal{F}\{W(x, y, z) A_{3D}(x, y, z)\}| \tag{227}$$

where $W(x, y, z)$ is either a cubic or a 3D-Tukey window function that selects the core of the autocorrelation. The power spectrum can be related to the modulus of the Fourier Transform of the object $O(x, y, z)$ via the following equation:

$$M(k_x, k_y, k_z) = |\mathcal{F}\{O(x, y, z)\}| = \sqrt{P(k_x, k_y, k_z)}. \tag{228}$$

In this description $O(x, y, z)$ is the three-dimensional object that we are interested in reconstructing of which we have calculated the Fourier modulus $M(k_x, k_y, k_z)$. We still miss the phase information $\Phi(k_x, k_y, k_z)$ which we aim to retrieve with a 3D implementation of the iterative phase retrieval algorithm described by Fienup [48]. We followed the procedure described by Bertolotti *et al.* [31] and Katz *et al.* [33] extending it for a three-dimensional problem solution. We describe the iteration steps according to the following numbered list. Starting from a random guess for the object $g_1(x, y, z)$ we input it in the iterative algorithm describing the k$^{th}$ step:

1. $G_k(k_x, k_y, k_z) = \mathcal{F}\{g_k(x, y, z)\}$

2. $\Phi_k(k_x, k_y, k_z) = \text{angle}\{G_k(k_x, k_y, k_z)\}$

3. $G_k'(k_x, k_y, k_z) = M(k_x, k_y, k_z) e^{i\Phi_k(k_x, k_y, k_z)}$

4. $g'_k(x, y, z) = \mathcal{F}^{-1}\{G_k'(k_x, k_y, k_z)\}$

5. Constrains of non-negativity and realness $g'_k(x, y, z) \rightarrow g_{k+1}(x, y, z)$

The guess $g_{k+1}(x, y, z)$ is calculated according to the model of the Hybrid Input-Output (HIO) algorithm for 5000 steps:

$$g_{k+1}(x, y, z) = \begin{cases} g'_k(x, y, z) & \text{for } (x, y, z) \notin \Gamma \\ 0 & \text{for } (x, y, z) \in \Gamma \end{cases} \tag{229}$$

followed by 1000 steps of Error Reduction algorithm:

$$g_{k+1}(x, y, z) = \begin{cases} g'_k(x, y, z) & \text{for } (x, y, z) \notin \Gamma \\ g_k(x, y, z) - \beta g'_k(x, y, z) & \text{for } (x, y, z) \in \Gamma \end{cases} \tag{230}$$

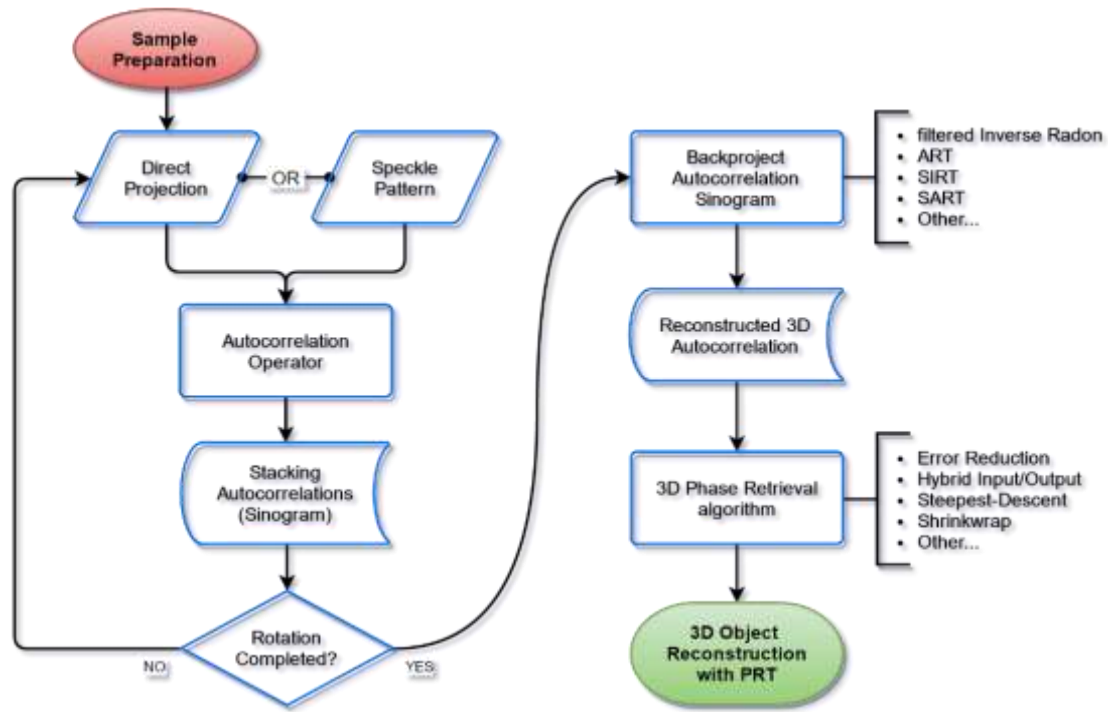

*Figure IV-11 | Flow chart describing the Phase-Retrieved Tomography method.* *The process is modular and allows room for different settings, both at the acquisition and reconstruction level.*

Where Γ is the region within $(x, y, z)$ violate the constrains of non-negativity and realness and $\beta$ is a parameter that controls the convergence properties of the algorithm. In our case $\beta = 0.9$ was a good choice. The entire algorithm was implemented in the Matlab environment using CUDA computation and was run on nVidia GPU. At the end of N iterations, $g_N(x, y, z)$ is the three-dimensional object reconstructed with the PRT method. In this way, we demonstrate for the first time that a three-dimensional phase retrieval algorithm is possible to be used for tomographic imaging.

### Phase-Retrieved Tomography – Flow Chart protocol

Finally, the protocol that we are going to propose for correctly reinterpreting misaligned datasets is schematically described in the flowchart of **Figure IV-11**. The sample will be placed into the setup described in the following section and, although the SPIM measurements are not strictly required, this reduces the out of focus contribution of the Average Intensity Projection (AIP). In the diagram, the Direct Projection block could be fed with different kind of projections obtained at a certain angle $\theta$: parallel bright field projections, AIP of fluorescence excited by SPIM or simple fluorescence signal exciting the whole sample. Speckle patterns produced by a sample hidden behind a scattering curtain are also a feasible choice, since it has been proved experimentally [31], [116] and numerically [85], [86], that the camera image of both the speckle pattern and the object itself share the same autocorrelation features. For each acquisition, we calculate the two-dimensional autocorrelation of the unfiltered camera image, stacking them to obtain the autocorrelation sinogram. After the whole rotation is accomplished, it is possible to backproject the autocorrelation sinogram, in order to obtain the three-dimensional autocorrelation of the whole object. It is worth noticing that several techniques can be used to backproject the sinogram [117], allowing for comparative studies of different approaches. The reconstructed three-dimensional autocorrelation then is used as starting point for a phase retrieval problem, which turns into the final reconstruction of the object.

### IV.2.3 EXPERIMENTAL METHODS

Since we proposed a totally new tomographic approach at theoretical, numerical and experimental level, this deserves a detailed description of the methodologies and methods used during the procedure for the final experimental measurements.

*Tumor spheroid generation.*

First of all, the spheroids were generated with the hanging drop method using the Perfecta3D 96-well hanging drop plates (3D Biomatrix, Ann Arbor, MI, USA) following manufacturer's instructions. Briefly, cell suspensions for hanging drop experiments were made by dissociating cells with 0.5% trypsin-EDTA (Gibco, Grovemont Cir, Gaithersburg, MD, USA). Cell density was estimated using a hemocytometer. Dissociated cells were centrifuged at 1200 rpm for 5 min at room temperature, re-suspended in growth medium and diluted to a final concentration of 12.5 cells/μl. A 50 μl cell suspension was dispensed into each well of the spheroid culture plate to achieve an initial seeding density of 625 cells/well. In order to prevent evaporation, 1% agarose was added to the peripheral reservoirs of the hanging drop plates. The growth media was exchanged every other day by removing 25 μl of solution from a drop and replacing with 25 μl fresh media into the drop.

*Spheroid preparation for SPIM imaging.*

4 days old T47D breast tumor spheroids were incubated at 37 °C with 1.5 μM DRAQ7™ (Biostatus, Leicestershire, UK) for 24 h prior to imaging. DRAQ7™ is a far-red membrane impermeable fluorescent DNA dye that selectively stains the nuclei in dead and permeabilized cells. Staining of the spheroid with the nuclear dye was performed by replacing 10 μl from each hanging droplet with 10 μl - 5x concentrated solution of the dye. Following, spheroids were transferred from the hanging drop plate to a microscope slide, washed twice with PBS and reconstituted in 100 μl CyGel Sustain (Biostatus, Leicestershire, UK) inside a cold room (4° C) to avoid rapid solidification of the CyGel. Then, the CyGel-embedded spheroid was transferred into a FEP tube (800 μm inner diameter, Bola, Germany) which was sealed with self-adhesive putty and then loaded on the SPIM instrument. The FEP tube containing the immobilized spheroid was embedded into a 37 °C water bath throughout the duration of the experiment to avoid liquefaction of the CyGel. The live spheroid was imaged in our custom SPIM setup using a diode laser for excitation (635 nm). The emission wavelength of the DRAQ7™ fluorophore peaks at 685 nm and, accordingly a 650 nm long pass filter was used to detect the fluorescence signal.

*Experimental SPIM-OPT Setup:*

All the images presented in this work were acquired with a combined SPIM/OPT setup, illustrated in **Figure IV-12**. It is composed of a custom single sided Selective Plane Illumination Microscope which is equipped with a LED illumination to perform Optical Projection Tomography as well. For SPIM various continuous wave diode lasers are being used. In this work, the output of a 635nm diode laser is used. The laser beam (colored with blue in **Figure IV-12**) is initially expanded (BE) and then is directed to a cylindrical achromat doublet (CL) through which it is focused in a horizontal line on the corner mirror (CM). After the mirror the formed light sheet is imaged through a 2x telescope (T) to the back focal plane of the illumination objective (IO) (Mitutoyo, Plan Apo, 10x/0.28, WD=34.0mm). The telescope is placed in such a way that two conjugate planes are formed on the mirror and the back focal plane of the objective, for a better and easier adjustment of the light sheet. The formed light sheet is established orthogonally to the detection axis, intersecting with the focal plane of the detection objective.

The emitted light (colored with green in **Figure IV-12**) is collected by a second 10x/0.28 detector objective (DO) (Plan Apo, Mitutoyo, Japan) and is projected through an apochromatic doublet tube lens (TL) (ITL200, Thorlabs) on a thermoelectrically cooled, electron multiplying CCD camera (1004x1002 pixels sensor, pixelsize: 8μm) (Ixon DV885, ANDOR Technology). Right after the objective an iris (ID) is placed in order to control the NA of the detection and thus define the depth of field and a filter wheel with appropriate fluorescence filters. For the DRAQ7 emission a 650 nm long-pass filter is used to acquire the signal. The sample is stabilized inside a FEP tube with a solidifying agent (CyGel), and then is mounted on the sample holder which has 4 degrees of freedom. Four motorized software controlled stages allow the micrometric translation along x, y, and z-axes and rotation around the vertical y-axis. For refractive index matching, the sample is inserted inside a chamber filled with water. In case wide field tomographic imaging is required instead of selective plane excitation and detection, the white LED Lamp is used to perform Optical Projection Tomography.

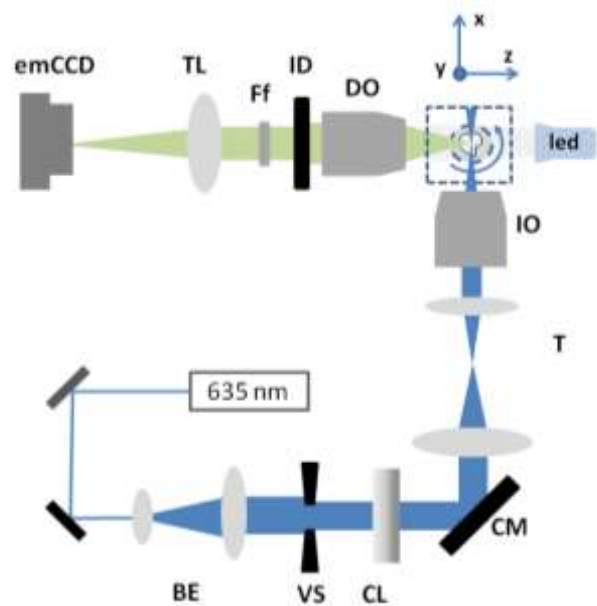

*Figure IV-12 | The combined SPIM-OPT setup. In blue the excitation line and in green the detection axis. The sample is placed into a bath (dashed square) and translated along the z axis to perform the SPIM scanning.*

*SPIM-OPT measurements.*

The experimental work presented here was entirely accomplished using the combined SPIM/OPT setup just described, in which the sample position can be software controlled along the three spatial directions and rotated along an axis perpendicular to the camera detection plane by motorized stages. The sample was imaged acquiring 180 Average Intensity Projections (AIP, separated with an angle step of 2°) in order to complete a full rotation. At each projection, the sample was scanned through the light sheet, having full width at half maximum of 7 μm, in steps of 20 μm while continuously recording with the camera.

For illumination, we used a continuous wave 635 nm diode laser. The light sheet was shaped by cylindrical optics and then was introduced vertically to the detection axis, having its central plane inside the focal plane of the 10x/0.28 infinity corrected detection objective (Mitutoyo, Japan). Finally for image acquisition a tube lens and an electron multiplying CCD (Ixon DV885, Andor Technology, Belfast, UK) were used. The camera has a resolution of 1004x1002 pixels and pixel size of 8 μm. The resulting pixelsize of the imaging system was 0.8 μm.

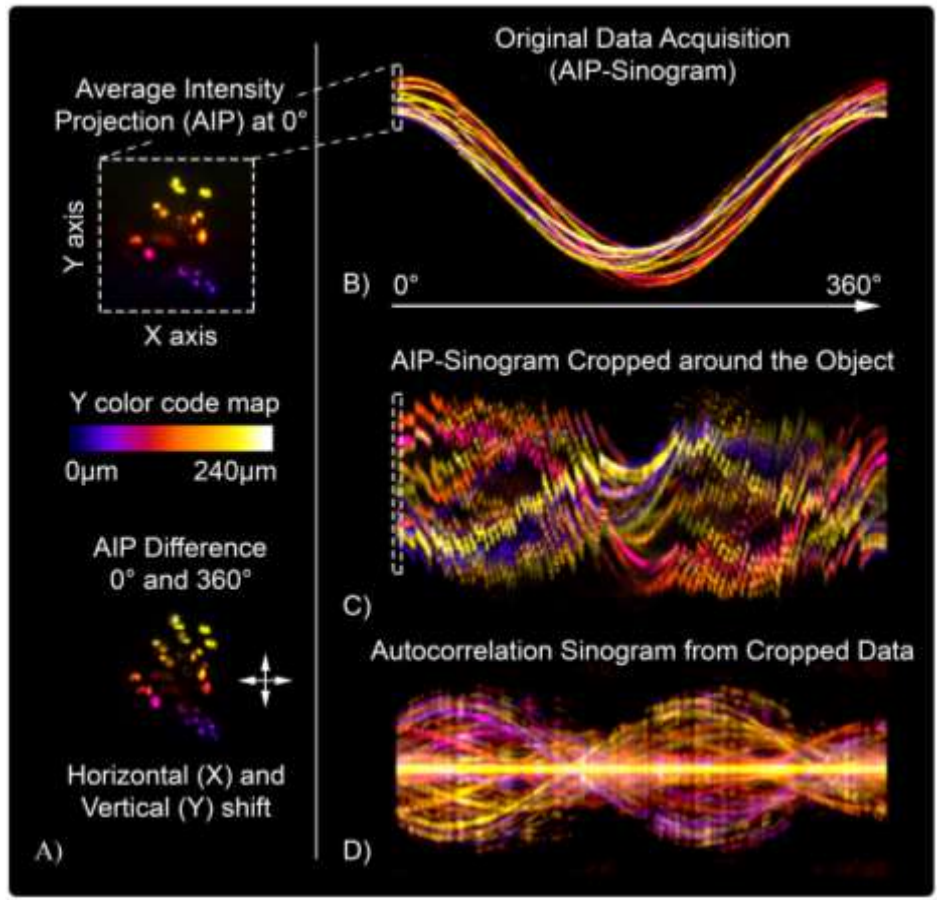

*Figure IV-13 | Sinogram based analysis of the acquired dataset. A, AIP at 0° and relative difference after a full 360° rotation. The bidirectional drift of the sample during the measurements is visible on the bottom of panel A, and implies a misaligned AIP-sinogram. B, Original sinogram of the SPIM-AIP measurements as a function of the angle, the color code that labels the Y-axis depth is the same as in panel A. It is worth noticing that the color of the sinogram turns toward blue-red at the end of the rotation, which means that the spheroid is slightly moving towards the bottom of the FEP tube. C, Cropped sinogram around the spheroid used to reduce the size of the reconstructed volume. It is worth noticing how the data is completely misaligned. D, Aligned autocorrelation sinogram calculated by using the data from the sinogram in C. The alignment of the dataset is achieved by simply calculating the autocorrelation for each AIP projection and stacking of the projections one after the other.*

### 3D Autocorrelation.

For each SPIM dataset at different angles, we calculated the Average Intensity Projection (AIP) of every frame. The images were cropped with a squared window of 300 pixels (field of view of 240 µm) containing the whole spheroid fluorescence signal. This results in a vibrational sinogram, impossible to backproject with standard approaches (**Figure IV-13** panel C). For each of the cropped AIP we calculated its autocorrelation with the Wiener-Khinchin theorem, stacking all of them in the same order. We thus obtain the autocorrelation sinogram (A-sinogram) which is always aligned. This data was backprojected with the inverse Radon transform function in MATLAB (using the default Ram-Lak filter), obtaining a cubic volume with side of 599 pixels. This volume is the three-dimensional autocorrelation of the specimen of interest. No further post processing of the data was performed.

*Phase Retrieved Tomography.*

The three-dimensional autocorrelation was used as an estimation of the Fourier modulus of the object to reconstruct, associating a random initial three-dimensional phase as starting point for the Phase Retrieval problem. The reconstructing window within the autocorrelation volume had the size of the object. A mixed Hybrid Input-Output (HIO) approach was used for 5000 steps followed by 1000 steps of Error-Reduction (ER). The program was implemented in MATLAB with GPU-CUDA extension and typical running time for the reconstruction was about one hour.

## IV.2.4 RESULTS

Once we have defined the methodologies and verified numerically and theoretically that our technique is able to retrieve a correct reconstruction from the calculation of the three-dimensional autocorrelation, we can move on to examine the experimental results that we obtained. As we said, we measured the fluorescence distribution of necrotic cells expressed in a human tumor spheroid, which we will present and analyze in the following, explaining how our technique has the potential to correctly reinterpret the measurement in such challenging experimental conditions.

### SPIM/OPT Phase-Retrieved Tomography

Employing PRT, we uniquely imaged the fluorescence distribution of the cell-death marker DRAQ7™ in a T47D human breast tumor spheroid with a diameter of about 200μm. A schematic of the experimental measurements performed in a combined SPIM/OPT setup is illustrated in **Figure IV-8**: per each rotation angle, we acquired a collection of tomographic slices by illuminating the sample with a light sheet perpendicular to the camera plane. We placed the specimen inside a Fluorinated Ethylene Propylene (FEP) tube and we measured it over a complete rotation, making sure the whole sample was scanned during each SPIM acquisition. In case the spheroid exited the field of view while rotating, we brought it back to the center by translating along the detection focal plane (x-axis), without being concerned for such a displacement. For each SPIM dataset, at every angle of rotation, we calculated the Average Intensity Projection (AIP), stacking them one after the other as it would have happened in an OPT experiment (AIP-sinogram).

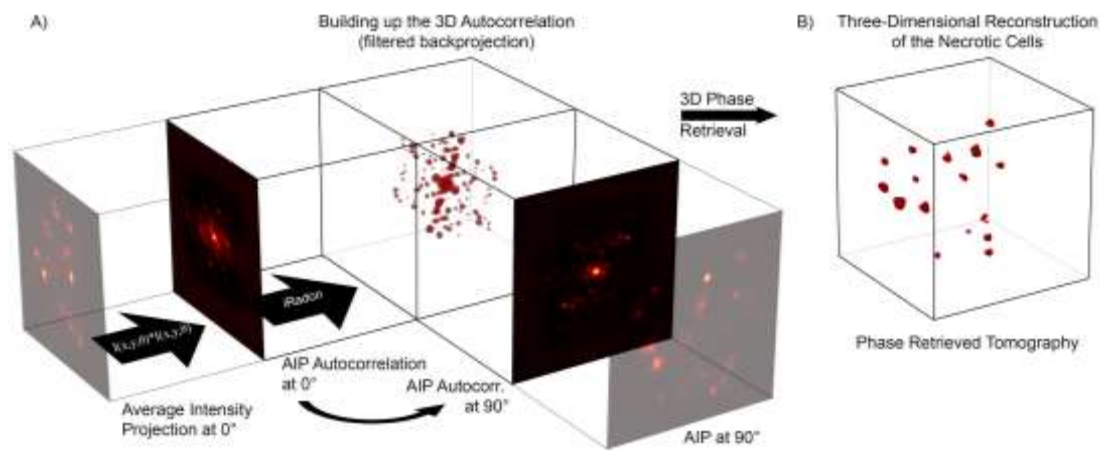

*Figure IV-14 | Schematics of the PRT approach. A, schematic showing the backprojection criteria. For each of the Average Intensity Projections (AIP) coming from every angle (Fig. 1) we calculate the autocorrelation images. These images are smeared along a volume in function of their angle of view, following the backprojection criteria of the filtered Inverse Radon transformation. The result of this is a volume that contains the three-dimensional autocorrelation information of the object we want to image. It worth to notice that the autocorrelations are always peaked in the center and symmetric respect this point. In panel B, the final reconstruction after the Phase-Retrieved Tomography (PRT). The 3D autocorrelation feed a Gerchberg-Saxton algorithm, which retrieves the phase information reconstructing the object with no artifacts due to misalignment.*

*Mesoscopic imaging of tumor spheroids*

The measurements were performed in the mesoscopic regime of light scattering, as can be observed in **Figure IV-9**. In fact, the sample is non-uniformly absorbing at visible wavelengths (bright field image in **Figure IV-9** A) and it is large enough to scatter the fluorescent light emitted at the side opposite to the camera detection line. Indeed, in **Figure IV-9** (panels B-E) it is notable that groups of fluorophores are highly blurred at certain angles, while these become evident after a rotation of 180°. Therefore, in these circumstances, it is crucial to exploit all the information coming from a full rotation of the sample to accurately retrieve all hidden fluorophore distributions. Additionally, in this regime of scattering and for such a small specimen, a classical sinogram-based reconstruction approach is strongly sensitive to mechanical vibrations and off-axis rotations. The sample drift is clearly visible when comparing the projections at 0° and 360° (bottom of **Figure IV-13** A) and yields a misaligned AIP-sinogram (**Figure IV-13** B). With our stage, the total diagonal drift after a complete rotation was 18.7 μm, 17.6 μm along the horizontal and 6.4 μm along vertical directions in the camera image plane, which is about 10% of the whole diameter of the spheroid. We believe that the two directions of the drift were the result of the combination of inaccuracy of the rotating stage and a slow gravity-induced fall of the sample in the CyGel environment. In this difficult scenario, using simple camera projections (AIP) will turn into a misaligned sinogram, which would have to be carefully post-processed in order to correct for such displacement error. In our experiment instead, we calculated the 2D autocorrelation of each AIP, stacking them in the same order and assuming the rotation step as known. The result of this calculation is a new autocorrelation sinogram (A-sinogram) which is always perfectly centered, regardless of where the spheroid was in the camera plane (**Figure IV-13** D). The simple backprojection of the A-sinogram, via inverse Radon transform, leads to the calculation of a volumetric dataset corresponding to the three-dimensional autocorrelation of the sample shown in **Figure IV-14** A. We prove this theoretically, by showing that the autocorrelation of the projection of an object is equal to its autocorrelation projected at the same angle. Moreover, we validated numerically this assumption by calculating the 3D autocorrelation of a commonly used test-object, a three-dimensional Shepp-Logan phantom, comparing it with what found after the backprojection of its A-sinogram. Then retrieving the phase information from such a volume, i.e. the three-dimensional autocorrelation of the real object, represents a typical phase retrieval problem with higher dimensionality [34], here for the first time employed for tomographic purposes.

The complexity of the algorithm poses computational challenges, which are substantially related to three-dimensional Fourier Transformations, and can be easily faced via parallel GPU implementation. In our specific case a normal GPU nVidia, 3-years old GeForce 780Ti with 2880 CUDA cores, could handle a volume up to $300^3$ voxels in direct space before running out of memory, a technical limitation that can be easily fixed in future algorithm developments. The result of the PRT imaging reconstruction is the volume shown in **Figure IV-14** B, which is the indirect reconstruction of the DRAQ7™ fluorescence distribution within the tumor spheroid.

*PRT validation*

To validate the PRT reconstruction we compared it with classical OPT reconstruction and multi angle SPIM measurements. **Figure IV-15** D-F shows quantitative results for the PRT technique by visualizing the color-coded AIP (hyperstack, in which the color indicates depth) of the whole dataset (from **Figure IV-14** B) seen from different angles. The classical Radon transform of the sinogram of **Figure IV-13** panel B, co-registered by finding its CoR, leads to inaccurate reconstructions (**Figure IV-15** A-C) in which we cannot distinguish any single necrosis, clearly demonstrating that classical approaches are not usable for this imaging regime. Similar low performance is achieved with stand-alone SPIM (**Figure IV-9** B-E), producing blurred images for groups of fluorophores situated deep inside the spheroid (underlined by dashed circles). Instead, PRT imaging, has achieved the retrieval of highly detailed fluorescent distributions (**Figure IV-15** D-E) with single cell resolution, visualizing even the cells that where blurred in SPIM measurements at various projection angles (dashed circles). The correct coloring sequence of the dataset, hyper-uniform stack visualization using the Fiji toolbox [118], efficiently displays correct distances between fluorophores compared to those of SPIM and OPT. This is a natural consequence of the correct data backprojection in the autocorrelation domain, which does not need any alignment procedures. The phase retrieval algorithm always returned comparable reconstructions, regardless of the initial and random starting guess for the phase, thus making the PRT results more efficient than the currently used techniques. Interestingly, PRT could also be used for the restoration of previously acquired datasets, which did not lead to any proper imaging due to system misalignments. We therefore believe that PRT can enter convincingly in the current biomedical imaging scene, in particular as an efficient tool for correct data reinterpretation in highly sensitive measurements.

## IV.2.5 DISCUSSION.

With this work we have presented, for the first time to our knowledge in biomedical imaging, the possibility to perform a three-dimensional autocorrelation reconstruction in combination with a phase retrieval algorithm to radically improve mesoscopic three-dimensional tomographic imaging of opaque biological specimens. We have validated our method on a human tumor spheroid successfully imaged in the mesoscopic regime by using a combined SPIM/OPT setup. In fact, we refer to this innovative approach as Phase-Retrieved Tomography (PRT) since it retrieves the phase related to the autocorrelation, and through the Fourier modulus, that of the entire object. The new approach is insensitive to specimen translational misalignment and stage drifts in all three spatial directions, only requiring prior knowledge of the rotational degree to correctly back-project the autocorrelations sinogram. We proved this theoretically and numerically, showing with a 3D Shepp-Logan phantom quantitative and robust reconstructions, even perturbing the sinogram with random shifts in two directions in the camera plane. Although the position of the retrieved object within the reconstruction volume is random, because of phase retrieval ambiguities [34] its signal distribution is always consistent with the real object, which implies that reconstructions are not affected by data collection misalignments. The experiments confirmed the same behavior and we have achieved exceptional results by collecting AIPs of full rotation in steps of 2°. Consistent reconstructions can also be achieved with bright field illumination in a classical OPT approach, as the spheroid depicted in **Figure IV-8** was in fact retrieved with PRT in rear bright field illumination, demonstrating the flexibility of the method with different acquisition techniques.

Of more interest are the results obtained by PRT which fully exploit the SPIM/OPT setup. To date, one of the disadvantages of SPIM is its poor resolution along the perpendicular direction with respect to the planar illumination (axial resolution). This can be improved with light sheet deconvolution [119] [120] or co-registration of stacks at different angles [121], both of which enhance the resolution along the third direction of the reconstruction, but share the disadvantages of being highly specific to each set of measures and sensitive to drifts. On the other hand, in stand-alone OPT, accurately recovering the Center of Rotation (CoR) of the specimen is still an unsolved problem. Although many approaches have been proposed for data post-alignment [108], they fail to provide a generalized method for removing artifacts due to misalignments. Our method manages to overcome all these issues by retrieving the object from several SPIM-AIP autocorrelations rather than direct projections, which allows the obtained autocorrelation sinogram to be inherently aligned (**Figure IV-13** panel D) and rotate around its center of symmetry, regardless of the original object's rotating axis position. With our method, the need to estimate CoR or RoI from the sinograms becomes redundant, leaving the user free to focus on pure, single measurements instead of aligning the rotational system and the acquired dataset. It is worth noticing that the autocorrelation of the projection at 0° always exactly matches the one at 360° even if the sample, while rotating, does not return to the original starting position. The A-sinogram is used to compute the object three-dimensional autocorrelation, borrowing this ability from the classical OPT approach. In such a way, it is possible to also overcome the OPT weakness, i.e. the error in tracking the object while rotating, fulfilling, thus, the inverse Radon transform requirements and leading to the calculation of a nearly exact three-dimensional autocorrelation. Moreover, other iterative inversion techniques, such as ART, SIRT, and SART [122] can be straightforwardly implemented for the autocorrelation reconstruction, offering further room for improvement and applications for PRT.

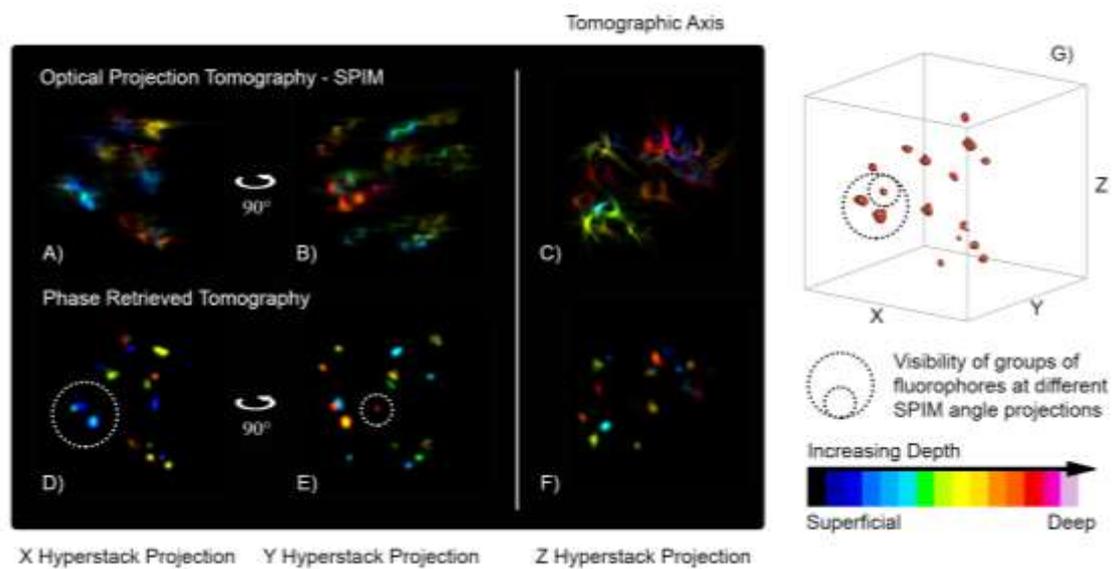

*Figure IV-15 | Imaging performance of the PRT technique. Hyperstack projections for the SPIM-OPT measurements (panels **A, B, C**) and the PRT reconstructions (panels **D, E, F**). The color code denotes the depth at which the fluorophore is located, starting from the first signal. It is important to notice that the two methods share the same coloring order and the same fluorophore distribution, validating our reconstruction. Moreover, PRT shows a fluorophore while SPIM cannot resolve it, due to its location on the other side of the spheroid (red to white in the color scale). The Z hyperstack suggests comparable resolutions between the two methods along the tomographic axis. Finally, panel **G** locates in space the groups of fluorophores dashed in panel **D-E**, showing that the small circle includes a cell that belongs as well to the bigger one.*

Quantitative results are presented in **Figure IV-15** where the PRT reconstructions are compared with classical OPT and where the color code represents the relative depth with respect to the first fluorescent signal. The artifacts due to misalignment of the data (**Figure IV-15** A-C) disappear in the PRT reconstruction, which convincingly demonstrates its tomographic capabilities, eliminating the need for data post-alignment (**Figure IV-15** D-F). SPIM hyper-stacks at four perpendicular angles as shown in **Figure IV-9** can also be compared to PRT reconstructions viewed from the same angles (x and y projections in the volume coordinates). The color-coded depth of each fluorophore validates the results of the PRT, which exhibits an enhanced and uniform resolution compared to that of SPIM. This is clearly demonstrated by the SPIM-AIPs underlined by dashed circles: in the mesoscopic regime, SPIM cannot resolve in all projections the presence of groups of fluorophores located on the opposite face of the spheroid, resulting in blurring and reduced non-uniform resolution. PRT, instead, clearly retrieves these objects (dash circles in **Figure IV-15**) by fully exploiting the information coming from multiple angle AIPs and exactly combining them in autocorrelation space for enhanced depth resolution. The novel method proposed here is beneficial in terms of correct reconstruction of relative depth and allows number estimation of the fluorophores distribution, important biological parameters to study tumor growth, cell clustering, viability and proliferation. It retrieves intensity distribution of cells with uniform resolution in three dimension, and in addition, it is robust and easy to implement in a regular SPIM/OPT setup. Furthermore, it can be applied to the regular OPT approach using a variety of illumination schemes (bright field, fluorescence) and can also be extended to other projection based approaches, such as Xray-CT or PET. In those cases, the entire specimen would have to lie within the camera's depth of field, otherwise defocused images would be acquired and used to calculate the autocorrelations. In our scanning approach, the plane illuminated is always in focus and its AIP is blurred only because of internal scattering, allowing, thus, better reconstructions. Finally, our method can accept further improvements in terms of effectiveness of the Phase Retrieval algorithms used which could potentially enhance even more its accuracy and its convergence rate, pushing the community towards developing faster three-dimensional implementations. At the moment, in fact, we use a common combination of Hybrid Input-Output (HIO) and Error Reduction (ER) retrieval methods but other choices can be explored [34] to efficiently treat the reconstructed 3D autocorrelation, such as sparsity-based prior information [123] or the interesting Oversampling Smoothness (OSS) method [47] to overcome noisy measurements or backprojection artifacts.

In this scenario, it is worth taking into account another key aspect of this work. Because of its design, the PRT protocol can potentially be implemented for imaging hidden three-dimensional specimens behind scattering curtains or around corners. Optical Imaging in such extreme conditions has been performed only with two-dimensional objects [31] [33] by exploiting the speckle's autocorrelation property which, within the memory effect [26] regime, is identical to the autocorrelation of the object itself. The conceptual idea behind these studies is based on the fact that the autocorrelation of the signal is preserved while being scrambled by the scattering media within a certain range. The produced speckle retains the autocorrelation properties of the object and it relates to it through the Fourier Transform (FT) modulus. The autocorrelation of the speckle produced in front of the turbid layer can feed a Gerchberg–Saxton algorithm [48] used to retrieve its Fourier phase, allowing the reconstruction of the hidden object. Although mathematically the phase retrieval process works particularly well at every dimensionality [34], except for the lack of uniqueness in 1D problems, for optical imaging purposes (to the best of our knowledge) it has been used only in 2D implementations. In principle, PRT can tackle the current lack of techniques for high-resolution 3D imaging of hidden objects, by exploiting speckle acquired at different angles rather than single projections, and performing accurate reconstructions of the embedded object following our proposed methodology. To conclude, we believe that with this work we have opened a new direction towards biomedical imaging inside diffusive media by virtue of the use of autocorrelation implemented in a tomographic fashion. A variety of experimental and algorithmic possibilities emerges from this study, such as the effectiveness, accuracy and reliability of 3D imaging of hidden objects, and the application in a wide range of biological applications. Our immediate target is to address these considerations and further apply phase-retrieved based tomography for biomedical imaging and beyond.

# DISCUSSION &
# FUTURE PERSPECTIVES

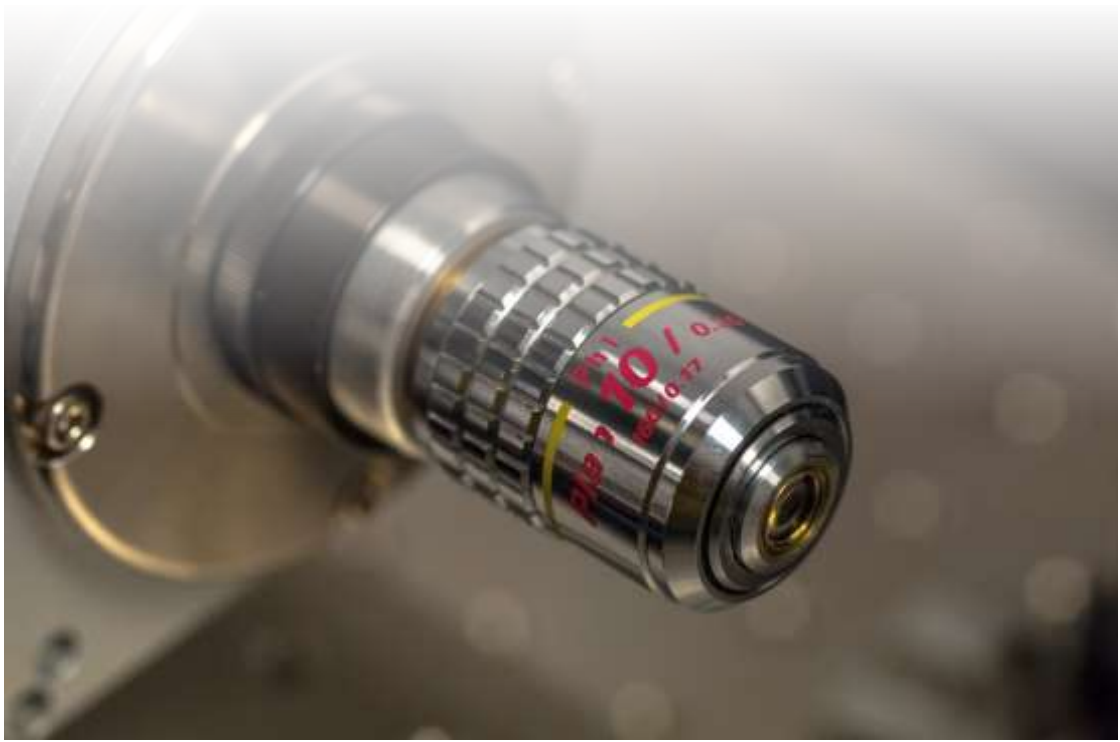

So far and throughout the whole text, we have examined extensively the main work carried on during this doctoral period. As discussed in details, the main research activity was focused on the study of light propagation through biological media, complemented with interesting excursus in the field of disordered photonics and turbid lens imaging. A quite comprehensive picture arose from this study, in particular for what concerns the biomedical imaging in the extreme scenarios described in the present text. On one side, we have discussed the case of optically transparent layers enclosed in highly scattering regions and, on the other hand, how it is possible to perform hidden tomography in highly scattering layers by recovering the phase connected with the autocorrelation of a seemingly information-less speckle pattern. The latter, made us reflect on the problem of the alignment of projection-based reconstruction schemes, that we cleverly approached in the autocorrelation space, thus resulting in the new Phase-Retrieved Tomography technique.

In the first one, we pointed out how the common trend of neglecting clear layers for imaging with Diffuse Optical Tomography affects the reconstruction ability. The Diffusive Equation used for the improving computational efficiency, suffers from too strict approximations if we aim at high-resolution neuroimaging, where the presence of the optically transparent Cerebral Spinal Fluid surrounding the brain makes the light-diffusion assumption to fail. If we aim at high resolution reconstruction in fact, we should make use of the more accurate Monte Carlo Photon Propagation to model the forward problem of the photon diffusion through the specimen of interest. A more accurate forward modelling in fact, will lead to a more efficient inverse fluorescence reconstruction as accurately pointed out through the whole *Chapter III*. We have shown this by approaching the simulations in a novel fashion, reproducing a virtual fluorescence DOT experiment, in which we tried to image fluorescent distribution (spherical tumor) located in different position within the brain. A previously existent Digimouse atlas model was re-segmented in order to introduce the optically clear layer surrounding the brain and then used for the simulations. We have discussed how the structural features of the CSF layer in a realistic case plays a very important role, comparing it with the results obtained in the case of more simplistic planar layer geometries. We pointed out that, although for planar layered structures neglecting the CSF could constitute still a good choice, when more complex atlases are used its structural geometry turns the diffusion into a complex range of differences at camera detection level. We have extensively characterized those differences, pointing out how it is not possible to normalize them, due to the fact that they also depend on the position of the fluorescent source within the brain. The use of the MC-PP, then, is strongly suggested in the case of diffused optical neuroimaging, in particular nowadays that modern GPU computation approaches the world of computed imaging techniques. MC methods, in fact, start to become a reasonable choice in terms of speed, reliability and the need of more accurate forward sampling, thanks to the parallel computational power offered by the use of modern graphic cards.

On the other hand, *Chapter IV* focused on the achievement of hidden three-dimensional tomography via the usage of phase retrieval methods applied to the autocorrelation of the signal detected. We already have discussed how hidden two-dimensional imaging was recently performed [31] [33], but the lack of a tomographic methodology pushed us to develop a complementary approach. We experimentally tested the method on a bi-dimensional sample, achieving hidden reconstruction only exploiting the information contained in the speckle pattern produced in front of the scattering layer. Then we created a three-dimensional virtual sample, that we numerically hid by scrambling its phase and taking into account the theory behind the memory effect. Optical memory effect, in fact, rules the possibility of hidden imaging and defines its reconstruction capability. We pointed out that hiding the object within this regime correspond mathematically to the convolution of the projection of the object with a random speckle pattern ideally generated by a point source scrambled in phase. This enabled us to reproduce a virtual hidden three-dimensional tomographic experiment that we used to test our reconstruction scheme. In fact, running independent PR would have not allowed the reconstruction of the hidden sinogram which is at the base of a correct tomographic reconstruction. This is due to the fact that the phase retrieval is an inverse problem that admits unique solutions only if we relax the positional information of where the object was: the same object located at different places in space would return exactly the same autocorrelation. We discovered instead, that by feeding the following PR algorithm with the previous reconstruction of the object would have returned a very nicely aligned sinogram, thus making us able to perform the hidden reconstruction without having any positional a-priori information. This important result made us reflect about the problem of the alignment of projection measurements, in which an unknown spatial drift would compromise the reconstruction abilities. By paying attention on the mathematical property of the autocorrelation to be always intrinsically centered in its space of definition, we proved that the inverse Radon transform of the projection's autocorrelation sinogram (instead of the projections itself) would have returned the three-dimensional autocorrelation of the specimen, without being affected by any misalignment. We proved this mathematically, numerically and experimentally, in which we have shown the possibility of correctly performing tomographic fluorescence reconstruction with absolutely no care for the alignment of the measurements. Even in this case we made an intense use of computational GPU paradigms, which allowed faster implementation for the new three-dimensional PR implementation.

Indirectly, we could say, we have examined in detail how the increasing power of modern computation paradigms influences convincingly the ability to perform more accurate imaging. It is worth noticing how everything we have presented in this thesis would have been impossible to perform with normal desktop solutions, say for example, 5 years ago. Parallel computation, before only confined to highly expensive cluster of processors or huge supercomputers, is getting available also for portable computer solutions. Everything we have been discussing so far, in fact, was run on normal gaming GPUs: the architecture of modern graphic cards greatly suits the needs for matrix operations and trivially parallelizable tasks. Monte Carlo Photon Propagation, in fact, is based on statistically independent photon runs, which makes it readily parallelizable, while the intensive use of Discrete Fourier Transformation behind the phase retrieval methods suits very well the matrix-like arrangement of the calculating cores of the most recent GPUs. It is not difficult to realize, then, why GPU parallel computation paradigms are convincingly entering the medical imaging scenario. The need for fast and complex reconstructions, approaching real-time volumetric imaging, the compactness of normal desktop solutions and its inexpensiveness naturally finds its location to be easily moved from the bench to the bedside. In the works carried on during this period, the use of graphic cards rather than CPUs speeded up the programs of at least 100 times and at the moment we write this thesis, the GPU that carried the most of the work is three years old. Numerous other improvements have been happening in the meantime, for what it concerns GPUs and code optimization, which makes potentially even faster the usage of the more recent graphic solutions.

Mostly for what it concerns the phase retrieval, the current will is to fully exploit the power of the GPU computation. Numerous code optimization can be implemented, not only regarding the type of PR approach to be used to achieve fast convergence, but in particular for what it concerns more efficient memory usage which in fact limits the size of the three-dimensional DFTs. At the moment, the memory of the GPU limits the maximum size of the reconstruction volume up to 600 pixels, which actually means 300 pixels of effective reconstructing region due to the oversampling requested by the autocorrelation. Currently, in fact, the code has been optimized and made faster by a 5x factor, compared to the first working release, by only fine tuning the memory usage. Not only these mostly technical tasks appeared in the to do list while working at these projects, but also more interesting research paths opened their ways starting from these considerations. We strongly believe, in fact, that the Phase-Retrieved Tomography [124] can find broad range of applications in the biomedical imaging field. Numerous methods can take advantage from this alignment-free technique, such for example bright field structural Optical Projection Tomographic measurements, X-ray computed tomography and hidden imaging itself. Although the idea behind the PRT was originated by working with hidden modalities [86], in fact, it was not used so far for hidden reconstructions. We are currently going to tackle this in a near future work, in which the creation of a specifically designed fluorescent sample hidden into an egg-like scattering layer will be used to prove what we believe to be achievable from numerical simulations. Moreover, as we said, PRT could also aid other projection-based techniques, such as bright field transmission OPT or X-CT: the advantage in this case would be clear, we do not have to care anymore about alignment of the detection scheme and neither on eventual mechanical misalignment. Nevertheless, there are hidden challenges in order to accomplish these tasks and some of them have been already individuated. Bright field OPT measurements, in fact, is well known to produce quite noisy measurements in particular in the region of the sample background, and the PR algorithms used so far are known to not behave well in noisy measurements. Moreover, the autocorrelation calculation requires a dark signal background, a requirement that is not satisfied neither in OPT, nor in XCT measurements. In this scenario, preliminary tests showed that a good approach could be constituted by the usage of the PR-Oversampling Smoothness method applied on grayscale-inverted measurements (in which the grayscale image inversion turns bright regions into dark). Further investigations are going to be carried on the near future, with the specific aim to further extend the applicability of PRT-based techniques.

Even more interesting can be the implications behind the study on the effect of clear layers in through-skull light propagation [49]. We have extensively discussed, in fact, that the signal transmitted through the head still retains strong information on the presence (and also on the thickness) of the transparent Cerebral Spinal Fluid layer. More interestingly, these differences are also connected with the thickness of such layer, how was pointed out while answering to one of the Reviewers' comment. The fact that the measurements are so strongly affected by the size of this clear region seems to weaken the message we wanted to send: in fact, to correctly model the light-tissue diffusion we have to know very accurately the anatomy of the specimen, which is an important a-priori information. However, this can be used at our advantage, because the light diffused still contains a precious information that can be cleverly exploited. In fact, we can think of approaching a new kind of measurement, aiming this time to obtain an estimation for the size of the transparent tissue even in the case of human head, for example making use of structural Near Infra-Red Spectroscopy (sNIRS) measurements, by only detecting a non-invasive light signal propagated through the head. This can have a significant impact, it is well known that Alzheimer's disease and other dementia-related phenomenon are always connected with strong neuronal loss, effectively caused by brain shrinking and increased quantity of the CSF as a function of the disease progression. Indirectly then, measuring the thickness of the CSF layer could aid the medical research by making possible to follow in time the disease evolution, something that can currently be studied only with more expensive, slow and risky repeated XCT measurements. Regarding this specific point, we have already accomplished a full set of Monte Carlo simulations and created a virtual atlas that models the evolution of a hypothetic patient affected by the AD disease. From preliminary numerical simulations, we found very interesting and promising results, confirming that it is somehow possible to effectively estimate such CSF thickness by only looking at the time resolved detector response.

In conclusion, then, it is important to recall how remarkable is the contribution of numerical methods in optical biomedical imaging at every level of usage. Simulations, feasibility studies, tomographic reconstructions, phase retrieval methodology cannot prescind the use of computers for correct reinterpretations of the mere signal detected. We do believe then, in parallel with the advances in experimental methodologies and the clever design of advanced laboratory setups, that accurate numerical methods will definitely play the most important role, in order to solve with the highest possible accuracy every optical imaging problems that we are going to encounter along our research path.

# Acknowledgements

So finally here we come, after three years that simply flied away in less than a snap, it is about time to wrap up and unleash all my immense gratitude to who- and what-ever contributed directly and indirectly to such an experience. It is indeed difficult to start thanking someone before the others, because it might sound as a preference among all the fantastic people I have met while my stay in this amazing Island. I will be impartial then, so I will start with Giannis, from whom I ultimately owe my presence here. I was somewhere else, doing something else but you trusted me until my last word that day of April, a thousand days ago. I remember it like it was yesterday, I was at the same time excited but also scared, my first shot simply didn't work out and I was not aware of what exactly was going to happen next. But here I found everything I needed, no what am I saying? More than what I was expecting! I found a supervisor from whom I have learned a lot, and not only scientifically, but also all of those shades that smoothly blended my working experience with Life. Coming to your lab every day for me was never ever a burden, I felt home, welcomed, surrounded by friends, not competitors. You gave me the opportunity to let me play the real game of being a scientist, something priceless that I believe not easy to learn in other places, where one might be cornered with his wings tied. Thank you Gianni, without you I might have left science long ago. Catching the occasion, I want to thank the whole IVIL group, starting from Thanasiss with whom a laugh was always behind the corner and to Stelio, your contagious serenity extinguished every possible wildfire. And to Vagelis (Liapis), which was the best officemate ever and that I miss so much; to the sparkling Stella, with whom I shared such funny moments and to Marilena, always ready to listen my complaining even if we sometimes misunderstood each other. To Giorgos (Tserevelakis), a friend that I admire not only as a scientist but also as a person. To Georgia (Giasafaki), my beloved student to whom I asked the impossible and every single time the mission was accomplished. It was a pleasure for me teaching you whatever I could, good luck with whatever the life reserves for you Gio'. To Krystalia, so patient to spend so many hours listening to my terrible stories in the lab, I believe sometimes more boring than the measurements themselves, to Ilias for having with me the best laughs ever and to Asier, that apparently does not want to leave our lab!

A "Mention of Honor" goes to Giorgos (Tsibidis) that with his sharp, unexpected, genial quips made me always explode in the loudest lough. Thank you for the best working breaks ever! Another big thank goes to Maria (Dimitriadi), the kindest person I have ever met in the world as well as the most efficient. Last but not least, the biggest hug goes to Diego, friend, colleague, confident, house and office mate, rear face of my same coin. Together we couldn't see any limit, maybe because sometimes they were so close that we didn't just care and passed over! I miss you amico mio, I wish we get close once again. Ah! I almost forgot, if something goes wrong it is my fault, yes, but you gotta fix it for it! Finally, I want to save another enormous hug to all my friends I've met along this path. I shoot in random order: Luca, Veronica, Vincenzo, Stefano, Graziano, Kostis, Tryfon, Giannis, Chiara, Erieta, Antonio, Giulia and all the others that I cannot write for shortness that with their irony dressed up this period making it the happiest in my life. Thank you guys, for sharing with me all those beautiful moments, but now let's be serious, where do we eat tonight? And how could I forget about my Cretan family? Giorgos, Maria and Vagelis, to you all goes my warmest hug, for letting me in your house with no hesitation, celebrating festivities as part of your family or hanging out in normal days in front of a cold raki. You have been always lovely with me and I will save your smiles always in my heart. Another special thank is for my Mum, Dad, Sister and Grandma, that regardless of the distance they finally made it to come here all together just to do not understand anything at my final speech. I unconditionally love you all. I want to thank also all the rest of my family, that could not assist but I always carry with me in whatever I do. My sweetest hug goes to Sara, always next to me even when awake until very late at night (like now). Without your patience and kindness I really could not make this happen. Preciosa. And to all my friends that I left in Italy (or around), I wish I could have dragged you all here! To tell you the truth, with hindsight, my desires were finally heard: Thank you Emiliano, I won't forget the surprise you made me by hopping into these days!

In the end but not of less importance, I want to thank Prof. Antonio Pifferi and Prof. Jorge Ripoll for hosting me at their labs, sharing with me their knowledge and skills. Both of the periods in those universities (Dipartimento di Fisica at Politecnico di Milano and Department of Bioengineering and Aerospace Engineering of Universidad Carlos III Madrid) were simply amazing experiences, where I met many fantastic people that I wish to meet again. I want also to kindly thank Prof. Maria Kafesaki for being my scientific supervisor and have fully supported my PhD and Prof. Chrysoula Tsogka for the inspiring discussions we had.

Lastly, I want to thank Crete, which with its amazing landscapes gifted me with an everyday purest feeling of peace.

This work was supported by the grants "Skin-DOCTor" (1778) implemented under the "ARISTEIA" Action of the "Operational Programme Education and Lifelong Learning" co-funded by the European Social Fund (ESF) and National Resources, the EU Marie Curie ITN "OILTEBIA" PITN-GA-2012-317526 and the H2020 Laserlab Europe (EC-GA 654148). We also acknowledge support from the Institute of Molecular Biology and Biotechnology of the Foundation for Research and Technology – Hellas for the use of the cell culture facilities.

# Credits

The figure opening the first chapter for the theoretical background for the light diffusion in tissue shows the hand of Sara San Roman, in which the white light of the LED flash has experienced absorption and scattering before to reach the camera sensor. The picture was taken with a Nikon D3 with a Sigma 105 mm f2.8 macro lens by Daniele Ancora.

The figure opening the second chapter is a defocused image of the experimental setup of Diego Di Battista. It is possible to appreciate various speckle patterns propagating through several lens-elements of which the setup was composed. The picture was taken with a Nikon D3 with a Nikkor 50 mm f1.8 lens lens by Daniele Ancora and used as cover for PhD thesis of Diego Di Battista.

The figure opening the third chapter is a volumetric rendering of the newly segmented Digimouse head in light-white, with the inclusion of the CSF surrounding the brain shown as a light-blue structure. The rendering is accomplished by Daniele Ancora with the freeware ray-tracing software Voreen version 4.4.

The figure opening the fourth chapter is a bright field image of the 120-grit ground glass diffuser (Thorlabs) that we used in our experiments. The image was acquired with the illumination iris fully closed, to enhance the effects due to the random aberration (diffraction at different wavelength) introduced by the rough surface of the layer. The picture was taken by Daniele Ancora with a Nikon D3 connected to a Nikon Diaphot Phase-Contrast Inverted Microscope and using a 40X Nikon objective lens.

The figure in the closing chapter is a photo of the camera and objective lens used for hidden imaging reconstruction (Hamamatsu ORCA-Flash4.0 V2 Digital CMOS camera, Nikon CF plan 10x/0.30). The picture was taken with a Nikon D3 with a Sigma 105 mm f2.8 macro lens by Daniele Ancora.

All the figures, at exception of where explicitly referenced, where created by Daniele Ancora. **Figure IV-12** was realized by Stylianos Psicharakis.